\documentstyle[graphics,fullpage,11pt]{article}
\def\ci{{l}}
\def\cI{{\cal I}}
\def\cJ{{\cal J}}
\def\cO{{\cal O}}
\def\cM{{\cal M}}

\def\cF{{\cal F}}

\def\cL{{\cal L}}
\def\cP{{\cal P}}

\def\diag{\,{\mbox{diag}\,}}
\def\Diag{\,{\mbox{Diag}\,}}
\def\span{{\mbox{span}\,}}
\def\supp{{\mbox{supp}\,\,}}
\def\Vol{{\mbox{Vol}\,\,}}
\def\cof{{\mbox{cof}\,}}
\def\Tr{\,{\mbox{Tr}\,}}
\def\min#1{\,{\mbox{min}\,(#1)}}
\def\max#1{\,{\mbox{max}\,(#1)}}
\def\Pbar{\bar{P}}

\def\figscale{1.0}

\renewcommand{\thesection}{\mbox{\Roman{section}}\,}

\title{Multiresolution analysis of electronic structure: semicardinal
and wavelet bases}

\author{T.A. Arias\\
Department of Physics\\
Massachusetts Institute of Technology\\
Cambridge Massachusetts}

\date{\underline{{\em Reviews of Modern Physics}, in press for January 1999}.}

\begin{document}

\maketitle

\begin{abstract}
This article reviews recent developments in multiresolution analysis
which make it a powerful tool for the systematic treatment of the
multiple length-scales inherent in the electronic structure of matter.
Although the article focuses on electronic structure, the advances
described are useful for non-linear problems in the physical sciences
in general.  Among the reviewed developments is the construction of
{\em exact} multiresolution representations from extremely limited
samples of physical fields in real space.  This new and profound
result is the critical advance in finally allowing systematic, all
electron calculations to compete in efficiency with state-of-the-art
electronic structure calculations which depend for their celerity upon
freezing the core electronic degrees of freedom.  This review presents
the theory of wavelets from a physical perspective, provides a unified
and self-contained treatment of non-linear couplings and physical
operators and introduces a modern framework for effective
single-particle theories of quantum mechanics.
\end{abstract}

\tableofcontents 

\listoffigures 

\listoftables

\section{Introduction} \label{sec:intro}

The focus of this review is the application of wavelet theory to
the determination of electronic structure.  Wavelet theory has at
its foundation a single, simple idea: {\em multiresolution
analysis}\cite{mallat,meyer,meyer2}, a relatively recent and
mathematically rigorous theory of the description of functions which
provides simultaneously for a homogeneous
underlying description of space and the capacity to control and vary
the resolution of this description at will.

Interest in multiresolution analysis and wavelet theory has mushroomed
dramatically since their introduction.  Over seventy-two monographs
have been written on the general subject of wavelets within the last
six years.  (For a comprehensive review of the literature of the field
prior to 1993, see
\cite{book11}.)  In addition to several recent introductory
texts\cite{book2,book1,strang,book3}, specialized monographs are now
available which discuss applications to such wide-ranging fields as
chemical engineering\cite{book5}, bio-medical engineering\cite{book6}
and applied science\cite{book8}, as well as the traditional areas of
application in mathematics\cite{book7} and signal
processing\cite{book9}.  In terms of recent monographs, those of the
greatest relevance to the present discussion describe the application
of wavelets methods to partial differential
equations\cite{book4,book10}.

\begin{figure}
\begin{center}
\scalebox{\figscale}{\scalebox{1.25}{\includegraphics{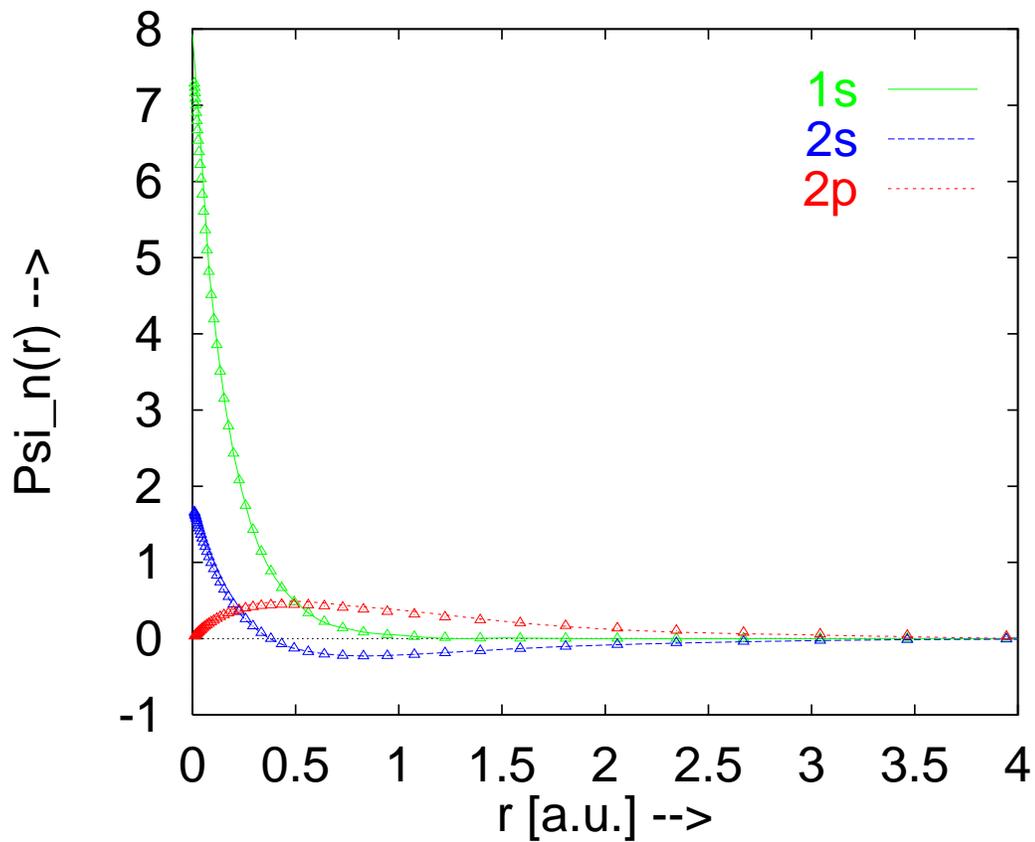}}}
\end{center}
\caption{Early electronic structure calculations using
multiresolution analysis\cite{mgras}: Kohn-Sham orbitals of the carbon
atom within the local density approximation from standard atomic
software (diamonds) and multiresolution analysis (curves).}
\label{fig:C}
\end{figure}

The well-known fact that the electronic wave functions in molecular
and condensed-matter systems vary much more rapidly near the atomic
nuclei than in interatomic regions calls for precisely the
capabilities of multiresolution analysis.  Figure
\ref{fig:C} illustrates the multiscale behavior of electronic wave
functions, using the carbon atom as an example.  The curves in the
figure show the Kohn-Sham orbitals of the atom as computed within the
local density approximation\cite{KohnSham} to density functional
theory\cite{HohenbergKohn}.  In the immediate vicinity of the nucleus
and its strong attractive potential, the electrons possess large
kinetic energies, as reflected by high spatial frequencies evident in
the orbitals.  In this example, the high-frequency ``core'' region
extends only approximately 0.5~Bohr radii out from the nucleus in the
case of carbon, beyond which the variations in the wave functions are
quite smooth.  Resolving the cusps in the $s$ states of this atom
requires a resolution on the order of 0.03~Bohr (corresponding to a
plane wave cutoff\cite{bible} of nearly 160,000 Rydberg).  To provide
this resolution uniformly throughout a computational cell of 8~Bohr on
a side would require a basis with 16~million coefficients.  The vast
majority of these basis functions would be wasted as they would serve
to provide unnecessarily high resolution outside the core region; only
about sixteen thousand functions would be needed to provide the
required resolution uniformly throughout the core region defined
above.

\begin{figure}
\begin{center}
\scalebox{\figscale}{\scalebox{1}{\includegraphics{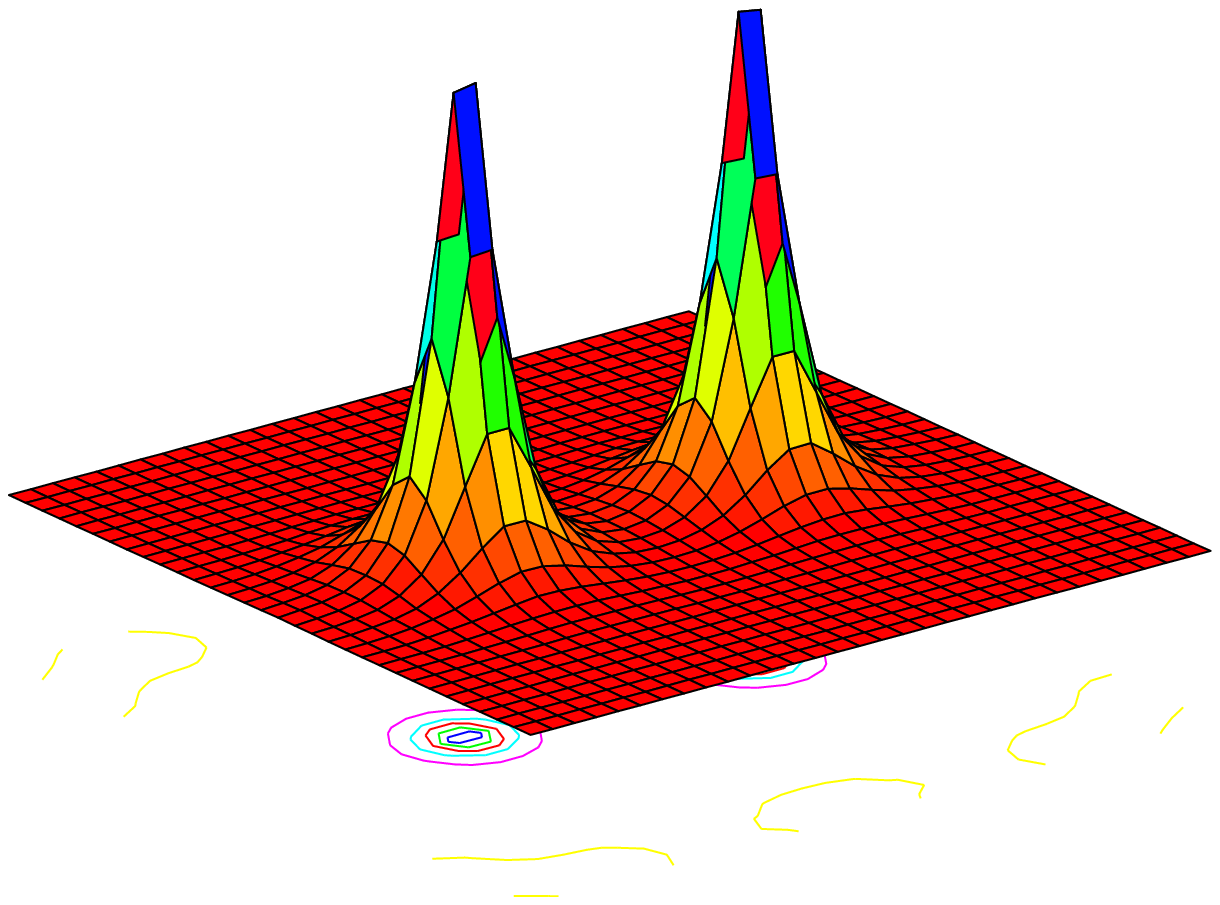}}}
\end{center}
\caption{$1\sigma$ Kohn-Sham orbital of N$_2$ in plane containing both
nuclei.}
\label{fig:wfsr}
\end{figure}

Multiresolution analysis allows us to add resolution precisely into
the core in a systematic, hierarchical manner.  The solid curves in
Figure~\ref{fig:C} come from the earliest reported application of
multiresolution analysis to self-consistent electronic structure
calculations in three dimensions\cite{mgras}.  Despite the high
resolution needed in the core, these multiresolution calculations
required fewer that three thousand basis functions and yet produce
results nearly indistinguishable from the output of atomic
calculations carried out at essentially infinite resolution by
exploiting spherical symmetry to produce an effective one-dimensional
problem and then using a very fine radial grid.

\begin{figure}
\begin{center}
\scalebox{\figscale}{\scalebox{0.75}{\includegraphics{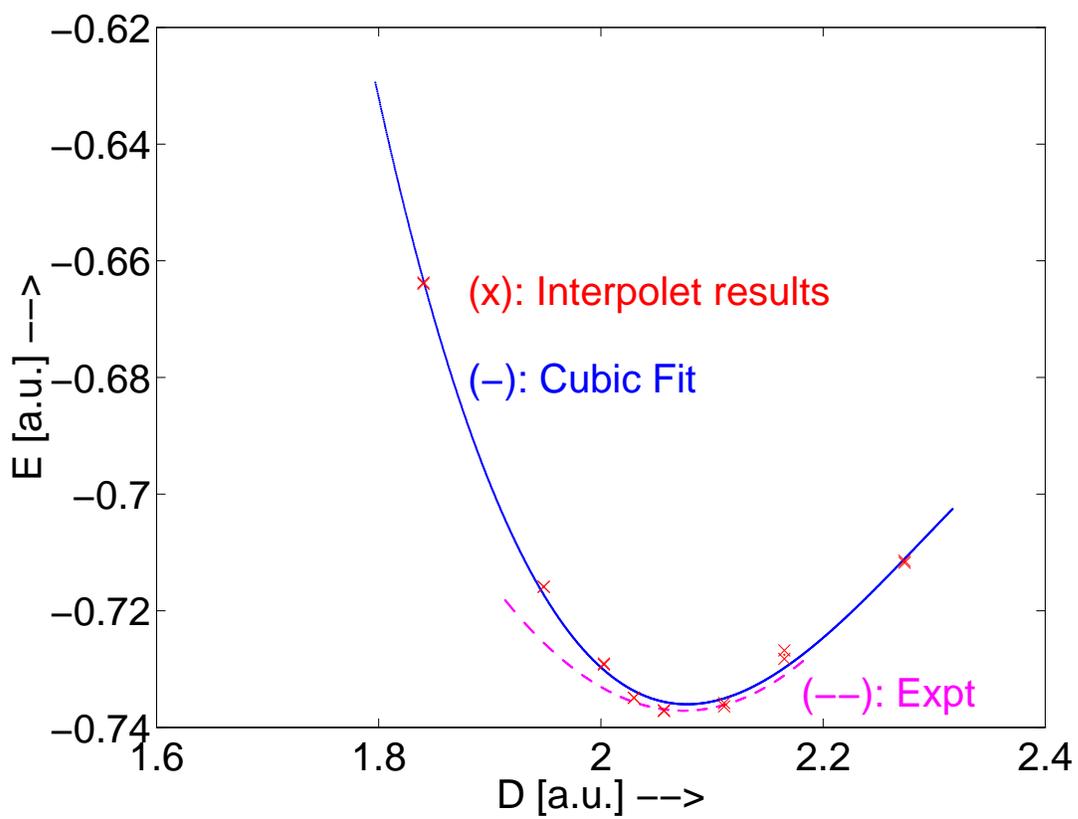}}}
\end{center}
\caption{Energy of N$_2$ as a function of inter-nuclear separation:
results of multiresolution analysis (crosses), cubic fit and quadratic
with experimental curvature and minimum (solid and dashed curves,
respectively).}
\label{fig:EvsR}
\end{figure}

Figures \ref{fig:wfsr}-\ref{fig:EvsR} present similar all-electron
calculations for the N$_2$ molecule\cite{aps}.  Figure \ref{fig:wfsr}
shows the lowest energy Kohn-Sham orbital.  The $1\sigma$ symmetry of
the state and the cusps near the nuclei are clearly visible.  Figure
\ref{fig:EvsR} shows results for the bond length and vibrational
frequency of the molecule.  Using the cubic fit in the figure, we find a bond length
and a spring constant
within $0.1$\% and $7$\% of experiment, respectively.  Both these and
the preceding atomic calculations were carried out in the same 8~Bohr
periodic cell at an effective resolution resolution of corresponding
to 16~million coefficients.  Both sets of calculations used the
$\ci=3$ interpolating scaling functions of the product form discussed
in Sec.~\ref{subsubsec:examples} and the basis restriction strategy
described in Sec.~\ref{sec:FrameMRA:Intro} with seven levels of
resolution.

The succeeding pages lay out the explicit details of how these and
other calculations of electronic structure have been performed using
multiresolution analysis.  Although the emphasis of this review is the
calculation of electronic structure, the techniques we describe are
widely applicable to other physical problems involving
coupled sets of linear and non-linear
partial differential equations.  Shortly after the initial reports of
the above applications to electronic structure calculations,
independent applications of the ideas of wavelet theory to physical
problems described by partial differential equations appeared in a
variety of areas, including
combustion\cite{FrohlichSchneider:94,FrohlichSchneider:97} and fluid
mechanics\cite{Vasilyev:97,CohenDanchin:98}.  General model problems
have also been
explored\cite{BeylkinKeiser:95,BertoluzzaNaldi:96,BeylkinKeiser:97}.
And, more recently, the solution of Poisson's equation has been
studied\cite{goedecker,JCP}.

The issue of multiple length-scales in electronic structure is not
new.  It has driven the development of a variety of techniques which
are now quite mature, including the linear muffin tin orbital (LMTO)
method\cite{lapwlmto}, the linearized augmented plane wave (LAPW)
method\cite{lapwbook}, the full potential LAPW (FLAPW)
method\cite{flapw} and the plane wave pseudopotential
approach\cite{bible}.  The first three methods use one type of basis
set inside of a set of spheres organized around the nuclei and another
type of basis set outside of the spheres.  The wave functions are then
matched at the spherical boundaries to determine the solution.  The
plane wave pseudopotential approach replaces the atomic core
with an effective potential manufactured to have similar scattering
properties.  While each of these approaches has had great success, none is
systematically improvable to complete convergence in a simple,
practical manner, and each requires great care and expertise in
the selection and construction of the atomic spheres or in the
development of appropriate pseudopotentials.  As a result, a general
method is still needed to obtain unambiguous results to a sufficient
accuracy to permit direct and systematic study of the relative accuracy
of competing density functionals and alternate theories of electronic
structure.

As illustrated briefly above and discussed in depth below,
multiresolution analysis provides a systematic approach which can
replace millions of grid points with merely thousands of basis
functions.  The development of the multiresolution analysis of
electronic structure therefore holds the promise of at last enabling
the systematic evaluation of different theories of electronic
structure at high precision.  In addition, the mathematical structure
of multiresolution analysis is sufficiently rich so that new
algorithms and techniques are constantly being developed, making it
probable that multiresolution approaches will prove to be not only
more accurate and systematic, but also more computationally efficient
than present approaches.

This review is organized as follows.  Section~\ref{sec:elecappr}
overviews the equations of density functional theory and gives an
extremely brief review of other modern, systematically improvable
approaches to the calculation of electronic structure.
Section~\ref{sec:FrameMRA} introduces a new basis-set independent
matrix language to express the equations of density functional theory.
Section~\ref{sec:MRATS} lays down the mathematical framework of
multiresolution analysis in a language suited for physical
applications in multiple dimensions, and Sec.~\ref{sec:semicbases}
gives specific examples of basis functions which fit into this
framework.
Secs.~\ref{sec:semicbases:fastalgs}-\ref{sec:inhomogridops} then
describe new methods which are needed to make the application of the
operators and transforms associated with these functions feasible in
physical calculations of complex systems.  Finally, the review
concludes in Sec.~\ref{sec:concs} with a few brief remarks.

\section{Electronic Structure} \label{sec:elecappr}

  \subsection{Kohn-Sham Lagrangian}

Over the last several decades, density functional theory has proven an
accurate, reliable and effective tool for predicting electronic
structure.  It has found application in such diverse areas as the
study of surfaces, point defects, melting, diffusion, plastic
deformation, disorder, catalysis, phase transitions and chemical
reactions.  For reviews see \cite{dft1,dft2,dft3,bible}.

In standard atomic units, $\hbar=m=e=1$, the equations of
density functional theory in the local density
approximation (LDA) \cite{KohnSham} are equivalent to finding the {\em
saddle} point of lowest energy of the Lagrangian functional,
\begin{eqnarray} \label{eqn:saddle}
\cL_{LDA}(\{\psi_i\},\phi) & = & \frac{1}{2} \sum_i f {\int{d^3r\, ||\nabla
\psi_i(r) ||^2}} + \int{d^3r \, V_{\mbox{ion}}(r) n(r)} \nonumber \\
 &  & +
\int{d^3r\,\,\epsilon_{xc}(n(r)) n(r)} -\Re \int{d^3r\,\phi(r) (n(r)-n_0)} \nonumber \\
&& - \frac{1}{8\pi} \int{d^3r\, ||\nabla \phi(r)||^2},
\end{eqnarray}
where the electron density is defined as
\begin{equation} \label{eqn:defnsaddle}
n(r)\equiv \sum_i f |\psi_i(r)|^2,
\end{equation}
the orbitals $\{\psi_i\}$ obey the orthonormality constraint
\begin{equation} \label{eqn:orthoconst}
\int d^3r\,\,\psi_i^*(r) \psi_j(r) = \delta_{ij},
\end{equation}
and $\Re(z)$ denotes the real part of the complex number $z$.
Here, $V_{\mbox{ion}}(r)$ is the
potential of each electron due to the presence of the nuclei (and core
electrons in the case of pseudopotential calculations) in the system,
and $\epsilon_{xc}(n)$ is the exchange-correlation energy per electron
in a uniform electron gas of density $n$.  For simplicity, in this
review we hold the occupation factors $f$ in 
(\ref{eqn:saddle}) and (\ref{eqn:defnsaddle}) fixed at $f \equiv 2$
to reflect the fact that two electrons of spin $1/2$ may 
occupy each orbital.  For calculations in periodic systems, we
introduce $n_0$, which corresponds to a positive charge background
neutralizing the electronic charge density.  The effect of this
background on the total energy is properly accounted when the Ewald
summation is used to compute the interionic interactions.

At the saddle point, the value of $\cL_{LDA}$ is the total
Kohn-Sham energy of the system\cite{KohnSham}, and the fields
$\left\{\psi_i(r)\right\}$ and $\phi(r)$ are the Kohn-Sham orbitals
(electronic wave functions) and the Hartree potential (the
electrostatic field arising from the mean electron density),
respectively.  Taking the real part of the integral coupling the
electron density $n(r)$ to the Hartree field $\phi(r)$ ensures that $\phi(r)$
is real at the saddle point.  This Lagrangian formalism for density
functional theory, introduced in \cite{JCP}, has the advantage over
the standard energy functional approach of rendering local in space
all couplings among the physical fields.  This not only dramatically
simplifies formal manipulations but also allows for the practical
strategy of a direct search for the saddle point to solve the
Schr\"odinger and Poisson equations simultaneously.

Three factors make locating the saddle point of (\ref{eqn:saddle})
challenging.  First, the Lagrangian deals with continuous fields
which must be describe in terms of a finite number of coefficients for
the purpose of calculation.  Second, 
the solutions we seek exhibit multiscale behavior.  Finally, the
Lagrangian couples the fields {\em non-linearly}, both through the
exchange-correlation energy density $\epsilon_{xc}(n) \cdot n$ and through
the term coupling the Hartree field and the electronic charge density,
$\phi(r)n(r)$.

A variety of systematically improvable approaches have appeared in the
literature to meet these challenges.  To place the development of
multiresolution analysis in context, we now give a brief overview of
these other approaches.  We shall not discuss the muffin-tin families of
approaches, which are not so closely related to multiresolution
analysis, beyond their brief mention in the introduction.

  \subsection{Systematic basis approaches} \label{sec:sysbasisappr}

      \subsubsection{Plane wave approach}

The plane wave approach is reviewed in detail in
\cite{bible}.  In this approach, the Kohn-Sham orbitals and the
Hartree potential are expanded in a discrete basis of plane waves
(complex exponentials) consistent with periodic boundary conditions.
The resulting discrete set of expansion coefficients for the orbitals
$\{\psi_i\}$ and Hartree potential $\phi$ are well-suited for
computation.

In a plane wave basis, differential operators are diagonal, making
their implementation particularly simple.  The remaining
couplings in the Lagrangian (\ref{eqn:saddle}) are the non-linear,
spatially local couplings.  When using a plane wave basis, one
implements these couplings on a point by point basis in real space,
using the fast Fourier transform (FFT) to convert efficiently between
the real space and the plane wave representations.

Plane wave calculations may be brought to convergence simply by
increasing a single parameter, the kinetic energy below which all
plane waves are included in the basis.  This makes systematic basis
set convergence studies straightforward.  However, the extremely high
resolution required near atomic nuclei combined with the uniform
resolution afforded by plane waves makes the direct
application of this method prohibitive for all but the lightest
elements.  The introduction of pseudopotential theory overcomes this
limitation, but at the cost of introducing the pseudopotential
approximation, which is uncontrolled.  One great advantage of
multiresolution analysis is that it maintains the regularity of plane
wave expansions while allowing variable resolutions and thus the
direct treatment of heavier elements.

Finally, even the wave functions in pseudopotential calculations at
times require significant resolution near ionic cores, particularly
when dealing with first-row elements or transition metals.  This opens
the exciting possibility of combining the pseudopotential approach
with multiresolution analysis, an issue which has begun to be
explored\cite{chou}.

      \subsubsection{Finite element and adaptive mesh approaches} 

There is an extensive literature dedicated to the finite element
approach, particularly in the fields of solid\cite{feb1} and
fluid\cite{feb2} mechanics.  See \cite{feb3} for a review of recent
developments, and \cite{feb4} for an introduction.

The application of the finite element approach to electronic structure
calculations began in the late eighties\cite{fe1} and has undergone a
recent revival\cite{fe2,fe3,fe4}.  As in the plane wave approach, the
finite element method expands the electronic orbitals in terms of a
set of basis functions.  By using localized basis functions
concentrated in the regions of space requiring high resolution, these
bases can provide a much more efficient description of electronic wave
functions.

Finite elements also represent an extremely efficient method for
dealing with non-linear interactions by providing, as do plane waves,
highly efficient rapid transforms.  For finite elements, these
transforms are based on two properties of the basis functions,
cardinality and interpolation\cite{feb3}.  Below, we shall discuss how
to construct multiresolution analyses which maintain these two highly
desirable properties.

One great difficulty with the application of finite elements to
electronic structure is that finite elements basis sets must be
uprooted and reformed as the nuclei move.  Each coefficient in a
finite element representation corresponds to the value or weight of a
function over the region of one basis function and thus cannot be
taken to be small where the electronic orbitals themselves are
non-negligible.  As a result, coefficients associated with the basis
functions which are uprooted as the atoms moves carry large values, and
this process must be managed with extreme care.  Multiresolution
analysis provides an elegant solution to this difficulty which we 
discuss briefly immediately below and in more depth in
Sec.~\ref{sec:FrameMRA:Intro}.

Another potential solution to variations in the basis as the atoms
move is provided by a branch of finite element methods particularly
attractive for the calculation of electronic structure, the
``Riemannian metric''\cite{gygi1,gygi2,gygi3,riem} or ``adaptive
curvilinear-coordinate''\cite{hamann1,hamann2,hamann3,hamann4,kaxiras,nakano,fe4}
approach.  This approach lays down the finite element mesh according
to a smooth mapping from a underlying cubic grid of points, thereby
ameliorating the problems of uprooting the finite element grid as the
atoms move.  Because it preserves the underlying cubic topology of the
grid, the Riemannian metric approach falls into the class of
``structured mesh'' methods, for which it is difficult to generate
very strongly graded meshes\cite{rank}, a limitation which
multiresolution analyses do not suffer.

      \subsubsection{Multigrid algorithms}

Multigrid algorithms provide an extremely effective means of solving
equations whose convergence is limited by a wide
range of length-scales.  This approach too has an extensive literature
associated with its application in a wide variety of fields.  For an
in-depth introduction, see \cite{mg1}.

Explorations of the multigrid approach as applied to electronic
structure calculations also date back to the late eighties\cite{fe1}
and have become much more common in the last few
years\cite{gygi3,mg2,mg3,mg5}.  Multigrid algorithms do not specify a
basis set and leave open the issue of how best to discretize physical
problems.  The mathematical structure of multigrid algorithms
parallels very closely the ideas of multiresolution analysis, and
multigrid algorithms can be applied directly or easily generalized to
the solution of differential equations expressed in wavelet
bases\cite{mgmra,dicle,dicle2}.

      \subsubsection{Multiresolution analysis}

Wavelet bases place functions of varying resolution on a
multiresolution grid while maintaining a uniform resolution throughout
all of space in the precise mathematical sense of multiresolution
analysis\cite{mallat,meyer,meyer2}.  The mathematical regularity of
the resulting basis leads to efficient fast transforms\cite{dau,chui}
and methods to apply differential operators\cite{BCR,JCP}.

In contrast to the expansion coefficients of a finite-element
expansion which reflect directly the
values of a function, the coefficients of a multiresolution analysis
separate information into different length-scales.  This subtle but
critically important difference means that, so long as a function
varies smoothly, the fine-scale coefficients will be quite small even
where the value of a function is quite large.  As described in
Sec.~\ref{sec:FrameMRA:Intro}, this means that, as the atoms move, one
may arrange for the changes in the basis to involve the truncation of
only coefficients which are small, thereby effectively providing a high
resolution throughout all of space with an extremely limited number of
coefficients.  This also means that no particular care is needed to
handle the regriding as the atoms move.

The first electronic structure calculations to use such a basis
employed a discrete frame of non-orthogonal Gaussian-Mexican hat
basis functions\cite{waveprl}.  This work established the efficacy of
such bases for representing electronic wave functions, but the
calculations were limited to one-electron systems.  The first reported
self-consistent, multiple-electron density functional
calculations\cite{mgras,aps} used the semicardinal bases described in
Sec.~\ref{sec:semicbases} and employed the analytically continued
conjugate gradient approach\cite{prldynam} to solve the Kohn-Sham
equations.  How Poisson's equation was solved in these calculations is
described in more detail in
\cite{JCP}.  Wei and Chou\cite{chou} carried out the first
calculations employing orthogonal Daubechies wavelets\cite{dau}.  This
work studied molecular dimers within the local density approximation
using Daubechies D6 wavelets, employed self-consistent iteration to
solve of the Kohn-Sham equations, and represented the first use of
pseudopotentials in the multiresolution analysis of electronic
structure.  Since that time, Tymczak and Wang\cite{tymczak} used
Daubechies D8 wavelets and introduced the innovations of dynamically
refining the basis and the use of the Car-Parrinello approach to solve
the Kohn-Sham equations.  Most recently, Goedecker and Ivanov
\cite{goedecker} have applied lifted wavelets\cite{sweldens:96} to the
solution of Poisson's equation.

Until recently, the primary bottleneck in multiresolution analysis
calculations of electronic structure had involved the performance of
transforms and the application of differential operators.  The
standard transforms associated with orthogonal wavelet bases require
or produce the values of the electronic orbitals on a uniform grid at
the finest resolution.  To use these transforms, one ``unpacks'' each
electronic orbital from its stored coefficients, operates on the
unpacked version, and then ``repacks'' the result.  Although this
gives great benefit in terms of the use of memory, processing the wave
functions in their highly redundant unpacked representation still
involves significant memory and also much wasted computation.  For
example, applying this approach to the nitrogen dimer calculations
described in the introduction expends hundreds of millions of floating
point operations to process each electronic orbital, each of which are
represented in terms of only six thousand coefficients.

Workers interested in electronic structure therefore have sought
different methods.  One approach has been to take techniques
from the wavelet literature such as the ``non-standard'' multiply
approach of Beylkin, Coifman and Rokhlin\cite{BCR}, which has been
applied to the solution of Poisson's equation in multiresolution bases
with the processing of some additional coefficients but still leading
to an efficient scheme\cite{goedecker}.
Workers also have developed new methods specifically for physical
calculations\cite{JCP}.  These new methods allow operators to be
applied to the electronic wave functions directly in their ``packed''
representation without processing any additional information and have
been shown to be several times more efficient than the non-standard matrix
approach in situations typical of electronic structure\cite{JCP}.  The
associated transforms have been shown to have the novel property that
the process of (a) unpacking the physical fields at a number of points
in space equal to only the number of packed coefficients, (b) coupling the
physical fields in any local, non-linear fashion at these points, and
(c) repacking the result always yields coefficients {\em identical} to
what would be obtained with a fully unpacked function on a grid of
{\em arbitrarily} fine resolution\cite{JCP}.  With these latest advances,
the field now stands poised to see the first applications of
multiresolution analysis to large-scale electronic structure
calculations.
\section{Multiresolution Analysis of Electronic Structure}
\label{sec:FrameMRA}

  \subsection{Multiresolution analysis and restriction}
\label{sec:FrameMRA:Intro}

Section \ref{sec:MRATS} reviews the mathematical structure of
multiresolution analysis in detail.  Here, we give a conceptual
overview sufficient to discuss the use of multiresolution analysis in
the calculation of electronic structure.  

\begin{figure}
\begin{center}
\scalebox{\figscale}{\scalebox{0.35}{\includegraphics{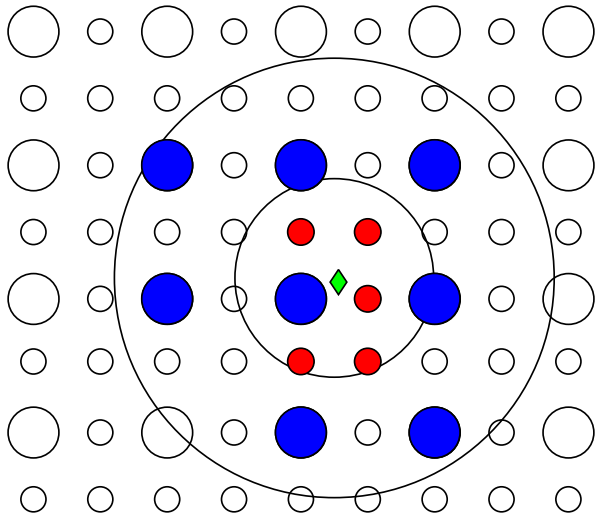}}}
\end{center}
\caption{Multiresolution analysis and the application of restriction to
electronic structure: coarse grid (larger circles), detail points of
finer grid (smaller circles), atomic nucleus (diamond), spheres of
resolution (large circles centered on the nucleus), basis functions
restricted from basis (empty circles), surviving basis functions
(filled circles).}
\label{fig:onoff}
\end{figure}

Figure \ref{fig:onoff} illustrates the concept of multiresolution
analysis and its application to electronic structure.  Throughout this
work we shall use the variables $Q$, $R$ and $P$ to denote different
levels of resolution and will take the coarsest and finest levels of
resolution in a given calculation to be $M$ and $N$, respectively.  As
\cite{mallat,meyer2,dau,chui,strang} describe, a multiresolution
analysis begins with a basis of coarse resolution $Q=M$, which
consists of basis functions laid out on a regular grid across the
region of interest (larger circles in the figure).  The {\em span} of
this set of functions, the vector space of all functions formed by
their linear combinations, is denoted $V_M$.  For reasons which will
become clear below, the basis functions of this coarse space are
referred to as the {\em scaling functions}.

The next conceptual step is to increase the resolution of the basis by
adding finer resolution functions at the points of a finer grid
(smaller circles in the figure).  The basis consisting of both the
original coarse scaling functions and the new finer functions now
spans $V_{M+1}$, a space of increased resolution $Q=M+1$.  The added
functions are referred to as the {\em detail functions} or the {\em
wavelets}.

One may continue adding finer levels of detail functions to reach the
final desired level of resolution, $Q=N$.  We shall designate as
$W_{Q+1}$ the space spanned by the detail functions which bring the
resolution from level $Q$ to the next level $Q+1$, so that
$V_{Q+1}=V_Q \oplus W_{Q+1}$.  (Note that some authors prefer to
designate the above detail space as $W_Q$, rather than $W_{Q+1}$.)
Here, as throughout this work, by the addition of vector subspaces
``$\oplus$'', we mean the space of all vectors which may be written as
a sum of a pairs of vectors, one from each subspace.  The space of all
functions which can be described by functions on all of the scales
included in the basis is thus
$$
V_N \equiv V_M \oplus W_{M+1} \oplus \ldots \oplus W_N.
$$

\begin{figure}
\begin{center}
\scalebox{\figscale}{\scalebox{1.0}{\includegraphics{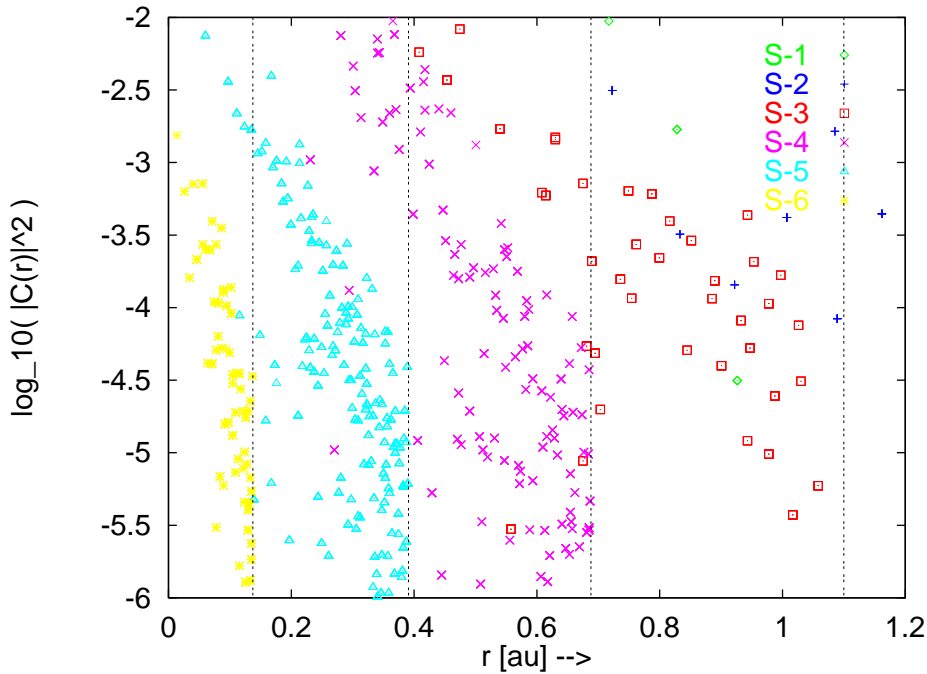}}}
\end{center}
\caption{Coefficients of the multiresolution analysis of the $1\sigma$
orbital of N$_2$: radii of restriction are denoted by vertical lines.}
\label{fig:wfsI}
\end{figure}

Figure~\ref{fig:wfsI} shows the behavior of the expansion coefficients
in such a multiresolution basis for the $1\sigma$ state of the
nitrogen molecule, which appeared in Figure~\ref{fig:wfsr} in the
introduction.  Figure~\ref{fig:wfsI} displays, on a logarithmic scale,
the magnitude of the expansion coefficients for this state as a
function of the two parameters characteristic of each basis function:
location in space $\vec r$ and resolution $Q$.  The three dimensional
location $\vec r$ is projected onto the one-dimensional horizontal
axis $r$ as the distance from the center of the basis function to the
nearest atomic nucleus.  The scales $Q$ of the basis functions are
coded by different symbols.  As evident in the figure, the separation
of information into different length-scales afforded by the
multiresolution analysis results in separate characteristic
exponential decay envelopes with distance from the nuclei for the
coefficients of each scale.  These envelopes illustrate the fact that
the finest scale coefficients need only be kept for basis functions in
the immediate vicinity of the nuclei.  This is precisely the behavior
which makes multiresolution analysis so attractive for the calculation
of electronic structure.

The strategy introduced by\cite{waveprl} to exploit this behavior is
illustrated in Figure~\ref{fig:onoff}: about each nucleus we draw a
set of successively inscribed spheres of appropriate radii for the
scales $M, M+1, \ldots, N$, and we keep in the basis only those
functions of a given scale which fall within the corresponding sphere.
The grid points whose associated functions appear in the final basis
set according to this prescription appear as filled circles in the
figure.  (Calculations with periodic boundary conditions generally
include all functions on the coarsest scale.)  We refer to this
process of selecting grid points and their associated functions as
{\em restriction}.  For other applications, it is not always known
{\em a priori} how to restrict the basis.  For adaptive restriction
approaches, the reader may wish to
consult\cite{liandrat,LiandratTchamitchian:90,BeylkinKeiser:97,BeylkinKeiser:97}.
Tymczak and Wang\cite{tymczak} have also developed an adaptive
restriction technique specifically for electronic structure
calculations.

Selecting the cutoff spheres so as to discard only coefficients below
a given tolerance gives a systematic procedure for working with a
dramatically reduced number of coefficients while maintaining a
description equivalent, for any given tolerance, to the full basis
$V_N$.  In contrast to finite element approaches, in multiresolution
analysis there are no basis set derivative corrections in the
Hellman-Feynman theorem because the basis functions of a
multiresolution analysis remain fixed in space as the atoms move.  The
only effect of the motion of the atoms is to turn on or turn off basis
functions whose coefficients are below the selected tolerance.  As a
result, the discontinuous effects from such on-off switching events are
controllable and generally quite small.  For example, in the calculations of
\cite{waveprl}, although the basis was chosen only to produce
correct total energies without regard to the calculation of forces,
the small jumps in the forces calculated from the Hellman-Feynman
theorem were only on the order of $1~meV/$\AA.

  \subsection{Algebraic structure}
\label{sec:Over} 

The main step in implementing a density functional calculation is to
express each term of the LDA Lagrangian (\ref{eqn:saddle}) in terms of
the coefficients $d_{\alpha}$ and $c_{\alpha,i}$ for the Hartree field
$\phi(r)$ and the Kohn-Sham orbitals $\{\psi_i(r)\}$, which appear in
the expansions
\begin{eqnarray}
\psi_i(r) & = & \sum_\alpha c_{\alpha,i} b_\alpha(r) \label{eqn:exps} \\
\phi(r) & = & \sum_\alpha d_\alpha b_\alpha(r), \nonumber
\end{eqnarray}
where $\{b_\alpha(r)\}$ is the basis set used in the calculation.
Once the Lagrangian is represented in terms of these coefficients, the
gradients of the Lagrangian with respect to $d_{\alpha}$ and
$c_{\alpha,i}$ may be calculated so as to locate the saddle point and
thereby determine the orbitals $\{\psi_i(r)\}$, potential $\phi(r)$
and total electronic energy $\cL$ of the system.  Several different
approaches for expressing the Lagrangian have been used in the
application of multiresolution analysis to electronic structure.  Here
we present an overview of these approaches and the operators and
transformations which they involve.  Explicit details and formulae are
given in Secs.~\ref{sec:LagrEner}-\ref{sec:KohnSham}.

The simplest terms in the LDA Lagrangian are the electronic kinetic
energy $T$ and the Hartree field self-energy $V_{H-H}$, the first and
final terms of (\ref{eqn:saddle}).  These are bilinear in the
coefficients and may be evaluated exactly in terms of two-center
integrals of the Laplacian operator between the basis functions.
Because the matrix elements of the Laplacian operator are known
explicitly for functions making up multiresolution analyses
(Sec.~\ref{sec:semicbases:matels}), all implementations of
multiresolution analysis to date use this simple two-center form to
evaluate $T$ and $V_{H-H}$\cite{waveprl,mgras,chou,tymczak}.

The electron-ion potential energy $V_{e-i}$, the second term in
$\cL_{LDA}$, is also bilinear in the expansion coefficients
$c_{\alpha,i}$, but the matrix elements of the ionic potential needed
to compute $V_{e-i}$ consist of three-center integrals, each involving
two basis functions and one ion, and thus require special care.  To
date, $V_{e-i}$ has been handled through
the introduction of a grid $G$ of points $p$ in real space.  Two
prescriptions for employing this grid exist.
The first prescription (used in \cite{mgras}) follows the spirit of
the energy functional and computes the electron-ion interaction as a
functional of the electron density.  This recipe first uses a forward
transform to determine the values of the wave functions on the grid.
From these, the single particle density $n(r)=\sum_i f |\psi_i(r)|^2$
is computed on the grid and the corresponding expansion coefficients
are determined through an inverse transformation.  From the resulting
expansion coefficients, the final step is to compute the total
potential energy in terms of known overlaps between the basis
functions and the ionic potential.  This approach has the advantage
that the grid need only be able to resolve the electron density
$n(r)$ and not necessarily the potential $V_{\mbox{ion}}(r)$, which
may vary much more rapidly.  We refer to this approach below as the
{\em energy functional prescription}.

The second prescription (used in \cite{chou,tymczak}) follows the
spirit of the effective Schr\"odinger equation for the Kohn-Sham
orbitals and applies the electron-ion interaction as a diagonal
operator in real space.  This recipe also begins with a forward
transform to evaluate the wave functions $\psi_i(r)$ on the points of
the grid.  It then applies the potential operator at each point $p$ of
the grid $G$ to produce $V_{\mbox{ion}}(p) \psi_i(p)$ and inverse
transforms the result to coefficient space.  Finally, the overlap of
$\psi_i^*(r)$ and $V_{\mbox{ion}}(r)\psi_i(r)$ is computed using the
known overlaps between the basis functions.  We refer to this approach
below as the {\em operator prescription}.

The fourth term of (\ref{eqn:saddle}) describes the coupling between
the electrons and the Hartree field, $V_{e-H}$.  This term is cubic in
the expansion coefficients and involves three-center integrals.  In
principle, a treatment in terms of direct three-point interactions is
possible using analytic results for the integrals of triple products
of scaling and detail functions developed in \cite{BeylkinKeiser:97}.
To date, this direct route has not been pursued in the calculation of
electronic structure.  Instead, one may exploit the fact that the
coupling $V_{e-H}$ has the same structure as the electron-ion coupling
and compute it in the same manner, either as a functional of the
electron density or as the result of the application of $\phi(r)$ as a
local operator.

The final remaining term, the third term in (\ref{eqn:saddle}), gives
the local density approximation to the total exchange-correlation
energy of the system, $E_{xc}$.  This is known only in terms of the
non-algebraic function $\epsilon_{xc}(n)$, which is tabulated in \cite{PZ},
and thus cannot be evaluated in terms integrals of products of basis
functions using the formalism of
\cite{BeylkinKeiser:97}.  For this term, there is no choice but to evaluate the
electron density on a grid $G$ and evaluate the tabulated function
$\epsilon_{xc}(n)$ on a point by point basis in real space.  Once this
is done, there are a variety of choices for how to use the
exchange-correlation field to determine the total exchange-correlation
energy $E_{xc}$.  One could proceed directly and evaluate the
exchange-correlation energy density per unit volume
$\epsilon_{xc}(n(p)) \cdot n(p)$ at each point and then integrate the
result numerically, which would introduce an additional class of
approximation in the evaluation of the Lagrangian.  What has been
done, instead, is to follow the same prescriptions as used in
computing the electron-ion interaction, following either the energy
functional prescription and computing $E_{xc}$ as the integral of the
product of $n(r)$ and $\epsilon_{xc}(n(r))$\cite{mgras} or following
the operator prescription and computing $E_{xc}$ as the expectation of
$\epsilon_{xc}(n(r))$ as an operator acting on the Kohn-Sham
orbitals\cite{chou,tymczak}.

  \subsection{Matrix language}
\label{sec:MatrixFramework}

To discuss the various approaches available for evaluating the LDA
Lagrangian in terms of the expansion coefficients for the physical
fields, we now introduce a matrix language for electronic structure
methods which are based on single-particle orbitals.  In analogy with
Dirac's bra-ket language, this matrix language is completely explicit
but keeps the expressions for physical quantities independent of the
details of the underlying basis set.  This allows us to discuss the
various strategies for applying multiresolution analysis to electronic
structure without obscuring the discussion with irrelevant details.
This new language is useful in its own right for both formal
manipulations and the development of highly portable, efficient
software.  Plane wave calculations performed with software developed
by translating this language directly into the C++ include
\cite{bars,APD}.

For simplicity, we consider below cases where the Fermi occupations
$f$ are constant and the sampling of the Brillouin zone is carried out
at the $\Gamma$ point.  More general cases including variable fillings
and multiple k-points may be worked out with some additional
complexity\cite{prldynam}.  For systems sufficiently large that the
Brillouin zone may be sampled at the $\Gamma$ point alone, the
formulae below may be used directly.

    \subsubsection{Fundamental basis-dependent operations} \label{sec:physops}

We begin by introducing two operators, $\diag$ and $\Diag$.  The
operator $\diag$ converts a matrix to a column vector containing the
elements along the diagonal of the matrix, and the operator $\Diag$
converts a column vector to a diagonal matrix with the components of
the vector placed along the diagonal.  In terms of components, for a
matrix $M$ and a vector $v$, these
operators are
\begin{eqnarray}
(\diag M)_\alpha & \equiv & M_{\alpha\alpha} \\
(\Diag v)_{\alpha\beta} & \equiv & v_{\alpha} \delta_{\alpha\beta}, \nonumber
\end{eqnarray}
respectively, where $\delta_{\alpha\beta}$ is the Kronecker
$\delta$.  Note that while $\diag \Diag v = v$ for any vector
$v$, $\Diag \diag M = M$ if and only if the matrix $M$ is diagonal.
Two identities involving these operators which we shall use freely are
\begin{eqnarray} 
(\diag M)^\dagger v & = & \Tr \left(M^\dagger \Diag v\right),
\label{eqn:diagids} \\ 
v^\dagger \diag M & = & \Tr \left( (\Diag v)^\dagger M \right). \nonumber
\end{eqnarray}

Next, it is useful to regard the Hartree field expansion coefficients,
the $d_\alpha$ from (\ref{eqn:exps}), as the components of a column vector
$d$ and the electronic wave function coefficients, the $c_{\alpha,i}$ from
(\ref{eqn:exps}), as the elements of a matrix $C$, each of whose
columns contains the expansion for a single electronic orbital,
$C_{\alpha i} \equiv c_{\alpha,i}$
For formal manipulations, it is convenient to define also
\begin{equation} \label{eqn:defP}
P \equiv f C C^\dagger,
\end{equation}
the representation of the single particle density matrix in the space
of basis functions $\{b_\alpha(r)\}$.

As Sec.~\ref{sec:Over} describes, evaluation of the Lagrangian
(\ref{eqn:saddle}) in terms of the expansion coefficients contained in
$d$ and $C$ requires knowledge both of the overlaps of the basis functions
among themselves and of their matrix elements through the Laplacian operator,
\begin{eqnarray} 
\cO_{\alpha\beta} \equiv \int d^3r\,b^*_\alpha(r) b_\beta(r)
\label{eqn:defO} \\ 
L_{\alpha\beta} \equiv \int d^3r\,b^*_\alpha(r) \nabla^2 b_\beta(r).
\label{eqn:defL}
\end{eqnarray}
The first two non-trivial basis-dependent operations which require
careful implementation are multiplication by the inner product matrix
$\cO$ and by the Laplacian matrix $L$, to which we refer below as {\em
application of the overlap operator} and {\em application of the
Laplacian operator}, respectively.
Section~\ref{sec:semicbases:matels} discusses how to compute the
required matrix elements, and Sec.~\ref{sec:semicbases:fastalgs}
describes efficient techniques for applying these operators.  Note
that in the special case of orthonormal bases\cite{chou,tymczak}, we
have simply $\cO=I$, where $I$ is the identity matrix.

The {\em forward transform} operation described in the previous
section converts the expansion coefficients of a function into
the values of the function on the points $p$ of the grid $G$ in real
space.  This operation simply amounts to multiplication of a column
vector containing the expansion coefficients by the matrix
\begin{equation} \label{def:cI}
\cI_{p\alpha} \equiv b_\alpha(p),
\end{equation}
whose $\alpha^{\mbox{th}}$ column consists of the values
of the $\alpha^{\mbox{th}}$ basis functions at all of the points $p$
of the grid.  In the case where different bases are used for the wave
functions and for the Hartree field\cite{chou,tymczak}, the columns of
$\cI$ consist of two subsets, one for each of the two basis sets.

As described in Sec.~\ref{sec:Over}, it is at times necessary to find
the expansion coefficients for a function from its values on the grid
$G$.  We denote this linear {\em inverse transform} operator as $\cJ$.
In implementations where the number of grid points equals the number of basis functions\cite{mgras,JCP,dicle}, the natural choice is to take $\cJ
\equiv \cI^{-1}$.  However, we maintain the distinction between $\cJ$
and $\cI^{-1}$ because in implementations where a full uniform grid
$G$ of sampling points is used\cite{chou,tymczak}, there are generally
more grid points $p$ than basis functions $b_\alpha(r)$, and $\cI$ and
$\cJ$ cannot be inverses.  In the case where more that one basis set
is used\cite{chou,tymczak}, formally, $\cJ$ computes the inverse
transform separately for each basis set.  In practice, however, as
seen below, only one inverse generally need be computed because $\cJ$
will be needed only on one basis set.

Finally, as we shall see in
Secs.~\ref{sec:LagrEner}~and~\ref{sec:KohnSham}, two further transforms
appear in the calculation of the gradients of the Lagrangian.  These
two {\em conjugate} transforms represent multiplication by the Hermitian
conjugates $\cI^\dagger$ and $\cJ^\dagger$ of the standard transforms.
The reader should bear in mind that the relation $\cI^\dagger = \cI^{-1} =
\cJ$ for the discrete Fourier transform is quite special.  Because
orthogonality with respect to integration does not ensure
orthogonality with respect to a discrete sampling of the functions the
above relation for the discrete Fourier transform is not generally
true, even for orthonormal bases.  In particular, $\cI^\dagger \ne
\cI^{-1}$ for the multiresolution bases of Daubechies wavelets used in
\cite{chou,tymczak}.

In summary, six non-trivial basis-set dependent operations are needed in
the calculation of electronic structure: the application of two
operators (the Laplacian and overlap operator) and
four transforms (forward, inverse, and the conjugates to each), which we
denote as $L$, $\cO$, $\cI$, $\cJ$, $\cI^\dagger$ and
$\cJ^\dagger$, respectively.

    \subsubsection{Identities}
\label{sec:MatrixFramework:Identities}

Although the action of the six operations $L$, $\cO$, $\cI$, $\cJ$,
$\cI^\dagger$ and $\cJ^\dagger$ depend on the specific choice of
basis, they obey several important identities of which we shall make
use.  In addition to their use in formal manipulations, these
identities provide a useful practical tool to verify the
implementation of the various operators.

The most important among these identities involve the constant
function.  To represent the constant function on the grid, we
introduce {\bf 1}, a column vector containing unity as each entry on
the grid:
\begin{equation} \label{eqn:def1}
{\bf 1}_p  \equiv  1.
\end{equation}
Plane waves, finite element bases, and proper multiresolution analyses
all have the property of being able to represent this function
exactly.  For these bases, we have
that for all points $r$, $\sum_\alpha (\cJ {\bf 1})_\alpha
b_\alpha(r) = 1$.  Evaluating this on
the points $p$ of the grid $G$ yields that, in particular, 
\begin{equation}
\cI \cJ {\bf 1}={\bf 1}.
\end{equation}
For other sufficiently descriptive bases, such as Gaussian bases, this
and the relations below should hold at least approximately in the regions
described by the basis.

There is a close relationship between the inner product matrix $\cO$
and the integrals of the basis functions.  The vector
\begin{equation}
s \equiv \cO \cJ {\bf 1},
\end{equation}
is a column vector containing the integrals of each of the basis
functions:
\begin{eqnarray*}
s_\alpha & \equiv & \left( \cO \cJ {\bf 1} \right)_\alpha \\
& = & \int d^3r\,\, b^*_\alpha(r) \left(\sum_\beta (\cJ {\bf 1})_\beta b_\beta(r)\right) \\
& = & \int d^3r\,\, b^*_\alpha(r). \\
\end{eqnarray*}
Thus, if $g$ is a vector of expansion coefficients, the integral
of the function represented by $g$ is
\begin{equation}
\int d^3r\, g(r) = \int d^3r \, \sum_\alpha g_\alpha b_\alpha(r) = s^\dagger g.
\end{equation}
From this, we may also derive the normalization condition
\begin{equation}
s^\dagger \cJ {\bf 1} = \int 1 \, d^3r = \Omega,
\end{equation}
where $\Omega$ is the volume in which the calculation is
carried out.

In solving Poisson's equation for the Hartree potential, care always
must be taken with the null space of the operator $L$.  Integrating the
identity $\nabla^2 1 = 0$ against the complex conjugate of each basis
function gives
\begin{equation}
L \cJ {\bf 1}=0.
\end{equation}
In periodic systems, where the only solution to Laplace's equation is
the constant function, the entire null space of the operator $L$
consists of the vector $\cJ {\bf 1}$.  Below, we use this to
determine the precise value of the compensating average density $n_0$
needed to avoid divergences in the Lagrangian in periodic calculations.

Finally, although there is no {\em a priori} relationship between
$\cI^\dagger$ and $\cJ$, an approximate relation exists when the grid
$G$ is uniform and of high resolution.  Under these conditions, we
have
\begin{equation} \label{eqn:OJId}
\cO \cJ \approx \omega \cI^\dagger,
\end{equation}
where $\omega$ is the volume per grid point.  To see this, consider
two arbitrary functions $g(r)$ and $h(r)$, where $g(r)$ is represented
by the vector $\tilde{g}$ of its expansion coefficients and $h(r)$ is
represented by the vector $h$ of its values on the points $p$ of the
grid $G$.  There are then two ways of approximating the integral $\int
d^3r\,g(r)^* h(r)$, both of which should give nearly the same result.
First, one could inverse transform $h$ to find appropriate
expansion coefficients and then evaluate the overlap in
coefficient space, giving the result $\tilde{g}^\dagger \cO \cJ h$.
Alternately, one could determine the values of $g$ on the grid by the
forward transform and then approximate the integral as the sum $\omega
(\cI \tilde{g})^\dagger h$.  Equating these two expressions for general
vectors $\tilde{g}$ and $h$ leads to (\ref{eqn:OJId}).

  \subsection{Lagrangian and energy functionals}
\label{sec:LagrEner}

As described above, the cornerstone of all density functional
calculations which use basis set expansions is the explicit expression
of the LDA Lagrangian (\ref{eqn:saddle}) in terms of the coefficients
$d$ and $C$ of the expansions (\ref{eqn:exps}).  The language and
operators defined in the previous section were created specifically so
that the expression of the Lagrangian is identical for most basis
sets, including plane wave, Gaussian orbital, and multiresolution
bases.
As described in Sec.~\ref{sec:Over} two common strategies exist for
expressing the Lagrangian in terms of the expansion coefficients in
multiresolution analyses, the energy functional prescription and the
operator prescription.  We now give explicit expressions for these two
prescriptions.

    \subsubsection{Energy functional prescription} \label{sec:LagrEner:EF}

      \paragraph{Lagrangian}

The only terms in the
Lagrangian (\ref{eqn:saddle}) which cannot be written directly as a
functional of the electron density are the electronic kinetic energy
and the Hartree field self-energy.  These two terms are simple
bilinear forms in the expansion coefficients and are best evaluated
exactly by the direct substitution of the expansions (\ref{eqn:exps})
into (\ref{eqn:saddle}), giving
\begin{eqnarray} 
T & = & -\frac{1}{2} \sum_i f \sum_{\alpha\beta} c^*_{\alpha,i}
L_{\alpha\beta} c_{\beta,i} = -\frac{1}{2} \Tr f C^\dagger L C =
\Tr\left( (-\frac{1}{2} L) P \right) \label{eqn:T} \\
V_{H-H} & = & \frac{1}{8\pi} \sum_{\alpha\beta} d^*_{\alpha}
L_{\alpha\beta} d_{\beta} = \frac{1}{8\pi} d^\dagger L d, \label{eqn:VHH}
\end{eqnarray}
where we have detailed the explicit conversion to matrix language and
made use of cyclic property of the trace to give a representation of
$T$ in terms of the density matrix $P$ defined in (\ref{eqn:defP}).

The remaining terms in the Lagrangian all may be written directly in
terms of the electron density.  In our matrix representation, each
column of the matrix $\cI C$ contains the values of one of the
electronic orbitals when evaluated on the grid.  The electron density
at each point $p$ of the grid is thus
\begin{eqnarray}
n(p) & = &  \sum_i f (\cI C)_{pi}^* (\cI C)_{pi} \nonumber \\
& = & \left(  f (\cI C) (\cI C)^\dagger \right)_{pp}. \nonumber 
\end{eqnarray}
From this, we may gather the real space charge density into the column
vector $n$ as
\begin{eqnarray}
n & = & \diag\left( \cI P \cI^\dagger \right). \label{eqn:defn}
\end{eqnarray}
Finally, the expansion coefficients for the function $n(r)$ are just
$\cJ n$.

Using this last result to evaluate $V_{e-i}$, we find
\begin{eqnarray} 
V_{e-i} & = & \int d^3r\,V_{\mbox{ion}}(r) n^*(r) \label{eqn:Vei1} \\
& = & (\cJ n)^\dagger v \nonumber \\
& = & \Tr\left( \cI^\dagger (\Diag \cJ^\dagger v) \cI P\right), \nonumber
\end{eqnarray}
where the vector
\begin{equation} \label{eqn:defv}
v_\alpha \equiv \int d^3r\,b^*_\alpha(r) V_{\mbox{ion}}(r)
\end{equation}
represents the ionic potential as its overlap with each function of
the basis.  (In conjunction with the appearance of $n_0$ and the use
of the Ewald summation, the zero wave-vector component of
$V_{\mbox{ion}}$ should be subtracted when evaluating these overlaps
in calculations with periodic boundary conditions.)  Although $n(r)$
is real, we introduced $n^*(r)$ above as a formal device to reduce the
number of complex conjugations appearing in the final expression.  The
conversion to density-matrix form in the last line of (\ref{eqn:Vei1})
may be carried out using the identities (\ref{eqn:diagids}).

The evaluation of the next term in the Lagrangian, the total exchange
correlation energy $E_{xc}$, is similar.  Given access to the
values of $n$ on the grid points, it is a simple matter to also
evaluate the exchange-correlation energy per particle
$\epsilon_{xc}(n(p))$ at those points.  Collecting these into the
column vector $\epsilon_{xc}(n)$, inverse transforming both this and
the charge density and taking the overlap in coefficient space gives
the final result,
\begin{eqnarray}
E_{xc}
& = & \int d^3r \epsilon_{xc}(n(r)) n^*(r) \label{eqn:Exc} \\
& = & (\cJ n)^\dagger \cO (\cJ \epsilon_{xc}(n)), \nonumber
\end{eqnarray}
where again we introduce the conjugation of $n$ to reduce the number
of complex conjugations appearing in the final result.  $E_{xc}$ may
also be converted to density-matrix form using (\ref{eqn:diagids}).

The final term remaining in (\ref{eqn:saddle}), $V_{e-H}$, also has
the overlap form
\begin{eqnarray} 
V_{e-H} & = & - \Re\left[\int d^3r\, \phi(r) \left( n(r) - n_0 \right)^*
\right] \label{eqn:VeH1} \\
& = & - \Re \left[ \left(\cJ (n-n_0 {\bf 1}) \right)^\dagger \cO d\right],  \nonumber
\end{eqnarray}
where ${\bf 1}$ is as defined in (\ref{eqn:def1}).  Again, the
conjugation of the density term has no effect but to yield the simplest
final expression.

To determine the proper choice of $n_0$ for periodic supercell
calculations, we note that the Hartree self-energy term $V_{H-H}$ has
no contribution from the projection of $d$ in the null space of $L$,
which Sec.~\ref{sec:LagrEner} showed lies along the direction $\cJ
{\bf 1}$.  Thus, in order for a saddle point to exist for the
Lagrangian (\ref{eqn:saddle}), there can be no coupling of this
component of $d$ to the electron density in (\ref{eqn:VeH1}).  Hence,
we must have $(\cJ (n-n_0 {\bf 1}))^\dagger
\cO \cdot \cJ {\bf 1}=0$.  With the identities of
Sec.~\ref{sec:MatrixFramework:Identities}, this means that $n_0 =
s^\dagger (\cJ n)/\Omega$, in accord with our interpretation of $n_0$
as the integral of $n(r)$ divided by the volume of the supercell.
With this result, $V_{e-H}$ is
\begin{equation} \label{eqn:VeH2}
V_{e-H}  =  - \Re\left[ n^\dagger \cJ^\dagger (\cO
-\frac{s s^\dagger}{\Omega} ) d\right]. 
\end{equation}

Putting the preceding results together, the final expression for the
LDA Lagrangian under the energy functional prescription is
\begin{eqnarray}
\cL_{LDA}(C,d)
& = & \Tr \left( f C^\dagger (-\frac{1}{2} L) C \right) + (\cJ
n)^\dagger v \label{eqn:Llda1Cn} \\
&& \ \ \ + (\cJ n)^\dagger \cO \cJ \epsilon_{xc}(n)  \nonumber \\
&& \ \ \  - \Re\left[ n^\dagger \cJ^\dagger (\cO -\frac{s s^\dagger}{\Omega} ) d\right]
+ \frac{1}{8\pi} d^\dagger L d \nonumber
\\
& = & \Tr \left( (-\frac{1}{2} L) P \right) + \Tr\left( \cI^\dagger
(\Diag \cJ^\dagger v) \cI P\right) \label{eqn:Llda1P}\\
&& \ \ \ + \Tr \left( \cI^\dagger \left(\Diag \cJ^\dagger \cO \cJ
\epsilon_{xc}(n)\right) \cI P \right) \nonumber \\
&& \ \ \  - \Tr \cI^\dagger \Re \left[ \Diag \cJ^\dagger (\cO
-\frac{s s^\dagger}{\Omega}) d\right] \cI P
+ \frac{1}{8\pi} d^\dagger L d, \nonumber
\end{eqnarray}
where $n$ is defined as in (\ref{eqn:defn}).  Here we have given two
forms.  The form (\ref{eqn:Llda1Cn}) is more efficient for evaluating
the value of the Lagrangian in terms of the coefficients $C$ and $d$.
The form (\ref{eqn:Llda1P}), which is expressed almost entirely in terms of
the density matrix, is most useful in formal manipulations which
determine the gradients of the Lagrangian with respect to the
electronic coefficients.

      \paragraph{Poisson's equation}

To determine the expansion coefficients of the Hartree field $\phi$,
one sets to zero the variation of the Lagrangian with respect to all
possible infinitesimal variations $\delta d$ in the Hartree field
coefficients.  Taking the real part of (\ref{eqn:VeH2}) explicitly,
this variation is
\begin{eqnarray}
\delta \cL_{LDA} & = & - (\delta d)^\dagger \left( \frac{1}{2}(\cO -
\frac{s s^\dagger}{\Omega}) \cJ n + \frac{1}{8\pi} L d \right)
\label{eqn:Pvar} \\
&& \ \ \ - \left( \frac{1}{2}(\cO - \frac{s s^\dagger}{\Omega}) \cJ n +
\frac{1}{8\pi} L d \right)^\dagger \delta d. \nonumber
\end{eqnarray}
The gradient with respect to the real and imaginary parts of $d$
are thus the real and imaginary components of
\begin{equation} \label{eqn:gradd1}
\partial_d \cL_{LDA} = - \left( (\cO -
\frac{s s^\dagger}{\Omega}) \cJ n + \frac{1}{4\pi} L d \right),
\end{equation}
respectively.  Setting this to zero, we arrive at the basis
independent representation of Poisson's equation,
\begin{equation} 
L d = 4 \pi (\cO - \frac{s s^\dagger}{\Omega}) \cJ n.  \label{eqn:Poi1}
\end{equation}
The positive sign of the right side of this equation reflects the
negative unit charge of the electron.

Care must be taken in solving this equation because, as described
above, $L$ has a null space along the direction $\cJ {\bf 1}$.
However, by the construction of (\ref{eqn:VeH2}), the right-hand side
of the equation has no projection in this space.  Thus, in the
solution
\begin{equation} \label{eqn:HartreeF1}
d = 4 \pi L^{-1} (\cO - \frac{s s^\dagger}{\Omega}) \cJ n,
\label{eqn:Hartree1}
\end{equation}
$L^{-1}$ is understood to mean inversion of the linear operator $L$ in
the sub-space orthogonal to the null space and leaving zero
projection along the null-space in the result.

      \paragraph{Energy functional}

Substituting the explicit result for $d$ into the Lagrangian
(\ref{eqn:saddle}) gives the following explicit expression for the LDA
energy functional $E_{LDA}(C)$,
\begin{eqnarray} \label{eqn:Elda1}
E_{LDA}(C) 
& = & \Tr \left( f C^\dagger (-\frac{1}{2} L) C \right) + 
(\cJ n)^\dagger v  \\
&& \ \ \ + (\cJ n)^\dagger \cO \cJ \epsilon_{xc}(n) \nonumber \\
&& \ \ \  + \frac{1}{2} n^\dagger \cJ^\dagger (\cO - \frac{s
s^\dagger}{\Omega}) (-4 \pi L^{-1}) (\cO - \frac{s s^\dagger}{\Omega})
\cJ n. \nonumber
\end{eqnarray}
Note that Hermitian symmetry has ensured that the Hartree term is
explicitly real.

    \subsubsection{Operator prescription} \label{sec:LagrEner:KS}

      \paragraph{Lagrangian}

The alternative prescription for expressing the Lagrangian, followed
in \cite{chou,tymczak}, is to view it as the sum of the expectations
of an energy operator $\hat \cL$ among the orbitals, $\cL_{LDA} =
\sum_i f \int d^3r\, \psi^*_i \hat \cL
\psi_i$.  The only term which cannot be taken to involve involve the
electrons in this way is the Hartree self-energy term, which is best
represented in the exact form (\ref{eqn:VHH}).  Note that we have
already expressed the electronic kinetic energy in its operator form
in (\ref{eqn:T}).

The remaining contributions to $\hat \epsilon$ are local operators in
real space and thus all take on the same form as does the electron-ion
interaction $V_{e-i}$.  The technique which Wei and Chou\cite{chou}
introduced to treat $V_{e-i}$ is to first compute $\cI C$, the values
of the wave functions on the grid points and then multiply the result
by a diagonal matrix containing the values of the potential at the
grid points.  This results in $(\Diag \tilde{v}) \cI C$,
where $\tilde{v}$ is the vector containing the values of the ionic
potential at the grid points,
$$
(\tilde{v})_p \equiv V_{\mbox{ion}}(p),
$$
where the tilde on $\tilde{v}$ distinguishes this vector of values
from the vector $v$ of overlaps defined in (\ref{eqn:defv}).  To
compute the energy, this result is transformed into expansion
coefficients by the operation of $\cJ$ and the overlaps with the
$\psi^*_i(r)$ are computed using the known overlaps between basis
functions.  The total of the resulting potential energy among all of
the orbitals is
\begin{eqnarray} 
V_{e-i} & = & \sum_i f \int d^3r\,\psi^*_i(r) \left( V_{\mbox{ion}}(r)
\psi_i(r) \right) \label{eqn:WCerror} \\
& = & \Tr f C^\dagger \cdot \cO \cdot \cJ \left( (\Diag \tilde{v}) \cI
C \right)
\nonumber \\
& = & \Tr f C^\dagger V C = \Tr V P, \nonumber
\end{eqnarray}
where $V\equiv \cO \cJ (\Diag \tilde{v}) \cI$.  A potential difficulty
with this approach is that the inherently asymmetric roles played by
$\psi^*$ and $\psi$ have resulted in a potential energy operator $V$
which is not Hermitian and thus may lead to complex energies.  Taking
the Hermitian part of $V$ corresponds to taking just the real part of
(\ref{eqn:WCerror}) and gives the proper form for the
electron-ion energy.  After some manipulation and using the fact that
$\tilde v$ is real, we have the following
equivalent forms for $V_{e-i}$
\begin{eqnarray}
V_{e-i}
& \equiv & \Re \Tr( \cO \cJ \Diag \tilde{v} \cI \cdot P) \label{eqn:manipwn}\\
& = & \tilde{v}^\dagger \Re \diag( \cI P \cO \cJ ) \nonumber \\
& = & \omega \tilde{v}^\dagger \tilde{n}, \nonumber
\end{eqnarray}
where we have defined
$$
\tilde{n} \equiv \Re \diag( \cI P \cO \cJ )/\omega,
$$
as an effective charge density, and $\omega$ as the volume per grid
point.  With these definitions, the electron-ion interaction takes on
precisely the appearance of a numerical integration in real space of
the product of potential and the {\em effective} electron density.
Indeed, from (\ref{eqn:OJId}) and (\ref{eqn:defn}), as long as a fine,
uniform grid is used, we have $\Re
\diag( \cI P \cO \cJ ) \approx \omega n$, so that $\tilde{n}$
approximates the physical electron density.  As we shall see, the
effective density $\tilde{n}$ replaces the physical density in all
energy expressions of the operator prescription.

The remaining terms from (\ref{eqn:saddle}) have the same
structure as $V_{e-i}$ and thus may be evaluated in the same way to yield
\begin{eqnarray*}
E_{xc} & = & \omega \epsilon_{xc}^\dagger \tilde{n} \\
V_{e-H} & = & - \omega \Re[ (\cI d)^\dagger (\tilde{n} - n_0 {\bf 1})].
\end{eqnarray*}
To determine $n_0$, we again ensure that there be no coupling between
the electron density and the
projection of $d$ along the null space of $L$: $(\cI \cdot \cJ
{\bf 1})^\dagger (\tilde{n} - n_0 {\bf 1}) = 0$.  This simplifies to
\begin{equation}
n_0 = \frac{{\bf 1}^\dagger \tilde{n}}{{\bf 1}^\dagger {\bf 1}},
\end{equation}
the mean value of $\tilde{n}$ among the points of the grid.  

We shall find in Sec.~\ref{sec:KohnSham:KS} that the the
exchange-correlation energy should also be evaluated using the
{\em effective density} $\tilde{n}$, so that the final expressions for the
Lagrangian in the operator prescription are
\begin{eqnarray}
\cL_{LDA}(C,d) 
& = & \Tr \left( f C^\dagger (-\frac{1}{2} L) C \right) + \omega \tilde{v}^\dagger
\tilde{n} \label{eqn:Llda2Cn} \\
&& \ \ \ + \omega (\epsilon_{xc}(\tilde{n}))^\dagger \tilde{n} \nonumber \\
&& \ \ \  -\omega \Re\left[d^\dagger \cI^\dagger \left( I - \frac{{\bf 1} {\bf
1}^\dagger}{{\bf 1}^\dagger {\bf 1}} \right) \tilde{n}\right] +
\frac{1}{8\pi} d^\dagger L d \nonumber \\
& = & \Tr \left( (-\frac{1}{2} L) P \right) + \Re \Tr \left( \cO \cJ
(\Diag \tilde{v}) \cI P \right) \label{eqn:Llda2P}\\
&& \ \ \ + \Re \Tr \left( \cO \cJ \Diag
(\epsilon_{xc}(\tilde{n})) \cI P \right) \nonumber \\
&& \ \ \  - \Re \Tr \left( \cO \cJ  \left( \Diag \Re \left[ (I -
\frac{{\bf 1} {\bf 1}^\dagger}{{\bf 1}^\dagger {\bf 1}})  \cI d
\right] \right) \cI P\right) +  \frac{1}{8\pi} d^\dagger L d.
\nonumber
\end{eqnarray}

      \paragraph{Poisson's equation}

Following the same procedure as (\ref{eqn:Pvar}) to compute the variation of
(\ref{eqn:Llda2Cn}) with differential 
changes in the Hartree field coefficients $\delta d$ gives, in the
operator prescription,
\begin{equation} \label{eqn:gradd2}
\partial_d \cL_{LDA} = - \omega \cI^\dagger \left( I - \frac{{\bf 1} {\bf
1}^\dagger}{{\bf 1}^\dagger {\bf 1}} \right) \tilde{n} +
\frac{1}{4\pi} L d.
\end{equation}
Thus,
\begin{eqnarray} \label{eqn:Poi2}
L d & = & 4 \pi \omega \cI^\dagger \left( I - \frac{{\bf 1} {\bf
1}^\dagger}{{\bf 1}^\dagger {\bf 1}} \right) \tilde{n}
\end{eqnarray}
is the representation of Poisson's equation in the orbital approach.
Note that the effective electron density $\tilde{n}$, rather than the actual
density $n$ is the source term.

In solving this equation for $d$, the inversion of $L^{-1}$ is again
understood to take place in the space orthogonal to the null space of
$L$, giving the result,
\begin{equation} \label{eqn:HartreeF2}
d=4 \pi \omega L^{-1} \cI^\dagger \left(I - \frac{{\bf 1} {\bf
1}^\dagger}{{\bf 1}^\dagger {\bf 1}}\right) \tilde{n}.
\end{equation}

      \paragraph{Energy functional}

With the result (\ref{eqn:HartreeF2}), the expression
for the LDA energy functional $E_{LDA}(C)$ in the operator prescription is
\begin{eqnarray} \label{eqn:Elda2}
E_{LDA}(C) 
& = & \Tr \left( f C^\dagger (-\frac{1}{2} L) C \right) + \omega \tilde{v}^\dagger
\tilde{n} \\
&& \ \ \ + \omega (\epsilon_{xc}(\tilde{n}))^\dagger \tilde{n} \nonumber \\
&& \ \ \ + \frac{1}{2} (\omega \tilde{n})^\dagger \left( I - \frac{{\bf 1} {\bf
1}^\dagger}{{\bf 1}^\dagger {\bf 1}} \right) \cI (-4 \pi L^{-1})
\cI^\dagger \left( I - \frac{{\bf 1} {\bf 1}^\dagger}{{\bf 1}^\dagger
{\bf 1}} \right) \omega \tilde{n}. \nonumber
\end{eqnarray}

  \subsection{Kohn-Sham equations} \label{sec:KohnSham}

In the preceding sections, we computed the gradient of the LDA
Lagrangian (\ref{eqn:Llda1Cn},\ref{eqn:Llda2Cn}) with respect to the
Hartree field coefficients to derive the Poisson equation.  In this
section, we consider the gradient with respect to the electronic
coefficients to determine the effective Schr\"odinger equation for the
electronic orbitals.

Before proceeding, we first note that as a consequence of the
Hellman-Feynman theorem, the expression for the gradient with respect
to the electronic coefficients of the LDA Lagrangian may also be used
for the gradient of the LDA energy functional, so long as one
substitutes the appropriate expression (\ref{eqn:HartreeF1} or
\ref{eqn:HartreeF2}) for the Hartree potential coefficients in terms
of the electronic coefficients $C$.  This follows from the fact that
the solutions (\ref{eqn:HartreeF1},\ref{eqn:HartreeF2}) ensure
$\partial_d \cL_{LDA}(C,d(C))=0$ and thus
\begin{eqnarray}
\frac{d}{dC} E_{LDA}(C) & = & \frac{d}{dC} \cL_{LDA}(C,d(C)) \label{eqn:HF} \\
& = & \partial_C \cL_{LDA}(C,d(C)) + \partial_d \cL_{LDA}(C,d(C))
\cdot \partial_C d(C) \nonumber \\ & = & \partial_C
\cL_{LDA}(C,d(C)). \nonumber
\end{eqnarray}
It thus suffices for us to determine the gradient of the Lagrangian
with respect to $C$.

This is done most efficiently by expressing all contributions to the
variation of $\cL_{LDA}$ in the form $\Tr M \delta P$ where
$M$ is a Hermitian matrix and $\delta P$ is the variation in the
density matrix defined in (\ref{eqn:defP}).  In terms
of the variations in $C$, such a variation takes the form
\begin{eqnarray} \label{eqn:PCThm}
\Tr (M \, \delta P)
& = &  \Tr \left( \left( f M C \right)^\dagger \, \delta C \right) +
\Tr \left( \delta C^\dagger \, \left(f M C\right) \right).
\end{eqnarray}
The contribution from such a variation to the total gradient with
respect to the real and imaginary parts of of $C$ are therefore the
real and imaginary components of $2 f M C$, respectively.

The final consideration when varying the wave function coefficients
is the orthonormality constraint (\ref{eqn:orthoconst}), which in
our matrix language becomes
\begin{equation} \label{eqn:conorth}
C^\dagger \cO C = I.
\end{equation}
The analytically continued functional approach\cite{prldynam} deals
with these constraints by introducing a set of orbital expansion
coefficients $Y$ which are {\em unconstrained} and may have any set of
overlaps,
\begin{equation} \label{def:U}
U \equiv Y^\dagger \cO Y.
\end{equation}
The coefficients $C$ are then defined as dependent variables through
the mapping $C=Y U^{-1/2}$, which ensures that the constraints
(\ref{eqn:conorth}) are always satisfied automatically, as easily
verified by direct substitution.

In terms of the independent variables $Y$, the density matrix 
appearing in the functionals
(\ref{eqn:Llda1P},\ref{eqn:Llda2P}) is
\begin{equation} \label{eqn:defP2}
P \equiv f C C^\dagger = f Y U^{-1} Y^\dagger.
\end{equation}
After some manipulation, the variations with respect to the $Y$ take the form
\begin{eqnarray} 
Tr M \delta P & = & \Tr M \delta(f Y U^{-1} Y^\dagger)\label{eqn:PYThm}\\
& = & \Tr \left( f \Pbar M Y U^{-1}
\right)^\dagger \delta Y + \Tr \delta Y^\dagger
\left( f \Pbar M Y U^{-1} \right), \nonumber
\end{eqnarray}
where we have used the relation $\delta (U^{-1}) = -U^{-1} (\delta U)
U^{-1}$ and defined
\begin{equation} \label{def:Pbar}
\Pbar \equiv (I- \cO Y U^{-1} Y^\dagger) = (I-\cO C C^\dagger),
\end{equation}
an idempotent projection operator onto the subspace spanned by the
unoccupied wave functions.  From these considerations, we have that the
contributions from such a variation to the gradient of $\cL_{LDA}$
with respect to the real and imaginary parts of the unconstrained
variables $Y$ are the real and imaginary parts of $2 f \Pbar M Y
U^{-1}$, respectively.

    \subsubsection{Energy functional prescription}

All but one of the dependencies of (\ref{eqn:Llda1P}) with the wave
functions are already explicitly in the form $\Tr M P$, so that their
contribution to the gradient may be evaluated immediately according to
(\ref{eqn:PCThm},\ref{eqn:PYThm}).  The one remaining dependency of
(\ref{eqn:Llda1P}) on the electrons is through the density dependence
of $\epsilon_{xc}(n)$.  This variation may also be cast into the form
of (\ref{eqn:PCThm},\ref{eqn:PYThm}).  To do so, we consider the
effect of this variation on $E_{xc}$ when expressed in the form
(\ref{eqn:Exc}).  Using the definition of $n$ (\ref{eqn:defn}) and one
of the
identities (\ref{eqn:diagids}), this leads to
\begin{eqnarray*}
n^\dagger \cJ^\dagger \cO \cJ \delta\epsilon_{xc}(n)
& = &
n^\dagger \cJ^\dagger \cO \cJ \left((\Diag \epsilon'_{xc}(n)) \delta n \right)\\
& = & \Tr \left[ \cI^\dagger \Diag\left( (\Diag \epsilon'_{xc}(n))
\cJ^\dagger \cO \cJ n \right) \cI \delta P \right].
\end{eqnarray*}

The total variation of (\ref{eqn:Llda1P}) as the wave functions vary
is thus,
$$
\delta \cL_{LDA} = \Tr \left(H_{sp} \delta P\right),
$$
where
\begin{eqnarray} 
H_{sp} & \equiv & -\frac{1}{2} L + 
\cI^\dagger  \Diag \left\{ \cJ^\dagger v + \cJ^\dagger \cO \cJ
\epsilon_{xc}(n)  \right.   \label{def:Hsp1} \\
&&  \ \ \ \ \ \  + \left(\Diag \epsilon'_{xc}(n)\right) \cJ^\dagger \cO
\cJ n \nonumber \\
&&
\left. \ \ \ \ \ \ - \Re \left(\cJ^\dagger (\cO-{ss^\dagger}/{\Omega})d\right)
 \right\} \cI \nonumber
\end{eqnarray}
is the energy functional prescription for the representation of the
single-particle Kohn-Sham Hamiltonian.  Using
(\ref{eqn:PCThm},\ref{eqn:PYThm}), the gradients which we seek are
therefore
\begin{eqnarray} \label{eqn:gradCY}
\partial_{C} \cL_{LDA}(C,d) & = & 2 f H_{sp} C \\
\partial_{Y} \cL_{LDA}(Y,d) & = & 2 f \Pbar H_{sp} Y U^{-1}, \nonumber
\end{eqnarray}
where $U$ and $\Pbar$ are defined in (\ref{def:U}) and
(\ref{def:Pbar}), respectively.

    \subsubsection{Operator prescription} \label{sec:KohnSham:KS}

The results for the orbital approach are directly analogous.  All of the
dependencies of the Lagrangian in the operator prescription
on the wave functions, except for that of
$\epsilon_{xc}$, are already explicitly in the form $\Tr M P$.  If we
take $\epsilon_{xc}$ as a function of $\tilde{n}$, then this variation
of $E_{xc}$ (second line of (\ref{eqn:Elda2})) is
\begin{eqnarray*}
\omega (\delta \epsilon_{xc}(\tilde{n}))^\dagger \tilde{n} 
& = & \omega (\Diag \epsilon'_{xc} \cdot \delta \tilde{n})^\dagger \tilde{n} \\
& = & \omega \left( (\Diag \epsilon'_{xc}) \tilde{n}\right)^\dagger \delta
\tilde{n}\\ 
& = & \Re \Tr \left( \cO \cJ (\Diag (\Diag \epsilon'_{xc} \cdot \tilde{n}))
\cI \delta P \right).
\end{eqnarray*}
This combines with the explicit dependency of the
exchange-correlation term in (\ref{eqn:Llda2P}) on $P$ to produce the total
exchange-correlation energy dependency in the form
\begin{eqnarray*}
\delta E_{xc} & = & \delta( \omega \epsilon_{xc}(\tilde{n})^\dagger \tilde{n})\\
& = & \Re \Tr \left( \cO \cJ (\Diag v_{xc}) \cI \delta P \right),
\end{eqnarray*}
where $v_{xc}$ is the usual exchange-correlation potential operator,
\begin{equation} \label{eqn:vxc}
v_{xc} \equiv \partial_{\tilde{n}} (\epsilon_{xc}(\tilde{n})\,{\tilde{n}} ) = \epsilon_{xc}'(\tilde{n})\,{\tilde{n}} + \epsilon_{xc}(\tilde{n}).
\end{equation}
Note that, as referred to in Sec.~\ref{sec:LagrEner:KS} this form is
obtained only when the exchange correlation energy per particle
$\epsilon_{xc}$ is evaluated as a function of the {\em effective
density} $\tilde{n}$, not as a function of the physical density $n$.

Combining this result for $E_{xc}$ with the remaining dependencies in
the operator prescription for the Lagrangian (\ref{eqn:Llda2P}) gives
the total variation in the form $\delta \cL_{LDA} = \Tr
H_{sp} \delta P$, where
\begin{equation} \label{def:Hsp2}
H_{sp} = -\frac{1}{2} L + 
\frac{1}{2}\left[ \cO \cJ \Diag \left\{
\tilde{v}+\left[ \left(\Diag
\epsilon_{xc}'(\tilde{n})\right)\,\tilde{n}+\epsilon_{xc}(\tilde{n}) \right]
-\Re \left[ (I - \frac{{\bf 1} {\bf 1}^\dagger}{{\bf 1}^\dagger {\bf 1}}) \cI
d \right] \right\} \cI + \mbox{h.c.}\right],
\end{equation}
and ``h.c.'' stands for the Hermitian conjugate of the entire matrix product
to the immediate left.
Again, once given this expression for $H_{sp}$, the expressions
(\ref{eqn:gradCY}) gives the gradients with respect to the wave
functions.

  \subsection{Solution techniques} \label{sec:Solutions}

\begin{figure}
\begin{center}
\scalebox{\figscale}{\scalebox{0.60}{\includegraphics{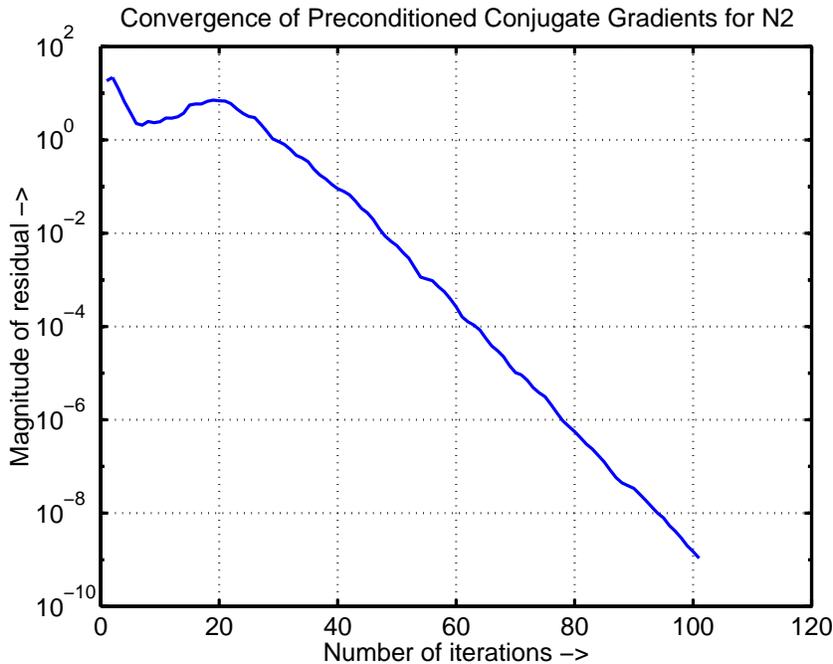}}}
\end{center}
\caption{Convergence of iterative conjugate-gradient solver used in
\cite{mgras} (results from \cite{JCP}).}
\label{fig:cg1}
\end{figure}

Locating the saddle point of the LDA Lagrangian (\ref{eqn:saddle})
corresponds to the simultaneous, self-consistent solution of the
Poisson and Schr\"odinger equations.  Although the Lagrangian
formulation allows for a direct search for the saddle point, most
calculations to date have solved Poisson's equation explicitly at each
iteration and then followed the gradients in (\ref{eqn:gradCY}) to
solve the Schr\"odinger equations using standard electronic structure
approaches.  The approaches applied to solve the Schr\"odinger
equations include conjugate gradient minimization with respect to
$Y$\cite{mgras} and iterative matrix diagonalization\cite{chou} and
Car-Parrinello simulated annealing\cite{tymczak} to find $C$.  The
solution of Poisson's equation in multiresolution analyses, on the
other hand, has garnered special attention.

To solve the Poisson (\ref{eqn:HartreeF2}), Wei and Chou\cite{chou}
introduced the innovation of using a separate Fourier representation
for the operator $L$ in the self-energy term of the Hartree field.
The interpretation of (\ref{eqn:HartreeF2}) when using such a
representation is the following procedure.  First, subtract the
average value from the effective density $\tilde{n}$, then perform a
discrete Fourier transform $\cI^\dagger$ of the result into Fourier
space, divide by the familiar factor of negative squares of wave
vectors $-q^2$ (ignoring the $q=0$ term), and scale by $4 \pi \omega$
to obtain the final result.  The disadvantage of this procedure is
that one must store and process data associated with an unrestricted
grid of points at the highest resolution.

\begin{figure}
\begin{center}
\scalebox{\figscale}{\scalebox{0.60}{\includegraphics{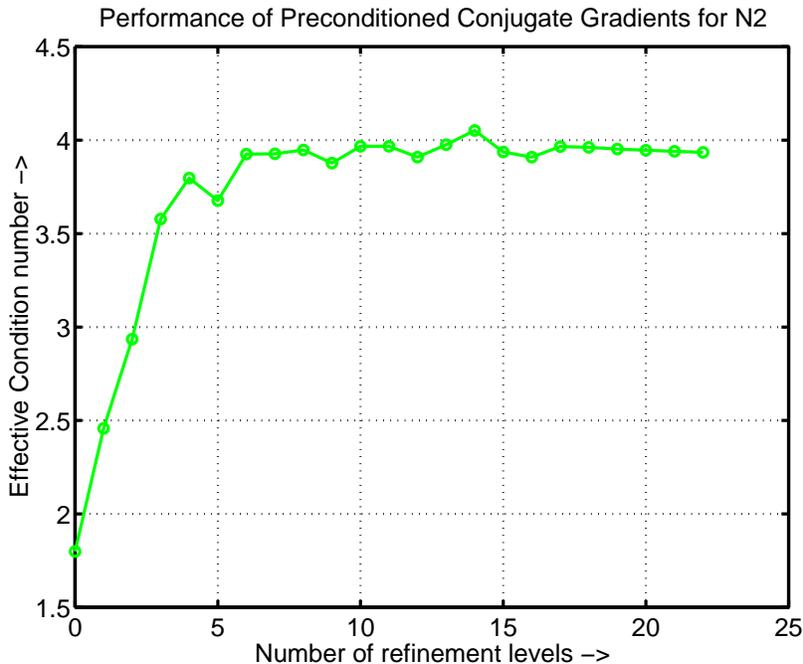}}}
\end{center}
\caption{Effective condition number of iterative conjugate-gradient
solver used in \cite{mgras} as a function of resolution. (Results from
\cite{JCP}.)}
\label{fig:cg}
\end{figure}

When working only with the points of a restricted grid, no such direct
solution is known.  Instead, iterative solutions have been used.  In
\cite{mgras}, a conjugate gradient solver with simple diagonal
preconditioning was used to solve Poisson's equation in a semicardinal
multiresolution analysis of the type described in
Section~\ref{subsubsec:semicmra}.  Figures~\ref{fig:cg1}~and~\ref{fig:cg}
show the results of a study of the performance of this
algorithm\cite{JCP}.  Figure~\ref{fig:cg1} shows the root mean square
magnitude of the residual as a function of iteration as this algorithm
is applied to the nitrogen molecule in a basis with seven levels of
refinement.  After an initial phase of about twenty iterations, the
convergence of the solution of Poisson's equation becomes nearly
perfectly exponential and reduces the residual by over ten orders of
magnitude in one hundred iterations, very good performance for a
system consisting of 14,000 degrees of freedom in which the Laplacian
operator has a nominal condition number of $2 \times 10^6$.

The slope of the exponential portion of the convergence curve in
Figure~\ref{fig:cg1} corresponds to a reduction in the error at each
iteration by 25\%.  Defining the effective condition number of 
the procedure as the inverse of this fractional improvement gives
an effective condition number $c \approx 4$.  Figure~\ref{fig:cg}
explores the behavior of this effective condition number as a function
of the number of refinement levels $k$.  Beyond about five refinement
levels, the effective condition number remains essentially constant,
implying that a constant number of iterations suffices to produce a
result of a given accuracy.  With the methods described in
Sec.~\ref{sec:homoops}, the computational work involved in each
iteration is linear in the number of coefficients $n_r$ in the
restricted multiresolution analysis.  For these problems, this simple
preconditioned conjugate gradient approach therefore produces the
solution to Poisson's equation with $O(n_r)$ operations.

Working with special linear combinations of the basis functions from
\cite{JCP} known as lifted wavelets\cite{sweldens:96}, Goedecker and
Ivanov have also used preconditioned conjugate gradients to solve
Poisson's equation and report similar convergence rates and an
$O(n_r)$ solution as well\cite{goedecker}.

Finally, multigrid techniques have been combined with multiresolution
analysis\cite{mgmra,dicle}.  This approach has also been shown to
produce $O(n)$ solutions of Poisson's equation in tests carried out in
one dimension\cite{dicle,dicle2}.

\section{Theory of Multiresolution Analysis} \label{sec:MRATS}

In this section we review multiresolution analysis, the conceptual
basis of wavelet theory.  Our presentation below departs somewhat in
notation and perspective from the traditional
discussion\cite{dau,chui,strang} to provide a presentation more suited
to physical calculations in multiple dimensions.

The basic idea behind multiresolution analysis is to provide a
framework for systematically increasing the resolution of a basis in
such a way as to always maintain a uniform description of space.  Each
stage of a multiresolution analysis doubles the resolution of the
basis in the precise mathematical sense described in
Secs.~\ref{sec:basisseq}~and~\ref{sec:MRA}.  In order for this to be
possible, the basis functions at the coarsest scale must be chosen
from among a special class of functions known as {\em scaling
functions}, whose explicit construction we describe in
Sec.~\ref{sec:scalingfunctions}.  To analyze a function into
contributions on separate scales and thus make possible the efficient
representation of electronic structure described in
Sec.~\ref{sec:FrameMRA}, a second type of basis function, known
as either as a {\em wavelet} or a {\em detail function}, is needed to
carry the finer scale information.  For a given choice of scaling
function for the coarsest scale, the use of only certain detail
functions results in the systematic doubling of resolution at each
stage in the multiresolution analysis.  Sections
\ref{sec:twoscalestart} and \ref{sec:waveletbasis} discuss this
doubling process, known as {\em two-scale decomposition}, and specify
the allowable classes of detail functions.  The last paragraph of
Section~\ref{sec:waveletbasis} sketches an alternate, more basis-set
oriented, viewpoint on two-scale decomposition which the reader may
find illuminating.  Once suitable scaling and detail functions are
decided upon, including them into the multilevel framework described
in Sec.~\ref{sec:MRA} completes the construction of the
multiresolution analysis.
Our review of multiresolution analysis concludes in
Section~\ref{sec:matrep} with the introduction a matrix representation
which is useful for developing and describing techniques for
electronic structure and other physical applications.

  \subsection{Bases of successive resolution} \label{sec:basisseq}

\begin{figure}
\begin{center}
\scalebox{\figscale}{\scalebox{0.80}{\includegraphics{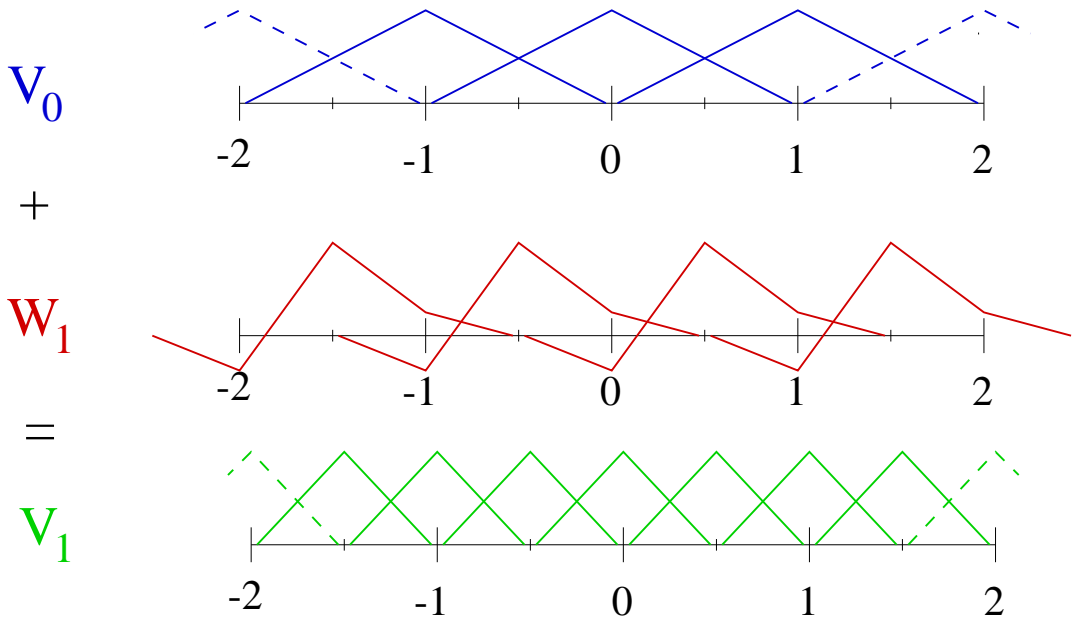}}} 
\end{center}
\caption{Bases involved in two-scale decomposition: coarse resolution
space ($V_0$), fine resolution space ($V_1$), detail space ($W_1$).}
\label{fig:MRA}
\end{figure}

The top panel of Figure \ref{fig:MRA} illustrates a typical coarse
resolution basis.  In a multiresolution analysis, this starting basis
consists of a set of $d-$dimensional functions $\phi(x-n)$ translated
to be centered on the points $n$ of $Z^d$, the cubic lattice of
integer spacing in $d-$dimensions.  The vector space of functions
which may be expressed as linear combinations of these basis functions
is denoted $V_0$.  All functions $F(x)$ in $V_0$ thus may be expanded
with expansion coefficients $F_{n}$ as
\begin{equation} \label{eqn:Ec}
F(x) = \sum_{n \in Z^d}{ F_{n} \phi(x-n) }.
\end{equation}

Starting from $V_0$, one approach to produce a set of bases of
successively higher resolution is to reduce the scale of space by
increasing powers of two, compressing both the lattice on which the
functions are centered and the basis functions themselves.  This
process defines a sequence of lattices $C_0, C_1, \ldots, C_N,
\ldots$, associated with a sequence of bases of increasing resolution,
$V_0, V_1, \ldots, V_N,
\ldots$,
\begin{eqnarray}
C_Q & \equiv & \left\{\frac{n}{2^Q}\right\}_{n \in Z^d}, \label{eqn:MRB}
\\
V_Q & \equiv & \span \left\{ \phi\left(2^Q (x-p)\right) \right\}_{p
\in C_Q} = \span \left\{ \phi(2^Q x-n) \right\}_{n \in Z^d}.  \nonumber
\end{eqnarray}
Note that the two representations for the basis functions of $V_Q$
given in the second line of (\ref{eqn:MRB}) are equivalent.  Below, we
shall use whichever form is most suitable.  The top and bottom panels
of Figure~\ref{fig:MRA} serve to illustrate the effect of going from
$V_0$ to $V_1$.
 
As we will see in
Secs.~\ref{sec:scalingfunctions}---\ref{sec:waveletbasis}, it is
possible under quite general circumstances to find additional basis
functions, referred to as {\em detail functions} or {\em wavelets},
which when combined with the basis functions for the space $V_Q$ span
the space of next higher resolution, $V_{Q+1}$.  The advantage of this
approach for increasing the resolution is that it clearly separates
the description into different length scales.  The original coarse
basis functions will continue to carry information about coarse scale
behavior, and the new functions carry the higher frequency details.
Accordingly, the space spanned by the detail functions is referred to
as the {\em detail space} $W_{Q+1}$.  The center panel of Figure
\ref{fig:MRA} illustrates the appearance of such a detail space.

All functions in $V_{Q+1}$ may be decomposed into the sum of one function
from the coarser space $V_Q$ and another from the detail space
$W_{Q+1}$, which means
\begin{equation} \label{eqn:TSD}
V_Q \oplus W_{Q+1}=V_{Q+1},
\end{equation}
where the addition is in this sense of combining vector spaces.  We
shall refer to this representation of the functions in $V_{Q+1}$ as
the sum of a function in $V_Q$ and a function in $W_{Q+1}$ as the {\em
two-scale representation} or the {\em two-scale decomposition}.  We
refer to the alternate representation of the same space functions in
terms of the compressed scaling functions of scale ${Q+1}$ as the {\em
single-scale} representation.

\begin{figure}
\begin{center}
\scalebox{\figscale}{\scalebox{1.2}{\includegraphics{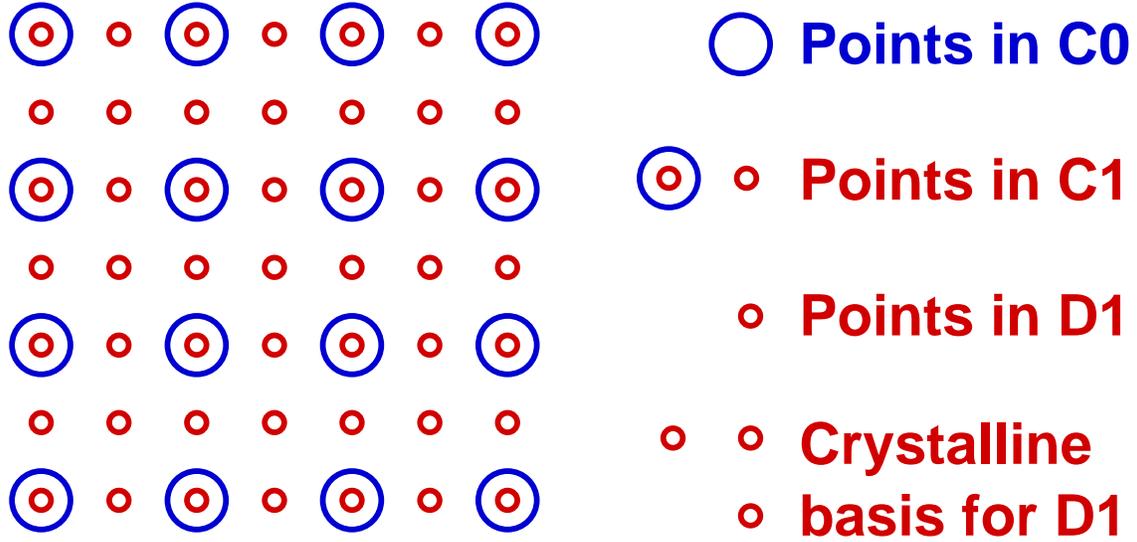}}} 
\end{center}
\caption{Lattices $C_0$, $C_1$ and detail ``crystal''
$D_1$ in $d=2$ dimensions.}
\label{fig:MRA2D}
\end{figure}

Clearly, for the single- and two-scale representations to be
equivalent, the spaces $V_{Q+1}$ and $V_Q \oplus W_{Q+1}$ must be of the same
dimension.  The basis spanning the finer space contains one function for
each point in $C_{Q+1}$, and the basis spanning the coarser space has one
function for each point in $C_Q$.  Therefore, the basis spanning $W_{Q+1}$
must consist of one function for each point in
\begin{eqnarray*}
D_{Q+1} & \equiv & C_{Q+1} - C_Q.
\end{eqnarray*}
In Figure~\ref{fig:MRA} for example, $C_0$ consists of all integer
points, $C_1$ contains all integer and odd half-integer points, and
$D_1$ consists of only the odd half-integer points.  Figure
\ref{fig:MRA2D} illustrates the appearance of these sets of points in
$d=2$ dimensions.

These various periodic sets of points appearing in the discussion of
multiresolution analysis may be described simply in the language of
crystallography.  The set $D_{Q+1}$ contains the ``decoration'' points
which when added to the lattice $C_Q$ produce the lattice $C_{Q+1}$.
$D_{Q+1}$ therefore is a crystal of points with a crystalline basis
containing $2^d-1$ points and with $C_Q$ as the underlying Bravais
lattice.
When dealing with a finite number of levels of resolution, $C_M
\subset C_{M+1} \subset \ldots \subset C_N$, it proves convenient to
have a notation for the set of points beyond a given level $Q$,
\begin{equation} \label{def:BP}
B_Q \equiv C_N-C_Q = D_{Q+1} \cup \ldots \cup D_N.
\end{equation}

  \subsection{Multiresolution analysis} \label{sec:MRA}

Given suitable detail spaces $W_{Q}$, iterative application of
two-scale decomposition can augment a space $V_M$ of fixed coarse
resolution to a space $V_N$ of {\em arbitrarily} high resolution,
\begin{eqnarray*}
V_N & = &  (V_0 \oplus W_1 \oplus \ldots \oplus W_M)\oplus W_{M+1} \oplus \ldots  \oplus  W_{N} \\
& = & V_M  \oplus  W_{M+1}  \oplus  \ldots  \oplus  W_{N},
\end{eqnarray*}
where $M<N$ in accord with the convention introduced in
Sec.~\ref{sec:FrameMRA:Intro}. 
Correspondingly, all physical functions may be described by a sequence
of evermore detailed components,
\begin{equation} \label{eqn:MRA}
f(x)  =  F_M(x)+G_{M+1}(x)+\ldots+G_{N}(x) 
\end{equation}
where $F_M(x)$ is a function in $V_M$ and the $G_Q(x)$ are functions from
the detail spaces $W_Q$.  We refer to the decomposition of a function
in this form as the {\em multiresolution analysis} of the function.

The analysis of functions in the form (\ref{eqn:MRA}) is what gives
rise to the great advantages of using multiresolution analysis.  As
illustrated previously in Figures~\ref{fig:onoff}~and~\ref{fig:wfsI}, the
detail component functions $G_Q(x)$ are physically relevant only
within the tiny volumes of the atomic cores and thus may be described
by a dramatically restricted subset of the basis.

  \subsection{Scaling functions} \label{sec:scalingfunctions}

For the condition (\ref{eqn:TSD}) to apply, each basis function of
the coarser basis $V_Q$, including in particular the function
$\phi(2^Q y)$ centered at origin, must be among the functions in the
space $V_{Q+1}$.  Thus, there must exist some set
of coefficients $c_{n}$ for which
$$
\phi(2^Q y)  =  \sum_{n \in Z^d} c_{n} \phi(2^{Q+1} y-n).
$$
This condition is universal and independent of the scale $Q$ as may be
seen by the substitution $x=2^Q y$, which re-scales the condition to
scale $Q=0$:
\begin{equation} \label{eqn:tsr}
\phi(x) = \sum_{n} c_{n} \phi(2x-n).
\end{equation}
Condition (\ref{eqn:tsr}) is known as the {\em two-scale relation}.
Those functions which satisfy (\ref{eqn:tsr}) are known as {\em
scaling functions}.  It easily may be verified that if $\phi(x)$
satisfies condition (\ref{eqn:tsr}), then all basis functions
$\phi(2^Q y-n)$ in $V_Q$ are also in the space $V_{Q+1}$, so that
indeed $V_Q
\subset V_{Q+1}$.

Because any multiresolution analysis must be based upon scaling
functions, it is useful to have a prescription for generating {\em all}
allowable scaling functions.  The two-scale relation (\ref{eqn:tsr})
expresses the function $\phi(x)$ as a convolution of the discrete
sequence $c_{n}$ and the continuous function $\phi(2x)$.  This
convolution takes the familiar form of a product in Fourier space,
\begin{eqnarray} 
\hat\phi(k)
& \equiv & \int
\frac{d^dx}{(2\pi)^d}\,\, e^{-ik \cdot x} \phi(x) \label{eqn:tsk} \\
& = & \int \frac{d^dx}{(2\pi)^d}\,\, e^{-ik \cdot x} \sum_n c_n
\phi(2x-n) \nonumber \\ 
& = & \left( \sum_{n} \frac{c_{n}}{2^d} e^{-i(k/2) \cdot n}
\right) \hat\phi(k/2) \nonumber \\
 & = & m_0(k/2)  \hat\phi(k/2), \nonumber
\end{eqnarray} 
where we have defined $m_0$, the {\em two-scale symbol} for the
scaling function $\phi(x)$, as
\begin{equation} \label{eqn:twoscalesymbol}
m_0(k) \equiv \sum_n \frac{c_{n}}{2^d}e^{-ik \cdot n}.
\end{equation}
By construction, the two-scale symbol is periodic on $2\pi C_0$, the
hyper-cubic lattice of lattice constant $2\pi$.  Note that the first
line of (\ref{eqn:tsk}) gives the normalization convention which we
shall use for Fourier transforms throughout this work.

The recursive nature of Eq. (\ref{eqn:tsk}) implies that the
Fourier transforms of scaling functions always take the form
\begin{eqnarray} 
\hat\phi(k) & = &  m_0(k/2) m_0(k/4) \ldots m_0(k/2^n) \ldots \hat
\phi(0) \nonumber \\
 & = & \left( \prod_{n=1}^\infty{m_0(k/2^n)} \right) \hat
\phi(0) \label{eqn:recurse}.
\end{eqnarray}
Conversely, substitution into the final lines of (\ref{eqn:tsk})
verifies that any function constructed from (\ref{eqn:recurse}) in
conjunction with any $2\pi C_0$-periodic function $m_0$ satisfying
$m_0(0)=1$ (so that the infinite product converges) results in a
proper scaling function.  Eq. (\ref{eqn:recurse}) therefore places the
set of acceptable scaling functions as in one-to-one correspondence
with $2\pi C_0$-periodic functions with value unity at the origin.
The interested reader may wish to consult \cite{dau} for a rigorous
discussion of technical mathematical details related to the infinite
product.

  \subsection{Two-scale decomposition theorem} \label{sec:twoscalestart}

With the allowable scaling functions $\phi(x)$ and hence the spaces
$V_Q$ defined, we now turn to the issue of what functions form appropriate
bases for the detail spaces $W_{Q+1}$.

Before specifying a basis for $W_{Q+1} \equiv V_{Q+1}-V_Q$, we first
determine what functions belong to this space by establishing the
nature of the excess freedom which exists in $V_{Q+1}$ over that which
exists in $V_Q$.  If $F$ and $f$ are two arbitrary functions from the
spaces $V_Q$ and $V_{Q+1}$, respectively, then
\begin{eqnarray*}
F(x) & = & \sum_{n \in Z^d} F_n \phi(2^Q x-n) \\
f(x) & = & \sum_{n \in Z^d} f_n \phi(2^{Q+1} x -n).
\end{eqnarray*}
These expressions are both convolutions of form similar to
(\ref{eqn:tsr}) and also give products in Fourier space:
\begin{eqnarray}
\hat F(k) & = & \left( \sum_{n} \frac{F_{n}}{2^{Qd}}
e^{-i(k/2^Q) \cdot n} \right) \hat\phi(k/2^Q) \nonumber \\
& \equiv & m_F(k/2^Q) \hat\phi(k/2^Q) \nonumber \\ 
& = & m_F(k/2^Q) m_0(k/2^{Q+1}) \hat\phi(k/2^{Q+1}),
\label{eqn:FkI}
\end{eqnarray}
where we have used the two-scale relation (\ref{eqn:tsk}); and
\begin{eqnarray}
\hat f(k) & = & \left( \sum_{n} \frac{f_{n}}{2^{(Q+1)d}}
e^{-i(k/2^{Q+1})\cdot n} \right) \hat\phi(k/2^{Q+1}) \nonumber \\
 & \equiv & m_f(k/2^{Q+1}) \hat\phi(k/2^{Q+1}), \label{eqn:fkI}
\end{eqnarray}
where $m_F(q)$ and $m_f(q)$ are two new $2\pi C_0$-periodic functions
defined from Fourier series constructed with the sequences $F_{n}$ and
$f_{n}$ respectively.  Thus, $F(x)$ and $f(x)$ are in $V_Q$ and
$V_{Q+1}$ if and only if their Fourier transforms are of the forms in
(\ref{eqn:FkI}) and (\ref{eqn:fkI}) with $2\pi C_0$-periodic functions
$m_F$ and $m_f$.

We can use these results to verify quickly that any function $F$ in
$V_Q$ is also in $V_{Q+1}$.  First note that if $F$ is in $V_Q$, then
there exists a suitable $2\pi C_0$-periodic function $m_F$ to
satisfy (\ref{eqn:FkI}).  Using this function, then clearly the choice
\begin{equation} \label{eqn:MRA1}
m_f(q/2) \equiv m_F(q) m_0(q/2),
\end{equation}
will cast $F$ in the form (\ref{eqn:fkI}).  All that remains to
confirm that this places $F$ in $V_{Q+1}$ is to show that the function
$m_f$ defined as in (\ref{eqn:MRA1}) is indeed always $2\pi C_0$-periodic, as
is easily verified using the fact that $m_0$ and $m_F$ both have $2\pi
C_0$-periodicity.

Naturally, it is not possible to describe a general fine-scale
function $f$ from $V_{Q+1}$ as a function $F$ from the coarser
space $V_Q$.  To write a function of the form (\ref{eqn:fkI})
in the form (\ref{eqn:FkI}), would require
\begin{equation} \label{eqn:MRA2}
m_F(q) \equiv m_f(q/2)/m_0(q/2),
\end{equation}
to be a $2\pi C_0$-periodic function.  The function $m_f(q/2)$,
however, can be any $4\pi C_0$-periodic function of $q$.  The failure of $f$
to be in $V_Q$ thus may be viewed as the inability of the $2\pi
C_0$-periodic function $m_F(q)$ to describe the full freedom present in
$m_f(q/2)$, a general $4\pi C_0$-periodic function of $q$.

As a hint as to how to capture the this full freedom, we note that the
Monkhorst-Pack theory of Brillouin zone sampling\cite{MP} shows that
any $4\pi C_0$-periodic function may be expressed as a sum of $2^d$
separate functions, each with a special $2\pi C_0$-periodicity.  The
particular $2\pi C_0$-periodicities chosen by Monkhorst and Pack are
those associated with points of the reciprocal lattice of $4\pi C_0$
which fall in the first Brillouin zone of the reciprocal lattice of
$2\pi C_0$, namely the set of k-points $k_i \equiv
\eta_i/2$ where
\begin{equation} \label{eqn:defeta}
\eta_i \in \{0,1\}^d.
\end{equation}
For instance, in $d=3$ dimensions, 
$$
\{\eta_i\}\equiv \{(0,0,0), (0,0,1), (0,1,0), (0,1,1),
(1,0,0), (1,0,1), (1,1,0), (1,1,1)\}.
$$
Thus, all $4\pi C_0$-periodic
functions of $q$ may be written as a linear combination of functions
$p_i(q)$ with the special $2\pi C_0$-periodicities described by the
condition that
$$
p_i(q+Q)=e^{-i Q \cdot (\eta_i/2)} p_i(q)
$$
for all $Q$ in $2\pi C_0$.
For convenience, we will adopt the notation that $\eta_0 = 0$ and will
index sums which omit this zero vector by $\alpha$, except for two 
cases noted explicitly in the text in Sec.~\ref{sec:wavelets}.

From the theory Monkhorst and Pack, we thus conclude that if we take
the $2\pi C_0$-periodic function $m_F(q)$, which has the same
periodicity as $p_0(q)$ above, and augment it with $2^d-1$ new
functions $m_{G_\alpha}(q)$ with the periodicities of the
$p_\alpha(q)$,
\begin{equation} \label{eqn:exG}
m_{G_\alpha}(q) \equiv \sum_{n \in C_0} G_{\alpha,n} e^{-i(n+
\eta_\alpha/2) \cdot q}
\end{equation}
for some set of Fourier coefficients $G_{\alpha,n}$, the collection of
these functions represents the same freedom present in the $4\pi C_0$-periodic
function $m_f(q/2)$.  Thus, for any set of {\em fixed} 
$2\pi$-periodic functions $m_\alpha(q)$ introduced to play roles
parallel to that of $m_0(q)$, we may always find
functions $m_{G_\alpha}(q)$ of the form (\ref{eqn:exG}) so that
\begin{equation} \label{eqn:tsdm}
m_f(q/2)  =  m_F(q) m_0(q/2) +  \sum_{\alpha=1}^{2^d-1}
m_{G_\alpha}(q) m_\alpha(q/2),
\end{equation}
provided the choice of the fixed functions $m_\alpha(q)$ is not
pathological.  Eq. (\ref{eqn:tsdm}), for instance, will not hold in
general if we choose $m_\alpha(q)=0$ for all $\alpha$.  For reasons
which will become clear in the next section, the $m_\alpha(q)$ are
called the {\em detail} or {\em wavelet} symbols.

The {\em Two-Scale Decomposition Theorem}~\cite{dau,chui} specifies
precisely the conditions under which the choice of $m_\alpha(q)$ is
not pathological so that $m_f(q/2)$ indeed may be decomposed in the
form (\ref{eqn:tsdm}).  Translating (\ref{eqn:tsdm}) by each of the
$2^d$ vectors $2\pi\eta_i$ and simplifying the result using the
periodicity properties of $m_F$ and $m_{G_\alpha}$, gives the
following system of equations:
\begin{eqnarray}
m_f(q/2) & = & m_F(q) m_0(q/2) +
 \sum_{\alpha=1}^{2^d-1} m_{G_\alpha}(q) m_\alpha(q/2) \label{eqn:syseq} \\ 
& \vdots &  \nonumber \\
m_f(q/2+\pi \eta_i) & = & m_F(q) m_0(
 q/2+\pi \eta_i) + \sum_{\alpha=1}^{2^d-1} (-1)^{\eta_i \cdot
 \eta_\alpha} m_{G_\alpha}(q) m_\alpha(q/2+\pi
 \eta_i) \nonumber \\ 
& \vdots &  \nonumber \\
m_f(q/2+\pi \eta_{2^d-1}) & = & m_F(q) m_0(q/2+\pi
 \eta_{2^d-1}) + \sum_{\alpha=1}^{2^d-1} (-1)^{\eta_{2^d-1}
 \cdot \eta_\alpha} m_{G_\alpha}(q) m_\alpha(q/2+\pi
 \eta_{2^d-1}). \nonumber
\end{eqnarray}
So long as this $2^d \times 2^d$ system is not singular, we may always
solve directly for the variable functions $m_F$ and $m_{G_\alpha}$
in terms the function $m_f$ we are to reproduce and the $2^d$ fixed
functions $m_i$ and thus accomplish the two-scale decomposition
(\ref{eqn:tsdm}).
Replacing $k \equiv q/2$, the final condition for two-scale
decomposition to obtain is therefore
\begin{equation} \label{eqn:MRAdet}
\left|\begin{array}{ccc}
m_0(k) & \ldots & m_{2^d-1}(k) \\
\vdots & & \vdots \\
m_0(k+\pi\eta_i) & \ldots & (-1)^{\eta_{2^d-1} \cdot  \eta_i}m_{2^d-1}(k+\pi\eta_i) \\
\vdots & & \vdots \\
m_0(k+\pi\eta_{2^d-1}) &  \ldots & (-1)^{\eta_{2^d-1} \cdot \eta_{2^d-1}} m_{2^d-1}(k+\pi\eta_{2^d-1})
\end{array}\right| \ne 0 \mbox{\ \ \ for all $k$}.
\end{equation}

We conclude this section by observing the implications of
(\ref{eqn:MRAdet}) for functions in the spaces $V_Q$ and $V_{Q+1}$.
Making the replacement $q=k/2^Q$ in (\ref{eqn:tsdm}), which we now
know holds whenever the determinant (\ref{eqn:MRAdet}) is non-zero,
multiplying through by $\hat \phi(k/2^{Q+1})$ and using
(\ref{eqn:FkI}), we obtain
\begin{eqnarray}
\hat f(k) & = & m_F(k/2^Q) m_0(k/2^{Q+1}) \phi(k/2^{Q+1}) +  \sum_{\alpha=1}^{2^d-1}
m_{G_\alpha}(k/2^Q) m_\alpha(k/2^{Q+1}) \phi(k/2^{Q+1}) \nonumber  \\
& \equiv & \hat F(k) + \hat G(k), \label{eqn:tsdg}
\end{eqnarray}
which decomposes a general function $f \in V_{Q+1}$ into a coarse
function $F$ in $V_Q$ and an additional function $G$, which carries
the finer scale, detailed information about $f$ and thus lies in the
space $W_{Q+1}$.

  \subsection{Wavelets} \label{sec:waveletbasis}

To determine the basis functions for $W_{Q+1}$, note that by its
definition (\ref{eqn:TSD}), the 
detail space contains precisely the functions $\hat G(k)$ from
(\ref{eqn:tsdg}).  Using the expansions (\ref{eqn:exG}) for the
functions $m_{G_\alpha}$, the Fourier transform of (\ref{eqn:tsdg}) gives
\begin{eqnarray}
G(x) & = & \sum_{n \in C_0} \sum_\alpha \left( 2^{Qd} G_{\alpha,n}
\right)
\psi_\alpha\left(2^Q (x-\frac{n+\eta_\alpha/2}{2^Q}) \right),
\end{eqnarray}
where
\begin{equation} \label{eqn:wk}
\hat \psi_\alpha(k/2^Q) \equiv m_\alpha(k/2^{Q+1}) \hat
\phi(k/2^{Q+1}).
\end{equation}
The detail space is therefore
\begin{eqnarray}
W_{Q+1} & \equiv & \span \left\{ 
\psi_\alpha\left(2^Q (x-\frac{n+\eta_\alpha/2}{2^Q}) \right)
\right\}_{n \in C_0, \alpha=1\ldots 2^d-1}, \label{eqn:MRW}
\end{eqnarray}
where the $\eta_\alpha$ are defined as the non-zero vectors in (\ref{eqn:defeta}).
The new functions $\psi_\alpha(x)$, defined in terms of the
detail symbols $m_\alpha$ from (\ref{eqn:tsdm}), when scaled and centered
appropriately, make up the basis for $W_{Q+1}$ and are thus the
detail functions, or wavelets.  Note that the extra phase factors in
(\ref{eqn:exG}) have ensured naturally that each detail function
$\psi_\alpha(x)$ is associated with a unique point in $D_{Q+1}$, as
anticipated in Sec.~\ref{sec:basisseq}.

The functional form of the detail functions is revealed by Fourier
expanding the $2\pi 
C_0$-periodic symbols $m_\alpha$,
\begin{equation} \label{eqn:defmq}
m_\alpha(q) \equiv \sum_{n \in C_0} \frac{d_{\alpha,n}}{2^d}
e^{-i q \cdot n},
\end{equation}
and transforming the $Q=0$ case of (\ref{eqn:wk}) to real space,
\begin{eqnarray}
\psi_\alpha(x) & = & \sum_{n \in C_0} d_{\alpha,n} \phi(2x-n), \label{eqn:dwlt}
\end{eqnarray}
Thus, as $V_Q \oplus W_{Q+1}=V_{Q+1}$ requires, each detail function from
$W_{Q+1}$ is a linear combination of scaling functions from $V_{Q+1}$.
The allowed choices for these linear combinations are those which
satisfy (\ref{eqn:MRAdet}) with the $m_\alpha(q)$ defined through
(\ref{eqn:defmq}).

An alternate route to two-scale decomposition is to note directly from
$V_Q \oplus W_{Q+1}=V_{Q+1}$ that the $\psi_\alpha(x)$ must be formed
from linear combinations of the scaling functions from $V_{Q+1}$.
Then, to ensure that the two-scale and single scale representations
are equivalent, one must show that each basis function of $V_{Q+1}$
may be expanded in terms of the basis functions from $V_Q$ and
$W_{Q+1}$.  When Fourier transformed, the system of equations which
must be solved to find the corresponding expansion coefficients is
precisely (\ref{eqn:syseq}), leading once again to the determinant
condition (\ref{eqn:MRAdet}) for the acceptable choices for the detail
functions.

  \subsection{Matrix language} \label{sec:matrep}

To develop a matrix language useful for the application of
multiresolution analysis to physical problems, we begin by developing
a language for the various descriptions of the functions $f(x)$ in
$V_N$ provided by multiresolution analysis.  There is the single-scale
representation,
$$
f(x) = \sum_{p \in C_N} (\vec F_N)_p \phi(2^N (x-p)),
$$
where we define $\vec F_N$ as a column vector whose components are the
coefficients associated with the scaling functions of $V_N$.  Note
that we index the components of $\vec F_N$ by the points $p$ on which the
basis functions are centered.  Alternately, we may represent $f(x)$
using the multiresolution analysis which starts from a scale $M \le
N$,
$$
V_N=V_M  \oplus  W_{M+1}  \oplus  \ldots  \oplus  W_N.
$$
This gives instead
\begin{equation} \label{eqn:FNM}
f(x) = \sum_{p \in C_M} (\vec F_{N:M})_p \, \phi(2^M(x-p)) +
\sum_{P=M+1}^N \,\, \sum_{p \in D_P} (\vec F_{N:M})_p \, \psi(2^{P-1}(x-p)),
\end{equation}
where, regardless of its scale, the expansion coefficient associated
with the point $p$ of the multiresolution representation spanning
scales $M$ through $N$ is indexed as simply $(\vec F_{N:M})_p$.  Because
the expansion coefficient associated with each point depends on the
scales present in the multiresolution analysis, it is critical to
include the subscript $N:M$ on the vector $(\vec F_{N:M})_p$ in order to specify the scales in the
multiresolution analysis.  Finally, we note that under this convention
the single-scale coefficients are also given by $\vec F_N \equiv \vec
F_{N:N}$.

Next, the two-scale relation relation allows us to write both the
scaling functions on scale $M$ and the detail functions on scale $M+1$
as linear combinations of scaling functions on scale $M+1$.  With
these replacements, (\ref{eqn:FNM}) becomes
\begin{eqnarray} 
f(x) & = & \sum_{p \in C_{M}} (\vec F_{N:M})_p  \sum_{n \in C_0}
c_n \phi(2^{M+1}(x-p)-n)) \nonumber \\
&& + \sum_{p' \in D_{M+1}} (\vec F_{N:M})_{p'} \sum_{n \in Z^d}
d_{\alpha(p'),n} \phi(2^{M+1}(x-p')-n))
 \nonumber \\
&& + \sum_{P=M+2}^N\ \sum_{p \in D_P} (\vec F_{N:M})_p\, \psi(2^{P-1}(x-p))
\nonumber
\end{eqnarray}
where the $\alpha(p')$ in the coefficients $d_{\alpha(p'),n}$ in the sum for
$D_{M+1}$ are defined so that $p' = p_0+\eta_{\alpha(p')}/2^{M+1}$ for some
$p_0$ in $C_M$.  By collecting terms, this expression
rearranges into the expansion for $f(x)$ in the $(M+1)^{\mbox{th}}$
scale representation,
\begin{eqnarray} 
f(x) & \equiv & \sum_{q \in C_{M+1}} (\vec F_{N:M+1})_{q}
\phi(2^{M+1}(x-q)) \nonumber \\
&& + \sum_{P=M+2}^N \sum_{p \in D_P} (\vec F_{N:M+1})_p\, \psi(2^{P-1}(x-p)). \label{eqn:FNMP1}
\end{eqnarray}
The linear map which this procedure represents between $\vec F_{N:M}$
and $\vec F_{N:M+1}$ is most compactly represented as a matrix
equation,
\begin{equation} \label{eqn:tsrM}
\vec F_{N:M+1} \equiv \cI_{M+1,M} \vec F_{N:M}.
\end{equation}

The {\em wavelet transform}, the recovery of the single-scale
representation $\vec F_N$ from the multiresolution representation
$\vec F_{N:M}$, is the result of cascading together 
these representations of the two-scale relation,
\begin{eqnarray}
\vec F_N \equiv \vec F_{N:N}
& = & (\cI_{N,N-1} (\cI_{N-1,N-2} \ldots (\cI_{M+1,M} \vec F_{N:M}))) \label{eqn:defINM} \\
& \equiv & \cI_{N:M} \vec F_{N:M}. \nonumber
\end{eqnarray}
Note that many authors, particularly in the field of signal
processing, refer to the operator $\cI_{N:M}$ as the ``inverse''
wavelet transform.  Because the primary mode of operation in physical
calculations is to treat the multiscale expansion coefficients as the
{\em independent} variables, in this work we refer to this process of
producing a function from its multiscale coefficients as the {\em
forward} transform, which conforms to our usage in
Sec.~\ref{sec:MatrixFramework}.

The transform operator appearing in (\ref{eqn:defINM}) may be defined
as connecting any two scales $P>Q$,
\begin{equation} \label{eqn:matMRA}
\cI_{P:Q} \equiv \prod_{R=P}^{Q+1} \cI_{R,R-1},
\end{equation}
where in this expression, as throughout this work, we adopt the convention that
non-commutative matrix products proceed in order from left to right
with the lower index of the product appearing leftmost in the product
and the upper index appearing rightmost.
A direct result of the definition (\ref{eqn:matMRA}) is that
$\cI_{M+1:M} \equiv \cI_{M+1,M}$.  Also,
\begin{equation}
\cI_{P:Q} = \cI_{P:R} \cI_{R:Q} \mbox{\ \ for ($P > R > Q$)}.
\end{equation}

Finally, to discuss {\em restricted} multiresolution analyses, where
functions are selectively removed from the basis as described in
Sec.~\ref{sec:FrameMRA:Intro}, we introduce projection matrices which
``zero out'' coefficients associated with the points not contained in
a given grid $G$,
\begin{equation} \label{def:cP}
(\cP_G)_{pq} \equiv \left\{ \begin{array}{cl}
\delta_{pq} & \mbox{if $p \in G$} \\
0 & \mbox{if $p \not \in G$}
\end{array}\right. .
\end{equation}
The product of two such projections is the projection associated
with the intersection of the associated grids, and, for disjoint
grids, the sum of two such projections is the projection
associated with the union of the two,
\begin{eqnarray} 
\cP_{G_1} \cP_{G_2} & = & \cP_{G_1 \cap G_2} \label{eqn:basePPids} \\
\cP_{G_1} + \cP_{G_2} & = & \cP_{G_1 \cup G_2} + \cP_{G_1 \cap G_2}. \nonumber
\end{eqnarray}
Most statements about the sets of points $C_Q$, $D_Q$, $B_Q$ defined
in Sec.~\ref{sec:basisseq} may be derived in terms of and translated into
identities involving the projections $\cP$.  Two facts of which we
shall make frequent use are that the finer grids contain the coarser
grids and that the detail points of level $P$ are contained only in
the grids of level $Q \ge P$.  In terms of projections, these
statements are
\begin{eqnarray}
\cP_{C_P} \cP_{C_Q} & = & \cP_{C_{\mbox{min $(P,Q)$}}}
\label{eqn:PPid} \\
&& \nonumber \\
\cP_{D_P} \cP_{C_Q} & = & \left\{\begin{array}{cc}
0 & Q<P \\
\cP_{D_P} & Q \ge P
\end{array}\right., \nonumber
\end{eqnarray}
respectively.  Also, because the linear map represented by
$\cI_{Q+1,Q}$ replaces scaling functions and detail functions on
scales $Q$ and $Q+1$ with scaling functions on scale $Q+1$, it acts
independently of the coefficients associated with scales beyond $Q+1$.
Thus,
\begin{equation} \label{eqn:IBQ1}
\cP_{B_{Q+1}} \cI_{Q+1,Q} = \cI_{Q+1,Q} \cP_{B_{Q+1}}  = \cP_{B_{Q+1}}
\end{equation}

\section{Bases for Multiresolution Analysis} \label{sec:semicbases}

Even with the restrictions of multiresolution analysis as laid down in
the previous section, significant freedom remains in the choice of the
scaling and detail functions.  Motivated by differing aspects of the
calculations, researchers in electronic structure have employed
different multiresolution analyses.  Ultimately, the optimal approach
may be to use different bases for different phases of the calculation.

The simplicity of working with orthonormal basis sets has led
several authors dealing with with electronic structure\cite{chou,tymczak}
to use the wavelets of Daubechies\cite{dau}.  Section
\ref{sec:wavelets} below reviews the construction of these bases.

The primary issue in the choice of bases in problems in the physical
sciences, however, is the ability of the basis to represent the
relevant physical functions, not necessarily the analysis of these
functions into separate frequency components.  While critical in
signal analysis, orthonormality is not as crucial in electronic
structure, as illustrated by the great success of the chemistry
community with the use of Gaussian bases\cite{gausschem}.  Several
authors therefore have exploited the freedom of not being confined to
orthonormal bases to improve the representation of physical
problems\cite{mgras,JCP,dicle,FrohlichSchneider:94,JCP,BertoluzzaNaldi:96,goedecker}.
The majority of these applications
use multiresolution analyses based upon the scaling functions of
Deslauriers and Dubuc\cite{DD}, which were developed independently in
several
fields\cite{Lemarie:91,ChuiShi:92,Donoho.unpublished:92,BeylkinSaito:92,BeylkinSaito:93,AldroubiUnser:93,Unser:93,mgras,Lewis:94}.
The great advantage of using these scaling functions is that they
function as finite element functions and thus provide both good
interpolation and transform properties.

Teter\cite{tetercom} was the first in the electronic structure
community to recognize the advantage of adapting the ideas of finite
elements to bases with a multilevel structure.  A key concept from
finite element theory is the property of {\em cardinality}, the
condition that each basis function have value zero at every point of
the finite element grid except for the one point with which it is
associated.  Unfortunately, it is impossible to maintain cardinality
in a multiresolution analysis because smooth basis functions from
coarse scales cannot oscillate so as to be zero on the grid points of
the finer scales.  Instead, Teter suggested the construction of bases
where cardinality is maintained for the points of coarser scales but
sacrificed for finer scales.  Section~\ref{subsec:defsemic} gives the
precise definition for such {\em semicardinal} bases, which developed
out of this idea, and describes the nature of multiresolution analyses
which are semicardinal.

Such semicardinal bases have the remarkable property that expansion
coefficients for a function may be extracted {\em exactly} from the
values of the the function on a finite set of points in space.  This
property makes these bases ideal for non-linear problems because
non-linear couplings such as those in the LDA Lagrangian are
computed in real-space on a finite grid and the corresponding
expansion coefficients must be extracted from the resulting values.
This highly desirable exact extraction property requires
the detail functions to have non-zero integral, and thus prevents such bases
from being wavelet bases in the technical sense described in
\cite{dau}.
Such semicardinal bases were used in \cite{mgras,aps,JCP} and are
described in detail in Sec.~\ref{subsec:defsemic} below.

Finally, Goedecker and Ivanov\cite{goedecker} have suggested that the
Poisson equation be solved with basis functions with zero integral and
higher multipole moments, as formed by linear combinations of cardinal
scaling functions according to the lifting scheme of
Sweldens\cite{sweldens:96}.  Because working with functions of zero
integral extends the range of the basis functions and disrupts the
exact reconstruction property which is so useful in treating
non-linear interactions, an intriguing possibility for future research
would be to use semicardinal basis for the non-linear phases of the
LDA calculations and a lifted basis for the solution of Poisson's
equation, should lifted wavelets prove superior in the solution of
Poisson's equation.  (See discussion in Sec.~\ref{sec:Solutions}.)

  \subsection{Orthonormal bases} \label{sec:wavelets}

Daubechies was the first to construct orthonormal bases of wavelets
with compact support\cite{dau}.  We now give a brief review of the
discussion in \cite{dau} within our present conventions and language.

The scaling functions on the coarsest scale of an orthonormal
multiresolution basis are themselves orthonormal.  This condition may
be re-scaled to the scale $Q=0$, where it becomes
\begin{eqnarray*}
\delta_{nm}
& = & \int d^dx\,\, \phi^*(x-n) \phi(x-m) \\
& = & \int d^dx\,\, (\int d^dk\,\, \hat \phi(k) e^{ik \cdot (x-n)})^* (\int
d^dk'\,\, \hat \phi(k') e^{ik' \cdot (x-m)} \\
& = & (2\pi)^d \int d^dk\,\, \phi^*(k) \phi(k) e^{ik \cdot (n-m)}
\mbox{\ \ \ for all $n,m$ in $C_0$}.  
\end{eqnarray*}
The entire reciprocal space of vectors $k$ may be partitioned into images of
the first Brillouin zone centered on the points $G$ of the reciprocal
lattice $2\pi C_0$.  Writing the integral in this way gives,
\begin{eqnarray*}
\delta_{nm}
& = & (2\pi)^d \int_{B.Z.} d^dk\,\, \sum_{G \in 2\pi C_0} \phi^*(k+G) \phi(k+G) e^{ik \cdot (n-m)},
\end{eqnarray*}
where we have used the fact $e^{iG \cdot (n-m)}=1$ for the reciprocal lattice
vectors $G$.  The fact that this is true for all $n,m$ in $C_0$,
combined with the Fourier space version of two-scale relation
(\ref{eqn:tsk}), gives
\begin{eqnarray}
(2\pi)^{-2d}
& = & \sum_{G \in 2 \pi C_0} \phi^*(k+G) \phi(k+G) \label{eqn:sG}\\
& = & \sum_{G \in 2 \pi C_0} \left( m_0(\frac{k+G}{2}) \phi(\frac{k+G}{2})
\right)^*  \left( m_0(\frac{k+G}{2}) \phi(\frac{k+G}{2})\right). \nonumber
\end{eqnarray}
Finally, defining $q\equiv k/2$ and re-expressing the sum over $G/2$ as a sum
over the reciprocal lattice and the decoration points $\pi\eta_i$ with $\eta_i$ defined in (\ref{eqn:defeta}), gives
\begin{eqnarray*}
(2\pi)^{-2d}
& = & \sum_{G' \in 2\pi C_0} \sum_i m_0^*(q+({G'}+\pi \eta_i))
\phi^*(q+({G'}+\pi \eta_i)) m_0(q+({G'}+\pi \eta_i))
\phi(q+({G'}+\pi \eta_i)) \\
& = & \sum_i  m_0^*(q+\pi \eta_i) m_0(q+\pi \eta_i) \sum_{G' \in 2\pi C_0}
\phi^*(q+({G'}+\pi \eta_i)) 
\phi(q+({G'}+\pi \eta_i)) \\
& = & \sum_i  m_0^*(q+\pi \eta_i) m_0(q+\pi \eta_i) (2\pi)^{-2d},
\end{eqnarray*}
where we have used the periodicity of $m_0$ and the first line of
(\ref{eqn:sG}).  We thus conclude that orthonormality among the scaling
functions places the following condition on the two-scale symbols,
\begin{equation} \label{eqn:m0orth}
\sum_{i=0}^{2^d-1}  m_0^*(q+\pi \eta_i) m_0(q+\pi \eta_i) = 1.
\end{equation}

In Sec.~\ref{sec:semicbases:functions}, we show how to construct
functions $M_0(q)$ which obey the constraint
$$
\sum_{i=0}^{2^d-1}  M_0(q+\pi \eta_i) = 1.
$$
Comparing with (\ref{eqn:m0orth}), we see that to construct
orthonormal scaling functions, one first finds such an $M_0(q)$ and
then takes the ``square root'' to find the appropriate two scale
symbol $m_0(q)$.  This procedure is described in \cite[Sec.~6.1]{dau},
which also tabulates the resulting two-scale coefficients.

Next, to construct orthonormal {\em detail} functions, one must
assure that the wavelet $\psi_\alpha$ for each decoration point
$\alpha=1\ldots2^d-1$ be orthogonal to the scaling
functions,
\begin{eqnarray}
0
& = & \int d^dx\,\, \phi^*(x-n) \psi_\alpha(x-m-\eta_\alpha/2)  \nonumber \\
& = & (2\pi)^d \int_{B.Z.} d^dk\,\, \sum_G  \hat\phi^*(k+G) \hat
\psi_\alpha(k+G) e^{ik \cdot (m-n) - i(k+G)\cdot \eta_\alpha/2}.\nonumber
\end{eqnarray}
Through an analogous analysis to that above, we will finally arrive at
\begin{equation} \label{eqn:m0maorth}
0 =  \sum_i  m_0^*(q+\pi\eta_i)
\left( e^{-i \pi \eta_i \cdot \eta_\alpha} m_\alpha(q+\pi\eta_i) \right).
\end{equation}
One must also ensure orthonormality among the
$\psi_\alpha(x)$,
\begin{eqnarray*}
\delta_{\alpha\beta} \delta_{nm} & = &
\int d^dx\,\, \psi_\alpha^*(x-n-\eta_\alpha/2)
\psi_\beta(x-m-\eta_\beta/2),
\end{eqnarray*}
which leads to
\begin{equation}  \label{eqn:mamaorth}
\delta_{\alpha\beta} = \sum_{i=0}^{2^d-1}
\left(e^{-i \pi \eta_i \cdot \eta_\alpha} m_\alpha(q+\pi\eta_i)\right)^*
\left( e^{-i \pi\eta_i \cdot \eta_\beta} m_\beta(q+\pi\eta_i) \right).
\end{equation}
Finally, we note that three orthonormality conditions
(\ref{eqn:m0orth}-\ref{eqn:mamaorth}) may be combined into one single
condition by simply
taking $\alpha$ and $\beta$ in (\ref{eqn:mamaorth}) to vary over
the full range $0 \ldots 2^d-1$.

In one dimension, Eq. (\ref{eqn:mamaorth}) represents three
independent conditions: orthonormality among the scaling functions,
orthonormality among the wavelets $\psi_1(x-n-1/2)$, and orthogonality
between these two sets of functions.  Given a two-scale symbol $m_0$
satisfying the orthonormality condition, one common choice for the
wavelet symbol $m_1$ which satisfies the remaining conditions in
(\ref{eqn:mamaorth}) is
\begin{equation} \label{eqn:m1d=1}
m_1(k) = m_0^*(k+\pi),
\end{equation}
as may be confirmed directly by substitution.  

For $d>1$ dimensions, a straight-forward way to satisfy the conditions 
(\ref{eqn:mamaorth}) is to use product functions formed from the symbols
$m_0$ and $m_1$ of an orthonormal, one-dimensional multiresolution analysis,
$$
m_\alpha(q) \equiv \prod_{e=1}^d m_{(\eta_\alpha)_e}(q_e),
$$
where $e$ varies over the Euclidean dimensions of space, $q_e$ and
$(\eta_\alpha)_e$ are the corresponding components of the respective
vectors, and, again, we let $\alpha$ vary over the entire range
including $\alpha=0$.  With this choice, the condition
(\ref{eqn:mamaorth}) factors and gives
$$
\sum_{i=0}^{2^d-1}
\left(e^{-i \pi \eta_i \cdot \eta_\alpha} m_\alpha(q+\pi\eta_i)\right)^*
\left( e^{-i \pi\eta_i \cdot \eta_\beta} m_\beta(q+\pi\eta_i) \right)
\hspace{2in}
$$\\
$$
\begin{array}{l}
 =  \prod_e \left[ \sum_{{\eta}=0}^1 
\left(e^{-i \pi {\eta} (\eta_\alpha)_e} m_{(\eta_\alpha)_e}(q+\pi {\eta})\right)^*
\left(e^{-i \pi {\eta} (\eta_\beta)_e} m_{(\eta_\beta)_e}(q+\pi {\eta})\right)
\right]\\
\\
 =  \prod_e \delta_{(\eta_\alpha)_e,(\eta_\beta)_e} = \delta_{\alpha,\beta},
\end{array}
$$
as is required.  This, coupled with the choice (\ref{eqn:m1d=1}), was
used to construct the multidimensional basis functions used in the
electronic structure calculations of \cite{chou,tymczak}.  The
particular choice of two-scale symbols $m_0$ in these works were the
D6 and D8 wavelets defined in \cite{dau}, respectively.

  \subsection{Semicardinal bases} \label{subsec:defsemic}

It is highly desirable in the calculation of electronic structure to
be able to determine the expansion coefficients of functions from
knowledge of their values on a grid of finite resolution.  If we
insist that at every level of resolution $Q$, the expansion
coefficients of a function depend only upon its values on
the grid $C_Q$, then, as we will now see, it is always possible to find a
multiresolution basis in which every basis function has the value zero
on all points of the grid of its corresponding scale, except for one,
where its value is unity.  We refer to such a basis as a {\em
semicardinal} basis.

The following definitions are useful in discussing such bases.  The
concept of semicardinality may be defined independently of
multiresolution analysis, so we couch these definitions in general
terms:
\begin{enumerate}
   \item A {\em hierarchical basis} is a set of functions, each
    associated with exactly one point $g$ from a hierarchy of nested grids
    $G_M \subset G_{M+1} \subset \ldots$.

   \item For such a set of grids, $H_{Q+1} \equiv G_{Q+1}-G_{Q}$.

   \item In a hierarchical basis, the {\em scale} of a point $g$ and
    it associated basis function is the least value of $Q$ for which $g$ is
    in $G_Q$.

   \item A function $f(x)$ is {\em cardinal} on a given grid $G$ if
   $f(g)=\delta_{gh}$ for all points $g$ and one point $h$, both in $G$.

   \item A {\em semicardinal} basis is a hierarchical basis in which
    each basis function is cardinal on the grid of its own scale.
\end{enumerate}
Note from the second and third definitions that when $Q=M$, $G_Q$ is
the set of all points from scale $Q$ and that when $Q>M$, $H_Q$ is the
scale of all points from scale $Q$.  In the language of
Secs.~\ref{sec:MatrixFramework} and
\ref{sec:matrep}, semicardinality is the condition that the matrix $\cI$,
whose columns contains the values of each basis function on the finest
grid, obeys two simple algebraic identities,
\begin{eqnarray}
\cP_{G_M} \cI \cP_{G_M} & = & \cP_{G_M} \label{def:semic} \\
\cP_{G_Q} \cI \cP_{H_Q} & = & \cP_{H_Q} \mbox{\ \ for $Q>M$}. \nonumber
\end{eqnarray}

Note that a multiresolution analysis is a hierarchical basis on the
grids $C_Q$, with the $D_Q$ playing the role of the $H_Q$.  As an
example, consider the multiresolution basis consisting of the basis
functions from the top two panels of Figure \ref{fig:MRA}.  The basis
functions for $V_0$ are associated only with points of $C_0$ but not
$C_1$ and thus are of scale $N=0$.  These functions are cardinal on
the grid $C_0$ and thus satisfy their condition for semicardinality.
The basis functions for $W_1$ are associated with the points in
$D_1=C_1-C_0$, and thus are of scale $N=1$.  But, they are not
cardinal on the grid $C_1$, and thus this two-scale basis fails to be
semicardinal.  For the basis to be semicardinal, we would have to
replace the functions centered on the odd half-integer points with
functions which are cardinal on $C_1$.

    \subsubsection{Exact extraction}

In a hierarchical basis on the scales $N:M$ (by our convention $N>M$),
the property of exact extraction means that for {\em all} scales $Q$, the
expansion coefficients on the scales $Q:M$ for a function $f(x)$ may be
determined from knowledge of the values $f(G_Q)$ of the function on
the grid $G_Q$ only.  This means, in particular that the map from the
expansion coefficients on scales $Q:M$ to the values of the function
$f(G_Q)$ is {\em invertible}.  We now show that this property of
obvious practical utility implies that such a basis may always be
chosen to be semicardinal.

Consider functions expanded from the coarsest scale $M$ up to a very
fine scale $R>>Q$.  By our assumption of the existence of an
invertible linear map, the {expansion coefficients} on the scales
$Q:M$ of any function with values $f(G_Q)=0$ must be zero.  Thus, the
space of functions satisfying $f(G_Q)=0$ must be a subspace of span of
the basis functions associated with the remaining points, $G_R-G_Q$.

Now, the space of functions $f(G_Q)=0$ has exactly one degree of
freedom for each point of $G_R-G_Q$ and thus has the same dimension as
the span of the basis functions associated with the same set of
points.  Having already established that the former space is contained
in the latter, we now see from their dimensionality that the two
spaces are in fact the same.

Specifically, this means that all functions $f(x)$ in the span of the
basis functions associated with $H_{Q+1}$ must all have zero value on
the next coarser grid, $f(G_Q)=0$.  Hence, the invertible linear map
which we have supposed to exit between the coefficients of scales
$Q+1:M$ and the values $f(G_{Q+1})$ induces an invertible map from the
span of the basis functions associated with the points $H_{Q+1}$ to
the sequences of values $f(G_{Q+1})$ for which $f(G_Q)=0$.  A basis
for this latter sequence space is the set of sequences which are zero
on all points of $G_{Q+1}$ but for one point in $H_{Q+1}$ where they
take the value unity.  Applying to each of these sequences the inverse
of the aforementioned induced map produces a basis for the span of the
functions of scale $Q+1$ satisfying the conditions of semicardinality.
Proceeding in this manner for all scales $N \le Q \le M$ produces a
semicardinal basis.

    \subsubsection{Algorithms} \label{subsubsec:semicops}

Although semicardinal bases have the desirable property of exact
extraction, without the additional structure of multiresolution
analysis, it is difficult to develop efficient methods for performing
the operations needed in physical calculations.

\begin{figure}
\begin{center}
\scalebox{\figscale}{\scalebox{0.55}{\includegraphics{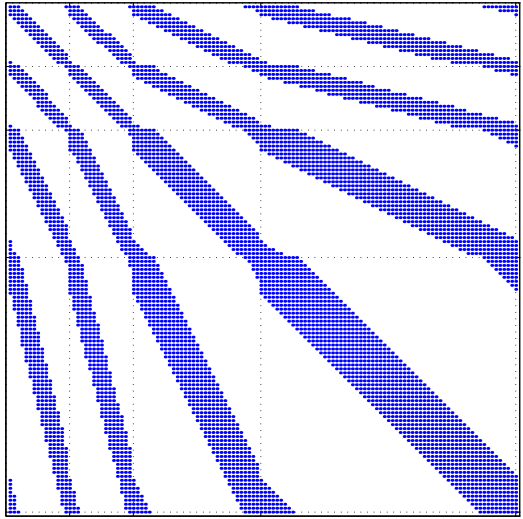}}}
\end{center}
\caption{Sparsity pattern of Hermitian operator in a
semicardinal basis on $C_0, C_1, \ldots$ in one dimension.
(Basis functions grouped into blocks according to scale from coarsest
to finest.)}
\label{fig:sppat}
\end{figure}

In particular, without the structure of multiresolution analysis, the
operations $\cL, \cO, \cI, \cI^\dagger$
must be applied directly as multiplication by the corresponding
matrices.  When working with the complete grids $C_0, C_1,
\ldots$ as defined in (\ref{eqn:MRB}), these matrices show a sparse,
fractal-like pattern. Figure~\ref{fig:sppat} shows the appearance of
this pattern in one dimension for $L$ and $\cO$ when the basis
functions are ordered first according to scale from coarsest to finest
and then within each scale by location in space.  A similar
pattern results for the matrices $\cI$ and $\cI^\dagger$ except that,
because of semicardinality, these matrices are either zero above or
below the diagonal, respectively.  

After the process of {restriction}, as defined in
Sec.~\ref{sec:FrameMRA:Intro} and illustrated in
Figure~\ref{fig:onoff}, little of this sparsity remains.  In
calculating electronic structure, the majority of functions surviving
the restriction will overlap with one of the nuclei.  The matrix
elements among these surviving functions thus consist of dense blocks
connecting all of the functions associated with each atomic nucleus.
Multiplication by matrix blocks of such size requires thousands of
operations per basis function and is therefore relatively inefficient.

The inverse transform $\cJ$ has the potential to require the solution
of a general system of linear system of equations and thus become even
more problematic.  Semicardinality, however, provides just enough
structure to allow the inverse transform to be performed in a direct
procedure requiring the same work as does $\cI$.
To see this, let $f \equiv f(G_N)$ be a column vector of the samples
of a function on the finest grid and $F \equiv \cJ f$ be the expansion
coefficients we seek.  Decomposing $F$ into contributions on different
scales and applying the identity $\cP_{G_M} \cI \cP_{H_Q} \equiv
\cP_{G_M} (\cP_{G_Q} \cI \cP_{H_Q}) = \cP_{G_M} \cP_{H_Q} = 0$, which
follows from the algebraic definition of semicardinality
(\ref{def:semic}) and the facts that $G_Q \subset G_M$ and $G_M \cap
H_Q$ for $Q>M$, gives the following result for the values of the 
function on the coarsest scale,
\begin{eqnarray} 
\cP_{G_M} f & = & \cP_{G_M} \cI F \label{eqn:singlescalerecon} \\
& = & \cP_{G_M} \cI (\cP_{G_M} + \cP_{H_{M+1}} + \ldots) F \nonumber \\
& = & \cP_{G_M} F. \nonumber
\end{eqnarray}
Thus, the expansion coefficients on the coarsest scale are just
the values of the function on the associated points.

Now, proceeding iteratively, suppose the coefficients of $F$ are known
up to scale $Q$ and consider the values of $f$ on the next scale.
Again using the conditions (\ref{def:semic}), we have
\begin{eqnarray*} 
\cP_{H_{Q+1}} f
& = & \cP_{H_{Q+1}} \cI (\cP_{G_Q} + \cP_{H_{Q+1}} + \cP_{H_{Q+2}}+
\ldots ) F  \\
& = & (\cP_{H_{Q+1}} \cI \cP_{G_Q} + \cP_{H_{Q+1}}) F,  \nonumber
\end{eqnarray*}
which gives the expansion coefficients on the next scale explicitly as
\begin{equation} \label{eqn:screcur}
\cP_{H_{Q+1}} F  =  \cP_{H_{Q+1}} f - \cP_{H_{Q+1}} \cI \cP_{G_Q} F.
\end{equation}
Combining (\ref{eqn:singlescalerecon},\ref{eqn:screcur}) gives
the following recursive procedure for computing $F=\cJ f$,
\begin{eqnarray}
F_M & \equiv & f; \label{alg:semicJ} \\
F_{Q+1} & = & F_Q - \cP_{H_{Q+1}} \cI \cP_{G_Q} F_Q; \nonumber \\
F & \equiv & F_N. \nonumber
\end{eqnarray}
Note that the total action of all of the operators $\cP_{H_{Q+1}} \cI
\cP_{G_Q}$ is just to connect once each of the columns of $\cI$ with
all points in the $H_{Q+1}$.  The implementation of (\ref{alg:semicJ})
is thus precisely as costly as the implementation of $\cI$.

Finally, note that the algorithm (\ref{alg:semicJ}) may be written
as the matrix product,
\begin{equation} \label{eqn:semicJprod}
\cJ = \prod_{Q=N-1}^{M} \left(I-\cP_{H_{Q+1}} \cI \cP_{G_Q}\right),
\end{equation}
from which the procedure to compute $\cJ^\dagger$ is easily
determined.

    \subsubsection{Semicardinal multiresolution analysis} \label{subsubsec:semicmra}

The key feature of semicardinality, that the expansion coefficients of
scale $Q$ for a function $f(x)$ may be extracted from the sample
values $f(C_Q)$, is incompatible with the use of standard orthogonal
or bi-orthogonal multiresolution analyses.  In bi-orthogonal
multiresolution analyses\cite{dau}, one considers both the basis of
scaling and detail functions and the {\em dual} to this basis.  The
{\em dual} to any basis is defined as those functions against which any
function $f(x)$ may be integrated to determine its expansion
coefficients in the original basis.  Orthogonal multiresolution
analyses are thus a special case of bi-orthogonal bases where the
basis is its own dual.  For semicardinal bases, the exact
extraction property implies that the dual of the functions of scale
$Q$ are linear combinations of Dirac $\delta$-functions centered on
the points of $C_Q$.  Thus, a semicardinal basis of smooth functions
can never be orthogonal, and, because Dirac $\delta$-functions are not
square-integrable, such a basis does not technically fit into the
bi-orthogonal wavelet framework.

Nonetheless, it is possible to construct multiresolution analyses
which are semicardinal.  Semicardinality, in fact, nearly completely
determines the allowable form for a multiresolution analysis.  To form
a semicardinal multiresolution analysis it is necessary and sufficient
that the scaling functions be cardinal and that each of the $2^d-1$
detail functions for scale $Q$ simply be the scaling functions of
scale $Q+1$ associated with the corresponding detail points.  While
this latter property may seem unusual at first, having only one type
of function in the multiresolution analysis proves to be of
considerable convenience\cite{JCP}.

To show that semicardinal multiresolution analyses must have this
form, consider first the requirements which semicardinality places on
the scaling functions.  The first semicardinality condition of
(\ref{def:semic}) states that the functions spanning the coarsest
scale $M$ must appear cardinal on the coarsest grid, which is $C_M$ in the case
of a multiresolution analysis.  Dilating this condition to the scale
$Q=0$, gives the condition that the scaling functions must be cardinal
on the integer grid,
\begin{equation} \label{def:semicscaling}
\phi(n-m)=\delta_{nm} \mbox{\ \ \ for $n,m$ in $C_0 \equiv Z^d$}.
\end{equation}
Section~\ref{sec:interpoletpcard} discusses the implications
of this for functions satisfying the two-scale relation.

Next, we turn to the detail functions.  The second
semicardinality condition of (\ref{def:semic}) states that for all $q
\in D_Q$ and $p \in C_Q$, the detail function associated with point
$q$ must have value $\psi_\alpha(2^{Q-1}(p-q))=\delta_{pq}$.
Enforcing this condition for all $p$ and dilating the condition to
scale $Q=0$, implies that for all $\alpha$,
$\psi_\alpha(n/2)=\delta_{n0}$ for $n \in Z^d$.  From this we find
that for the detail coefficients of (\ref{eqn:dwlt}) that
\begin{eqnarray} 
d_{\alpha,n} & \equiv & \sum_{k \in Z^d} d_{\alpha,k} \delta_{nk} = \sum_{k \in Z^d} d_{\alpha,k} \phi(n-k) \label{def:semicwlt} \\
  & = & \psi_\alpha(n/2) = \delta_{n0}. \nonumber
\end{eqnarray}
Therefore, as stated above, the detail functions $\psi_\alpha(x) = \sum_{k
\in Z^d} d_{\alpha,k} \phi(2x-k) = \phi(2x)$ are just the scaling
functions from the next finer scale.  For a general scale $Q$ this
means that the detail functions for the points $D_{Q+1}$ are just the
scaling functions on the corresponding points in the single-scale
representation of $V_{Q+1}$.

Finally, we now verify that such a basis satisfies the {\em two-scale
decomposition theorem}.  Eq.  (\ref{def:semicwlt}) implies that
$m_\alpha(q)=1/2^d$ for all $\alpha$ and $q$.  Thus the determinant
from (\ref{eqn:MRAdet}) becomes
\begin{eqnarray} 
\mbox{det} & = & (\frac{1}{2^d})^{2^d-1} \left|\begin{array}{ccccc}
m_0(q) & \ldots & 1 & \ldots & 1 \\
\vdots &  & \vdots & & \vdots \\
m_0(q+\pi\eta_i) & \ldots & (-1)^{\eta_\alpha \cdot
\eta_i}  & \ldots & (-1)^{\eta_{2^d-1} \cdot 
\eta_i} \\
\vdots &  & \vdots & & \vdots \\
m_0(q+\pi\eta_{2^d-1}) & \ldots & (-1)^{\eta_\alpha \cdot
\eta_{2^d-1}} & \ldots &
(-1)^{\eta_{2^d-1} \cdot \eta_{2^d-1}}
\end{array}\right| \nonumber \\
& = & \pm (\frac{1}{2^d})^{2^{d-1}} \sum_i m_0(q + \pi \eta_i), \label{eqn:MRAILdet}
\end{eqnarray}
where, we have used the fact, derived in the appendix, that each
cofactor in the determinant expansion along the first column has the
value $\pm (2^d)^{2^{d-1}-1}$, with a fixed sign for each cofactor.
For cardinal scaling functions, we have from (\ref{eqn:summ0}) in
Sec.~\ref{subsubsec:genconstruct}, that the sum over the $m_0$
appearing in (\ref{eqn:MRAILdet}) is unity.  Thus, as required for
two-scale decomposition, the determinant is never zero.  The fact that
we find a simple constant for the determinant is a direct reflection
of the compact nature of the dual basis.

  \subsection{Interpolating, cardinal scaling functions} \label{sec:semicbases:functions}

Because the previous section has already determined the form which the
detail functions of a semicardinal multiresolution analysis must take,
it remains now only to specify the scaling functions of such
multiresolution analyses.  Accurate calculations in physical systems
require that the reconstructions of physical functions from sample
values interpolate well the behavior of those functions between the
sampling points.  Additionally, in order to limit the processing
required for each expansion coefficient, in practice the basis
functions should have the minimal spatial extent possible.  These
conditions represent two additional constraints beyond those we have
already imposed, cardinality for exact recovery of expansion
coefficients and the two-scale relation to sustain multiresolution
analysis.  As we shall show in Sec.~\ref{subsubsec:genconstruct}, the
four constraints of (1) the two-scale relation, (2) cardinality, (3)
minimal support, and (4) interpolation almost uniquely specify the
scaling functions.  In the interest of brevity, we shall refer to such
interpolating, cardinal scaling functions as ``interpolets.''

Once the two-scale coefficients of the interpolets are determined, the
only other information needed to perform physical calculations are the
matrix elements of integral-differential operators between scaling
functions and the values of the scaling functions in real-space.
Sections~\ref{sec:semicbases:matels} and
\ref{subsubsec:values} describe the
determination of these quantities, and Sec.~\ref{subsubsec:examples}
gives examples of specific functions, with tables displaying all
information needed in physical calculations.

    \subsubsection{Construction} \label{subsubsec:genconstruct}

We now discuss the mathematical implications of the four constraints
outlined above.  Because we now consider the very specific class of
interpolating cardinal scaling functions, we shall denote these
functions as $\cI(x)$, in place of the generic $\phi(x)$.

\paragraph{Two-scale relation} 

As discussed in Sec.~\ref{sec:scalingfunctions}, we may consider the
entire allowable class of scaling functions by considering only
functions constructed in Fourier space via
\begin{equation} \label{eqn:Itsss}
\tilde\cI(k) = \left( \prod_{j=1}^\infty m_0(k/2^j) \right) \cI(0)
\end{equation}
with an arbitrary $2\pi C_0$--periodic function $m_0(q)$ satisfying
$m_0(0)=1$. 

Each of the remaining conditions thus becomes a condition on
the acceptable two-scale symbols $m_0(q)$.  Once $m_0(q)$ is known, the
two-scale coefficients $c_n$
\begin{equation} \label{eqn:Its}
\cI(x) = \sum_{n} c_{n} \cI(2x-n),
\end{equation}
used to form the operators
$\cI_{P+1,P}$ from Sec.~\ref{sec:matrep},
may be recovered from (\ref{eqn:twoscalesymbol}).

\paragraph{Cardinality} \label{sec:interpoletpcard}

Cardinality is the condition that the scaling functions $\cI(x)$ obey
\begin{equation} \label{eqn:Cardinal}
\cI(n) = \delta_{n,0}, \mbox{\ \ for all $n \in C_0$}.
\end{equation}

To transform this into a constraint on the $c_n$, and thus 
on the $m_0(q)$, we dilate the two-scale relation by a factor of two,
\begin{equation} \label{eqn:dtsr}
\cI(x/2) = \sum_{n} c_{n} \cI(x-n),
\end{equation}
and evaluate the result on the points $x=m$ in $C_0$,
\begin{equation}
\cI(m/2) = \sum_{n} c_{n} \cI(m-n) = \sum_n c_{n} \delta_{mn} = c_m. \label{eqn:Icn}
\end{equation}
Comparing with (\ref{eqn:Cardinal}), cardinality determines all
``even'' terms in $c_n$,
$$
c_{2n}=\delta_{n0}, \mbox{\ \ \ for $n$ in $C_0$}.
$$ 

To convert this into a condition on the two-scale symbol, we remove
all ``odd'' frequency components of $m_0(q)$ leaving untouched the
``even'' frequency components by periodizing the sum
(\ref{eqn:twoscalesymbol}) on the lattice $\pi C_0$.  Cardinality thus
implies that the two-scale symbol obey
\begin{equation} \label{eqn:summ0} 
\sum_{i=0}^{2^d-1} m_0(k+\pi\eta_i) = 
 \sum_{n \in 2C_0} 2^d \frac{c_{n}}{2^d}e^{-ik \cdot n} = 1,
\end{equation}
which we used in (\ref{eqn:MRAILdet}) to confirm that semicardinal
multiresolution analyses indeed satisfy the two-scale decomposition
theorem.

\paragraph{Minimal support}

From (\ref{eqn:Icn}) we see that cardinal scaling functions extend at
least as far as the range of non-zero elements in the sequence
$c_{n}$.  We thus limit our discussion to sequences with the shortest
possible length.  Correspondingly, the $m_0(q)$ are to be constructed
as the {\em finite} trigonometric polynomial of the {\em lowest} order
which satisfies all other conditions on the scaling functions.

\paragraph{Interpolation}

By ensuring the equivalence of the multiscale and single-scale
representations, multiresolution analysis simplifies the consideration
of how well a multiresolution basis interpolates physical functions to
the consideration of interpolation for the {\em single-scale}
representation on the finest scale of the analysis.  On the finest
scale, cardinality of the scaling functions gives
\begin{equation} \label{eqn:recon}
\tilde f_N(x) = \sum_{p \in C_N} f(p) \cI\left(2^N(x-p)\right)
\end{equation}
as the function $\tilde{f}(x)$ in $V_N$ which matches exactly the
sample values $f(C_N)$.  To ensure that this estimate interpolate
$f(x)$ to order $\ci$ between the sample points, we insist that it
reproduce {\em exactly} all polynomials up to degree
$\ci$.  By linearity, to do this, we need only impose the
reconstruction of all multinomials $\prod_i x_i^{\ci_i}$ where $\sum_i
\ci_i \le \ci$ and $0 \le \ci_i$.  Rescaling this condition to the scale
$Q=0$, it becomes
\begin{equation} \label{eqn:interp}
\prod_i x_i^{\ci_i} = \sum_{n \in C_0} (\prod_i n_i^{\ci_i}) \cI(x-n).
\end{equation}

To determine the constraints which this condition places on $m_0(q)$,
we first consider the constraints which it places on $\tilde \cI(k)$.
Condition (\ref{eqn:interp}) is a convolution leading to the familiar
product form in Fourier space
\begin{eqnarray}
\int \prod_i x_i^{\ci_i} \frac{e^{-ik\cdot x}\,d^dk}{(2\pi)^d} & = &
\tilde{\cI}(k) \sum_{n \in C_0} (\prod_i n_i^{\ci_i}) e^{-ik \cdot n} \nonumber \\
\partial_{\{\ci_i\}} \delta^{(d)}(k) & = & \tilde{\cI}(k) \cdot (2\pi)^d 
\sum_{n \in C_0} \partial_{\{\ci_i\}} \delta^{(d)}(k-2\pi n),
\label{eqn:deltas}
\end{eqnarray}
where $\delta^{(d)}(q)$ is the $d$-dimensional Dirac $\delta$-function and
$\partial_{\{\ci_i\}}$ denotes the mixed partial derivative $\prod_i
\partial_{x_i}^{\ci_i}$, and we have used the identity
$\sum_n e^{-ik\cdot n} = (2\pi)^d \sum_{n} \delta^{(d)}(k-2\pi n)$.
Eliminating the singularities at non-zero points in $2\pi C_0$ from
the right-hand side of (\ref{eqn:deltas}) requires a high degree of
regularity in $\tilde \cI(k)$ at these points.  Integrating the
derivatives of the $\delta$-functions by parts reveals that
(\ref{eqn:deltas}) implies the conditions
\begin{equation} \label{eqn:InterpCond}
(2\pi)^d \partial_{\{\ci_i\}} \tilde \cI(2\pi n) = \left( \prod_i
\delta_{\ci_i,0} \right) \delta_{n,0} \mbox{\ \ for $\sum_i \ci_i
\le \ci$}.
\end{equation}

\begin{figure}
\begin{center}
\scalebox{\figscale}{\scalebox{0.75}{\includegraphics{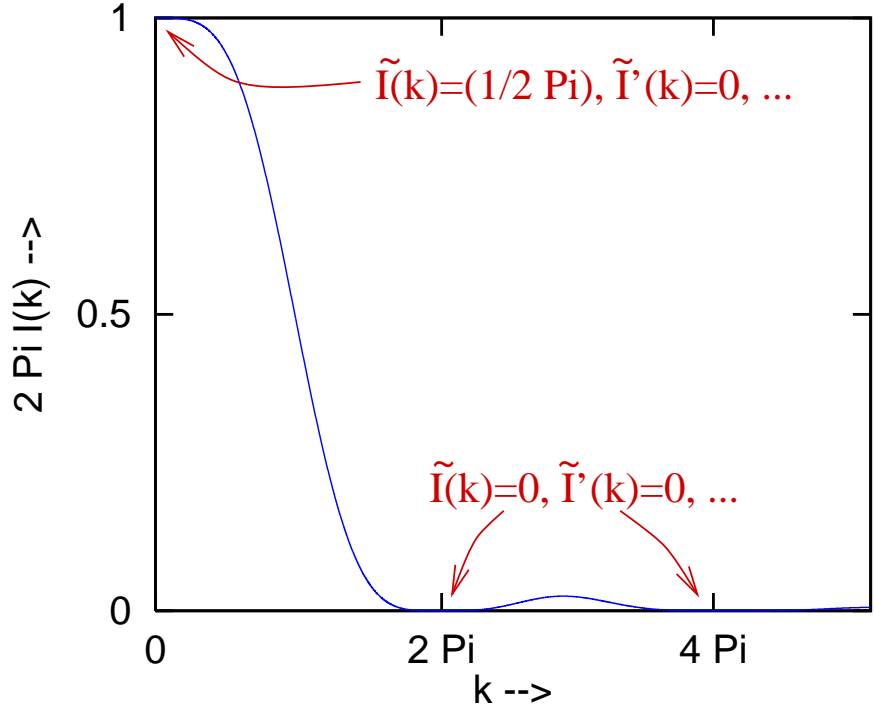}}}
\end{center}
\caption{Fourier transform of an interpolating, cardinal scaling function.}
\label{fig:FTinta}
\end{figure}

Figure~\ref{fig:FTinta} illustrates these conditions on $\tilde
\cI(k)$ for one dimension.  One important result is that
interpolation sets the normalization of the of the scaling functions.
The $\ci=0$ condition at $n=0$ is just $\tilde \cI(0)=1/(2\pi)^d$, or
equivalently,
\begin{equation} \label{eqn:intnorm}
\int d^dx\,\,\cI(x)=1.
\end{equation}
The other
conditions at $n=0$ are that the higher order integral moments up to
order $\ci$ be zero,
$$
\int d^dx\,\,(\prod_i x_i^{\ci_i}) \cI(x) = 0 \mbox{\ \ for $\sum_i \ci_i
\le \ci$}.
$$
The conditions at the {\em non-zero} points in $2\pi C_0$ are most
simply expressed as the condition that $\tilde \cI(k)$ have an
$\ci^{\mbox{th}}$-order zero at these points.

The $n=0$ conditions of (\ref{eqn:InterpCond}) constrain the behavior
of $m_0(q)$ near $q=0$.  With a finite number of non-zero $c_{n}$ from
the condition of minimal support, we are assured that $m_0(q)$ is
analytic.  Recalling that $m_0(0)=1$, we have that near $q=0$,
\begin{equation}
m_0(q)=1+O(q^\beta),
\end{equation}
for some leading order $\beta$.
Eq. (\ref{eqn:Itsss}) then ensures that $\tilde \cI(k)$ has a similar
analytic structure near $k=0$,
\begin{eqnarray}
\tilde \cI(k) & = & \tilde \cI(0) \prod_{n=1}^\infty \left( 1+O\left(
(k/2^n)^\beta \right) \right) \nonumber \\ & = &  \tilde \cI(0)
\left(1 + O\left(k^\beta\right) \right).
\end{eqnarray}
Thus, we satisfy all $n=0$ conditions from
(\ref{eqn:InterpCond}), including the zeroth-order normalization
condition, so long as we take
\begin{equation} \label{eqn:normprod}
\tilde\cI(k) = \frac{1}{(2\pi)^d} \prod_{j=1}^\infty m_0(k/2^j),
\end{equation}
for some $m_0$ satisfying
\begin{equation} \label{eqn:InterpCondm0}
\partial_{\{\ci_i\}} m_0(0) = \prod_i {\delta_{\ci_i,0}}
\mbox{\ \ \ for all $\sum_i \ci_i \le \ci$}.
\end{equation}

\begin{figure}
\begin{center}
\scalebox{\figscale}{\scalebox{0.75}{\includegraphics{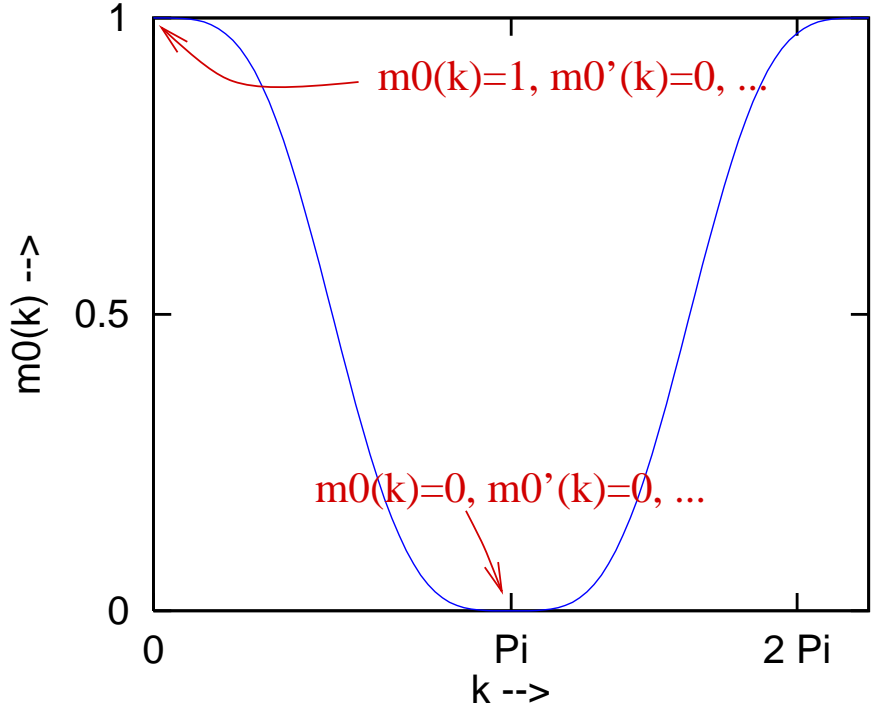}}}
\end{center}
\caption{Two-scale symbol $m_0$ for an interpolating, cardinal scaling
function.}
\label{fig:FTintb}
\end{figure}

The final conditions concern $\tilde \cI(k)$ at the non-zero points of
$2\pi C_0$.  From the form of the product (\ref{eqn:normprod}), we see
that if $\tilde \cI(k)$ is to have an $\ci^{\mbox{th}}$ order zero at
$G$, then the orders of the zeros of $m_0(q)$ among the points $q=G/2,
G/4, \ldots$ must sum to give at least this order.  Because all
non-zero points in $\pi C_0=2\pi C_0/2$ may be expressed in terms of
the product of some non-negative power of two with some vector in $\pi
D_1$, if $m_0(q)$ has an $\ci^{\mbox{th}}$ order zero at each of the
points $\pi D_1$, then $\tilde \cI(k)$ will have the appropriate
analytic structure at the non-zero points of $2\pi C_0$.  Noting the
$2\pi C_0$-periodicity of the $m_0(q)$, the final conditions imposed
by interpolation on the two scale symbol are thus
\begin{equation} \label{eqn:InterpCondmpi}
\partial_{\{\ci_i\}} m_0(\pi \eta_\alpha) = 0
\mbox{\ \ \ for all $\sum_i \ci_i \le \ci$ and $1 \le \alpha \le 2^d-1$},
\end{equation}
with the $\eta_\alpha$ defined as in (\ref{eqn:defeta}).
Figure~\ref{fig:FTintb} illustrates these conditions for one
dimension.

An interesting alternative to (\ref{eqn:InterpCondmpi}) is to note
that one could ensure the proper behavior in $\tilde \cI(k)$ by
instead placing $\ci^{\mbox{th}}$ order zeros in $m_0(q)$ at the
points of $\pi D_2$.  But, this and other such choices demand higher degrees
of oscillation in $m_0(q)$ and would to require longer non-zero
sequences in $c_n$.  Also, condition (\ref{eqn:InterpCondmpi}) has the
distinct advantage that when combined with condition
(\ref{eqn:InterpCondm0}), the two conditions automatically imply
cardinality, condition (\ref{eqn:summ0}).

To minimize the support of the resulting functions, we thus consider
scaling functions constructed according to (\ref{eqn:normprod}) with
$m_0(q)$ satisfying the conditions
(\ref{eqn:InterpCondm0},\ref{eqn:InterpCondmpi}) and which therefore
satisfy all of our constraints.  For a given degree of interpolation
$\ci$, the conditions (\ref{eqn:InterpCondm0},\ref{eqn:InterpCondmpi})
are equivalent to a system of simple linear equations for the
two-scale coefficients $c_n$.  The most compact set of coefficients
satisfying these conditions gives the most compact
$\ci^{\mbox{th}}$-order interpolating cardinal scaling function, or
``interpolet''.  Section~\ref{subsubsec:examples} details these
equations and their solution in several useful cases.

    \subsubsection{Overlaps} \label{sec:semicbases:matels}

We now present a method for determining matrix elements among the
scaling functions directly from the two-scale coefficients $c_n$.  As
many of these results are general, we revert temporarily to the
notation $\phi(x)$ for the scaling functions.  Once we begin to use
specific properties of interpolating cardinal functions, we shall
return to the notation $\cI(x)$.  The approach below has been
developed by several authors in one dimension and is described in
\cite{strang}.  We here give the appropriate generalizations for multiple
dimensions.

The specific information needed to apply the overlap matrices $\cO$
and $L$ when using the procedures of Sec.~\ref{sec:semicbases:fastalgs} is
the set of matrix elements among the functions of the finest scale,
\begin{equation} \label{eqn:defM}
{M}_{pq} \equiv  \int d^dx\,\phi^*(2^N(x-p)) \hat M \phi_m(2^N(x-q))
\mbox{\ \ \ for $p,q$ in $C_N$},
\end{equation}
with $\hat M$ being the the identity operator $\hat 1$ (for $\cO$) and
the Laplacian $\nabla^2$ (for $L$).  To compute these elements we
exploit the fact that both $\hat 1$ and $\nabla^2$ are {homogeneous operators}.

The {\em homogeneous operators of order $h$} are those which may be
written in Fourier space as a linear combination of multinomials of
order $h$,
\begin{equation} \label{eqn:defhomoop} 
\hat{\tilde M} = M^{(h)}(k) \equiv \sum_{\sum_{j=1}^d h_j=h} m_{\{h_j\}} \prod_j k_j^{h_j}
\end{equation}
for some set of coefficients $m_{\{h_j\}}$.  In real space, these
operators are just
\begin{equation}
\hat M = M^{(h)}\left(\frac{\nabla}{i}\right),
\end{equation}
where $i \equiv \sqrt{-1}$.  Taken together, the operators of order
$h$ thus form a linear space of dimension equal to the corresponding
number of multinomials, ${(d-1+h)!}/{[(d-1)!h!]}$.
The multinomial coefficients $m_{\{h_j\}}$ for any such operator may
be extracted simply by acting with the operator on the multinomial
functions of order $h$,
\begin{equation} \label{eqn:getcj}
m_{\{h_j\}} = \frac{\hat M (\prod_j x_j^{h_j})}{\prod_j{h_j!}}.
\end{equation}
Finally, we note that the identity operator $M(k)=1$ belongs to the
one-dimensional space of operators of order $h=0$ and that the
Laplacian operator $M(k)=-\sum_{j=1}^d k_j^2$ comes from the space of
operators of order
$h=2$, which in $d=3$ dimensions has dimension
six.

For any such homogeneous operator, the required matrix elements (\ref{eqn:defM})
for any scale may be related to a single universal set of matrix
elements,
\begin{eqnarray}
M_{pq} & \equiv & \int d^dx \,\,\phi^*(2^N(x-p)) M^{(h)}\left(\frac{\nabla}{i}\right)
\phi(2^N(x-q)) \nonumber \\
& = & \int d^dx \,\,\phi^*(2^N(x-p)) 2^{Nh} \left. \left( 
M^{(h)}\left(\frac{\nabla_u}{i}\right) \phi(u) \right) \right|_{u=2^N(x-q)} \nonumber \\
& = & 2^{N(h-d)} (\vec M)_{2^N(p-q)}, \label{eqn:mNto0}
\end{eqnarray}
where we have changed integration variables to give an expression in
terms of overlaps on scale $Q=0$ only, and where
\begin{equation} \label{def:vecM}
(\vec M)_{n \in C_0} \equiv \int d^dx\,\, \phi^*(x-n) \hat M \phi(x).
\end{equation}

Now, to determine $\vec M$ in terms of the two-scale coefficients, we
generate a recursion by using the two-scale relation to write
$\phi(x)$ as a sum of scaling functions from scale $Q=1$ and then using
(\ref{eqn:mNto0}) to re-express the resulting set of overlaps back in
terms of $\vec M$,
\begin{eqnarray*}
(\vec M)_{p \in C_0}
& \equiv & \int d^dx\,\, \phi^*(x-p) \hat M \phi(x) \nonumber \\
& = & \int d^dx\,\, \left( \sum_{m \in C_0} c_m^* \phi^*(2(x-p)-m)\right)
\hat M
\left( \sum_{n \in C_0} c_n \phi(2x-n) \right) \nonumber \\
& = & \sum_{m,n  \in C_0} c_m^* c_n  2^{h-d} \vec M_{2p+m-n},\nonumber
\end{eqnarray*}
which can be rearranged as
\begin{eqnarray*}
2^{-h} (\vec M)_{p  \in C_0} & = & \sum_{q \in C_0} \left( 2^{-d} \sum_{n \in C_0} c_{q-2p+n}^* c_n \right) (\vec M)_q. \label{eqn:eigenO}
\end{eqnarray*}
Thus, for each homogeneous operator of order $h$ for which the vector
of matrix elements $\vec M$ is well-defined, $\vec M$ is an
eigenvector of eigenvalue $\lambda=2^{-h}$ of the two-scale matrix
\begin{equation} \label{eqn:eigenMat}
{\tilde M}_{pq} \equiv 2^{-d} \sum_{n \in C_0} c_{q-2p+n}^* c_n.
\end{equation}
We therefore expect the eigenspectrum of ${\tilde M}$ to consist of
clusters of degenerate eigenvalues of value $2^{-h}$, with degeneracies equal
to the dimensionality of the space of operators of order $h$ and with
eigenvectors containing the overlaps for $h^{\mbox{th}}$ order
operators.

Determining the overlaps for a specific operator requires additional
information in order to select the appropriate vector from the
corresponding degenerate subspace of ${\tilde M}$.  In the case of
interpolating cardinal scaling functions this information comes from
the integral normalization of the scaling functions
(\ref{eqn:intnorm}) and the interpolation condition
(\ref{eqn:interp}).  To ensure that a given eigenvector represents the
correct operator, we verify the values of the multinomial coefficients
using the extraction formula
(\ref{eqn:getcj}), 
\begin{eqnarray}
m_{\{h_j\}}
& = & \int d^dx\,\, \cI^*(x)\, m_{\{h_j\}} \label{eqn:geteigenvec} \\
& = & \int d^dx\,\, \cI^*(x) \cdot \frac{\hat M (\prod_j x_j^{h_j})}{\prod_j{h_j!}} \nonumber\\
& = & \int d^dx\,\, \cI^*(x) \cdot \frac{1}{\prod_j{h_j!}} \hat M \left(
\sum_n \left( \prod_j n_j^{h_j} \right) \cI(x-n) \right) \nonumber \\
& = & \sum_n \left( \prod_j \frac{n_j^{h_j}}{h_j!} \right)
(\vec M)_{-n}. \nonumber
\end{eqnarray}
Thus, to determine the vector of overlaps $\vec M$ for an operator of
order $h$ with coefficients $m_{\{h_j\}}$ one simply forms the linear
combination of eigenvectors of (\ref{eqn:eigenMat}) with eigenvalue
$2^{-h}$ that satisfies the conditions (\ref{eqn:geteigenvec}).
Because the space of such vectors has one dimension for each
multinomial condition in (\ref{eqn:geteigenvec}), this completely
determines the vector $\vec M$.

    \subsubsection{Real-space values} \label{subsubsec:values}

Two approaches are available to compute the values of
the interpolets in real space from their two-scale
coefficients $c_n$.  The traditional approach, appropriate
for any type of scaling function, is to note that
the two-scale relation (\ref{eqn:tsr}), when evaluated at the points $x
\equiv m/2^{P+1}$ in $C_{P+1}$, gives 
\begin{equation} \label{eqn:rvI}
\phi(m/2^{P+1}) = \sum_{n \in C_0} c_{n} \phi((m-2^Pn)/2^P),
\end{equation}
a direct formula for the values $\phi(C_{P+1})$ in terms of the values
$\phi(C_P)$.  Using this, one may proceed iteratively to compute the
$\phi(C_Q)$ at any desired level of resolution.  To determine the
values $\phi(C_0)$ needed to initialize the process, one may generate
a self-consistency relation by evaluating the two-scale
relation (\ref{eqn:tsr}) on the points $x=m$ of $C_0$,
\begin{eqnarray*}
\phi(m) & = & \sum_{n \in C_0} c_{n} \phi(2m-n) \\
& = & \sum_{q \in C_0} c_{2m-q} \phi(q).
\end{eqnarray*}
The sequence $\phi(C_0)$ thus may be identified as the
eigenvector with eigenvalue unity of the matrix ${\tilde C}_{mq}
\equiv c_{2m-q}$.  For cardinal scaling functions, of course,  the
cardinality condition (\ref{eqn:Cardinal}) prescribes explicitly that the
$\cI(C_0)$ be a discrete Kronecker $\delta$-function
centered on the origin.

For cardinal scaling functions, a more convenient approach exists
which does not make large strides through the data-set as does
(\ref{eqn:rvI}).  Because $V_0 \subset V_P$, $\cI(x)$ may be written
exactly in terms of scaling functions on scale $P$.  Moreover, from
the interpolation property (\ref{eqn:recon}), the proper linear
combination for doing this is just
\begin{equation} \label{eqn:selfI}
\cI(x)=\sum_{p \in C_P} \cI(p) \cI(2^P(x-p)).
\end{equation}
Letting $p=m/2^P$, evaluating this relation on the points
$x=n/2^{P+1}$ where $n \in C_0$, and using (\ref{eqn:Icn}) to relate
the values $\cI(C_1)$ to the $c_n$ gives
\begin{eqnarray} 
\cI(n/2^{P+1})
& = & \sum_{m \in C_0} c_{n-2m} \cI(m/2^P), \label{eqn:selfIr} 
\end{eqnarray}
which again may be used iteratively to generate any $\cI(C_Q)$ from
the known values $\cI(C_0)$.  A variant of this approach exists for
orthonormal functions and is described in \cite[Ch. 6.5]{dau}.

    \subsubsection{Examples} \label{subsubsec:examples}

The simplest interpolating cardinal scaling function we shall
consider is the first-order interpolet in one dimension.
For $\ci=1$ and $d=1$, the conditions
(\ref{eqn:InterpCondm0},\ref{eqn:InterpCondmpi}) on the two-scale
symbol are just $m_0(0)=1$, $m_0(\pi)=0$ and $m_0'(0)=m_0'(\pi)=0$.
The general appearance of such a function (Figure~\ref{fig:FTintb})
leads quickly to the {\em ansatz}, which may be easily verified, that
the choice $m_0(q)=\cos^2(2q)=(1+\cos(q))/2$ satisfies these
conditions.  From this two-scale symbol, we have immediately the two
scale coefficients are $c_0=1$, $c_{\pm 1}=1/2$, and $c_n=0$ for
$|n|>1$, as listed in Table~\ref{tbl:d1L1}.

Using (\ref{eqn:selfIr}) to compute the real-space values of $\cI(x)$,
we see that for this interpolet the value on each detail point of
$D_{P+1}$ is given simply by the average of the values on the
neighboring points of $C_P$.  Using the notation $\{p\}$ to indicate a
complete shell of points related by cubic symmetry, so that simply
$\{p\} = \pm p$ in this one-dimensional case, Table~\ref{tbl:d1L1}
gives the results of this process.  As evident from the table, this
first-order interpolating function is the piece-wise linear
``triangle'' scaling function sketched in Figure \ref{fig:MRA}.
Finally, Table~\ref{tbl:d1L1} also lists the vectors of matrix
elements $(\vec M^{\{h\}})_n$ computed according to the prescription
(\ref{def:vecM}) of the previous section.  For the values with $n<0$,
note that $\vec M^{\{h\}}_{-n}=(-1)^h \vec M^{\{h\}}_n$ in this and
all cases below.

\begin{table}
\begin{center}
\begin{tabular}{|c||c|c|} \hline
$n$ & $0$ & $\{1\}$ \\ \hline \hline $c_n$ & $1$ & $\frac{1}{2}$ \\
\hline
\end{tabular}\\
\vspace{0.2in}
\begin{tabular}{|c||c|c|c|c|c|c|c|c|c|} \hline
x & $0$ & $\{\frac{1}{4}\}$ & $\{\frac{1}{2}\}$ & $\{\frac{3}{4}\}$ &
$\{1\}$ \\
\hline \hline
$\cI(C_0)$ & $1$ & & & & 0 \\ \hline $\cI(D_1)$ & & & $\frac{1}{2}$ &
& \\ \hline $\cI(D_2)$ & & $\frac{3}{4}$ & & $\frac{1}{4}$ & \\ \hline
\hline $\cI(C_2)$ & $1$ & $\frac{3}{4}$ & $\frac{1}{2}$ &
$\frac{1}{4}$ & $0$ \\ \hline
\end{tabular}\\
\vspace{0.2in}
\begin{tabular}{|c||c|c|} \hline
$n$ & $0$ & $1$ \\ \hline \hline
$\vec M^{\{0\}}$  & $\frac{2}{3} $ &
$\frac{1}{6}$ \\ \hline
$\vec M^{\{1\}}$  & 0 & $\frac{1}{2}$ \\
\hline
$\vec M^{\{2\}}$ & $-2$ & $1$ \\
\hline
\end{tabular}
\end{center}

\caption{First-order interpolating cardinal scaling function in one
dimension: non-zero two-scale coefficients ($c_n$), function values ($\cI$),
and overlaps of $\partial^h/\partial x^h$ ($M^{\{h\}}_n$, as defined
in (\ref{def:vecM})).}
\label{tbl:d1L1}
\end{table}

\begin{figure}
\begin{center}
\scalebox{\figscale}{\scalebox{0.6}{\includegraphics{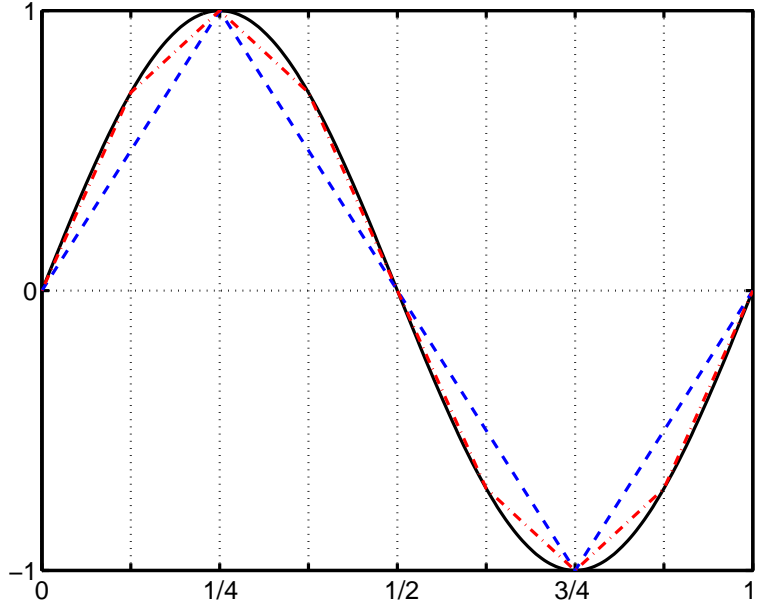}}} 
\end{center}
\caption{Interpolation with first-order interpolating scaling
functions: $\sin (2\pi x)$ (solid curve), representation in $V_2$ and
$V_3$ (dashed and dot-dashed curves, respectively).}
\label{fig:sined1L1}
\end{figure}

\begin{figure}
\begin{center}
\scalebox{\figscale}{\scalebox{0.6}{\includegraphics{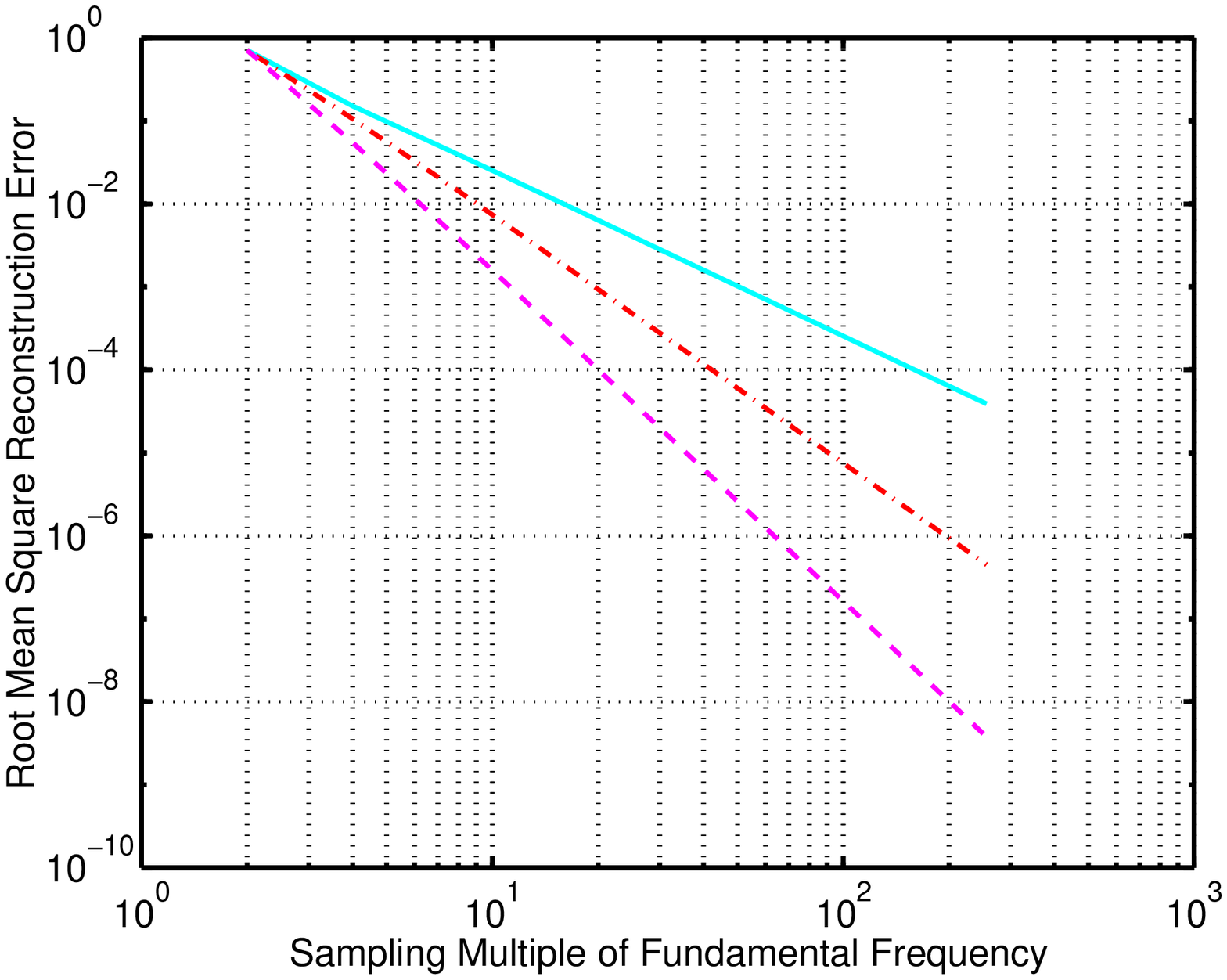}}} 
\end{center}
\caption{Root mean square error in reconstructing the sine function
as a function of the number of samples per period: first- (solid
line), second- (dash-dotted line) and third- (dashed line) order
interpolating scaling functions, exhibiting exponents of -2, -3 and
-4, respectively.}
\label{fig:convsine}
\end{figure}

Figure (\ref{fig:sined1L1}) illustrates the estimates $\tilde f_Q(x)$
for the function $f(x) = \sin(2 \pi x)$ afforded by these first-order
functions for scales $Q=2$ and $Q=3$.  The bases for $V_Q$ 
provide simple linear interpolation between the sample points $f(C_Q)$,
and as required by the interpolation condition (\ref{eqn:interp}),
clearly will reproduce exactly the constant and linear functions.  As a
quantitative illustration of the linear nature of this interpolation,
Figure \ref{fig:convsine} shows on a logarithmic plot the root
mean-square error in reproducing $\sin(2 \pi x)$ as a function of the
sampling rate $2^Q$, as $Q$ varies from one through eight.  As
expected for linear interpolation, to leading order, the error is
second-order and the data fall along a line of slope $-2$ in the plot.

To construct higher-order interpolets, we note that in one dimension
Eq.~(\ref{eqn:summ0}) ensures that the conditions on $m_0(q)$ at
$q=\pi$ will be satisfied automatically whenever the conditions
(\ref{eqn:InterpCondm0}) at $q=0$ are satisfied.
In terms of the two-scale coefficients, these latter conditions are
\begin{equation} \label{eqn:ccond1d}
\frac{1}{2} \sum_n n^\alpha c_n = \delta_{\alpha,0} \mbox{\ \ \ for $0
\le \alpha \le \ci$}.
\end{equation}

\begin{figure}
\begin{center}
\scalebox{\figscale}{\scalebox{0.6}{\includegraphics{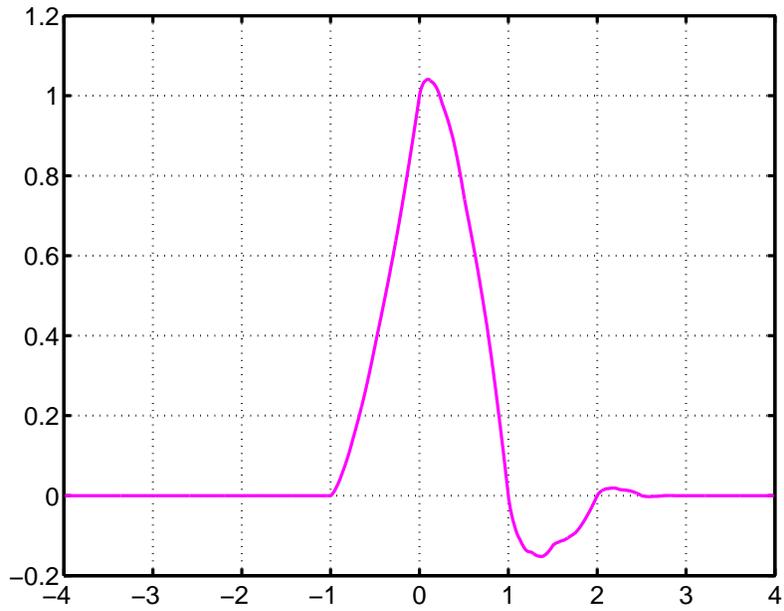}}} 
\end{center}
\caption{Second-order interpolet in one dimension.} \label{fig:d1L2}
\end{figure}

For second-order scaling functions this gives three independent linear
equations,
\begin{eqnarray*}
c_{-1}+1+c_1+c_3 & = & 2 \\ -c_{-1}+c_1+3c_3 & = & 0 \\ c_{-1}+c_1+3^2
c_3 & = & 0.
\end{eqnarray*}
Table~\ref{tbl:d1L2} lists the two-scale coefficients which solve this
system and the values of the corresponding scaling function on the
points of $C_1$.  Figure \ref{fig:d1L2} shows the appearance of the
second-order interpolet in real-space.  The asymmetry of the function
is a direct result of the lack of symmetry in the sequence $c_n$.
Figure~\ref{fig:convsine} shows the reconstruction of the sine
function for these interpolets also.  The reconstruction is correct to
second order and the leading-order error is third-order, as expected.

\begin{table}
\begin{center}
\begin{tabular}{|c||c|c|c|c|} \hline
$n$ & -1 & 0 & 1 & 3 \\ \hline \hline $c_n$ & $\frac{3}{8}$ & 1 &
$\frac{3}{4}$ & $-\frac{1}{8}$ \\ \hline
\end{tabular}\\
\vspace{0.2in}
\begin{tabular}{|c||c|c|c|c|c|c|c|c|c|} \hline
x & $-1$ & $-\frac{1}{2}$ & $0$ & $\frac{1}{2}$ & $1$ & $\frac{3}{2}$
& $2$ & $\frac{5}{2}$ & $3$ \\ \hline \hline $\cI(C_1)$ & 0 &
$\frac{3}{8}$ & $1$ & $\frac{3}{4}$ & $0$ & $-\frac{1}{8}$ & 0 & 0 &
0\\ \hline
\end{tabular}\\
\vspace{0.2in}
\begin{tabular}{|c||c|c|c|c|} \hline
$n$ & $0$ & $1$ & $2$ & $3$ \\ \hline \hline
$\vec M^{\{0\}}$ & ${\frac {247}{295}}$ & ${\frac {517}{4720}}$ & $-{\frac {17}{590}}$ & ${\frac {3}{4720}}$\\
\hline
$\vec M^{\{1\}}$ & $0$ & ${\frac {11}{16}}$ & $-\frac{1}{10}$ & ${\frac {1}{240}}$\\
\hline
$\vec M^{\{2\}}$ & $-{\frac {30}{11}}$ & ${\frac {397}{264}}$ & $-{\frac {5}{33}}$ & ${\frac {1}{88}}$ \\
\hline
\end{tabular}

\end{center}
\caption{Second-order interpolet in one
dimension: non-zero two-scale coefficients ($c_n$), function values ($\cI$),
and overlaps of $\partial^h/\partial x^h$ ($M^{\{h\}}_n$, as defined
in (\ref{def:vecM})).}
\label{tbl:d1L2}
\end{table}

\begin{table}
\begin{center}
\begin{tabular}{|c||c|c|c|} \hline
$n$ & $0$ & $\{1\}$ & $\{3\}$\\ \hline \hline $c_n$ & $1$ &
$\frac{9}{16}$ & $-\frac{1}{16}$ \\ \hline
\end{tabular}\\
\vspace{0.2in}
\begin{tabular}{|c||c|c|c|c|c|c|c|c|} \hline
x & $0$ & $\{\frac{1}{2}\}$ & $\{1\}$ & $\{\frac{3}{2}\}$ & $\{2\}$ &
$\{\frac{5}{2}\}$ & $\{3\}$ \\ \hline \hline
$\cI(C_1)$ & $1$ &
$\frac{9}{16}$ & 0 & $-\frac{1}{16}$ & $0$ & $0$ & $0$ \\ \hline
\end{tabular}\\
\vspace{0.2in}
\begin{tabular}{|c||c|c|c|c|c|c|} \hline
$n$ & $0$ & $1$ & $2$ & $3$ & $4$ & $5$ \\ \hline \hline
$\vec M^{\{0\}}$  & ${\frac {56264}{70245
}}$ & ${\frac {19253}{140490}}$ & $-{\frac {2827}{70245}}$ & ${\frac {6283}{
2247840}}$ & $-{\frac {16}{210735}}$ & $-{\frac {1}{6743520}}$ \\
\hline
$\vec M^{\{1\}}$ & $0$ & ${\frac {3659}{5280}}$ & $-{\frac {731}{6930}}$ & ${\frac {481}{73920}}$ & $-{\frac {4}{10395}}$ & $-{\frac {1}{665280}}$ \\
\hline
$\vec M^{\{2\}}$ & $-\frac{20}{9}$ & $\frac{9}{8}$ & $0$ &
$-\frac{1}{72}$ & 0 & 0 \\
\hline
\end{tabular}
\end{center}
\caption{Third order interpolating cardinal scaling function in one
dimension: non-zero two-scale coefficients ($c_n$), function values ($\cI$),
and overlaps of $\partial^h/\partial x^h$ ($M^{\{h\}}_n$, as defined
in (\ref{def:vecM})).}
\label{tbl:d1L3}
\end{table}

\begin{figure}
\begin{center}
\scalebox{\figscale}{\scalebox{0.6}{\includegraphics{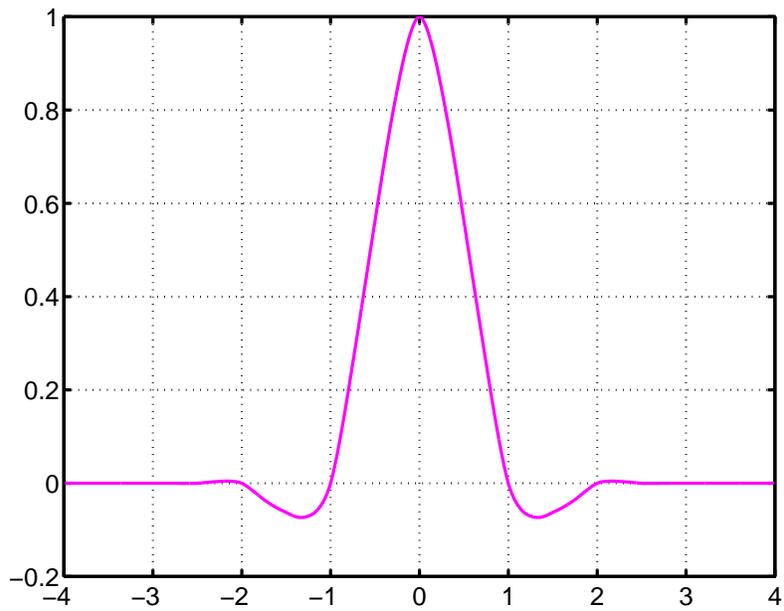}}} 
\end{center}
\caption{Third-order interpolet in one
dimension.}
\label{fig:d1L3}
\end{figure}

The final function we consider in one dimension is the
third-order interpolet, which was used in the calculations in
\cite{mgras,aps,JCP} to produce the results shown in
Figures~\ref{fig:C}, \ref{fig:wfsr}, \ref{fig:EvsR} and
\ref{fig:wfsI}.  For odd orders of interpolation, the odd-order
conditions from (\ref{eqn:ccond1d}) are always equivalent to the
condition that the interpolet be symmetric, $c_{n}=c_{-n}$.  The
remaining conditions for the third-order case are just
\begin{eqnarray*}
c_1+c_3 & = & \frac{1}{2} \\ c_1+9c_3 & = & 0
\end{eqnarray*}
Table \ref{tbl:d1L3} gives the resulting two-scale coefficients,
function values on $C_1$, and vectors of matrix elements for
homogeneous integral-differential operators.
Figure~\ref{fig:d1L3} shows the function in real space, and Figure
\ref{fig:convsine} shows that the leading error in
reconstructing the sine function has the expected fourth-order
behavior.  In one dimension, for each odd order, the minimally
supported interpolets are just the scaling functions of Deslauriers
Dubuc \cite{DD}.

\begin{table}
\begin{center}
\begin{tabular}{|c||c|c|c|c|} \hline
$n$ & $000$ & $\{001\}$ & $\{011\}$ & $\{111\}$ \\ \hline \hline $c_n$
& $1$ & $\frac{1}{2}$ & $\frac{1}{4}$ & $\frac{1}{8}$ \\ \hline
\end{tabular}
\end{center}
\caption{Two-scale coefficients for first-order interpolating cardinal
scaling function in $d=3$ dimensions (product form).}
\label{tbl:d3L1}
\end{table}

In higher dimensions, the simplest way to construct interpolets is to
take the two-scale symbol to be an outer product $m_0(q) \equiv
\prod_{i=1}^d m_0^{(1)}(q_i)$ of the two-scale symbol $m_0^{(1)}(q)$
for the one-dimensional interpolet of the corresponding order.  The
conditions (\ref{eqn:InterpCondm0},\ref{eqn:InterpCondmpi}) then
factor into $d$ separate one-dimensional conditions such that if
$m_0^{(1)}(q)$ is taken as the two-scale symbol for an
$\ci^{\mbox{th}}$-order interpolet in one dimension, $m_0(q)$ will be
the two-scale symbol for an $\ci^{\mbox{th}}$-order interpolet in $d$
dimensions.  Because the two-scale coefficients are the Fourier series
coefficients of $m_0(q)$, the two-scale coefficients for the
$d$-dimensional interpolet are the outer product of those for the
generating one-dimensional interpolet, $c_n \equiv \prod_{i=1}^d
c^{(1)}_{n_i}$.  Finally, from the two-scale relation
(\ref{eqn:normprod}), the scaling function corresponding to $m_0(q)$
is just the outer product of the one-dimensional generating
interpolet, $\cI(x) \equiv \prod_{i=1}^d
\cI^{(1)}(x_i)$.  As an example, Table~\ref{tbl:d3L1}
gives the two-scale coefficients for the first-order interpolet
in $d=3$ dimensions.

Although the outer-product prescription may always be used to generate
higher-dimensional interpolets of very simple form, the resulting
functions are generally not as compact as possible.  To illustrate the
fact that working in a full three-dimensional framework can generate
more compact functions, we conclude this section by constructing a
third-order interpolet in $d=3$ dimensions which is {\em not} an outer
product of three one-dimensional functions.

\begin{table}
\begin{center}
\begin{tabular}{|c||c|c|c|c|c|c|c|c|c|c|} \hline
$n$ &\{000\} & \{001\} & \{011\} & \{111\} & \{003\} & \{013\} &
\{113\} & \{033\} & \{133\} & \{333\} \\ \hline \hline
$c_n$ (3d) & 1 & $\frac{9}{16}$ & $\frac{5}{16}$ &$\frac{11}{64}$ &
$-\frac{1}{16}$ & $-\frac{1}{32}$ & $-\frac{1}{64}$ & 0 & 0 & 0 \\
\hline $c_n$ (pd) & 1 & $\frac{9}{16}$ & $\frac{81}{256}$ &
$\frac{729}{4096}$ & $-\frac{1}{16}$ & $-\frac{9}{256}$ &
$-\frac{81}{4096}$ & $\frac{1}{256}$ & $\frac{9}{4096}$ &
$-\frac{1}{4096}$ \\ \hline
\end{tabular}
\end{center}
\caption{Two-scale coefficients for third-order interpolating cardinal
scaling functions in $d=3$ dimensions: full $d=3$-dimensional construction
(3d), product form (pd).}
\label{tbl:d3L3}
\end{table}

For the third-order interpolet in three dimensions, Eqs.
(\ref{eqn:InterpCondm0},\ref{eqn:InterpCondmpi}) represent a total of
eighty linear conditions on the two-scale coefficients.  Considering
functions with full cubic symmetry reduces this to just seven
conditions, as we now show.  The reflection symmetries about the
coordinate planes, $c_{n_1,n_2,n_3}=c_{\pm n_1,\pm n_2,\pm n_3}$,
ensure that the first- and third-order conditions, as well as the
conditions for the off-diagonal second-order multinomials $x_1x_2$,
$x_1x_3$ and $x_2x_3$, from (\ref{eqn:interp}) are all satisfied,
leaving only the zeroth-order condition and the three diagonal
second-order conditions, $x_1^2$, $x_2^2$ and $x_3^2$.  This set of
four conditions has implications for $m_0(q)$ at $q=0$ and the seven
points $\pi \eta_i$.  By the permutation symmetries, among these eight
total sets of implications, only those for $\eta$ at $(000)$, $(001)$,
$(011)$, $(111)$ need be enforced explicitly.  The cardinality
condition (\ref{eqn:summ0}), moreover, implies that one of these sets
is redundant, so that we need not impose the conditions at $(011)$.
The zeroth order conditions at the three remaining points appear as
the first three conditions in (\ref{eqn:gen3d}).  For the diagonal
second-order conditions, symmetry renders the three conditions at each
of the two points $(000)$, $(111)$ equivalent.  One condition from
each of these points appears in (\ref{eqn:gen3d}).  At the final
point, $(001)$, the conditions for $x_1^2$ and $x_2^2$ are also
equivalent, and so only one of these appears in (\ref{eqn:gen3d}).
Finally, the $x_3^2$ condition at $(001)$ makes up the last of the
seven conditions in
\begin{equation}\label{eqn:gen3d}
A\left(\begin{array}{c}
c_{\{000\}} \\ c_{\{001\}} \\ c_{\{011\}} \\ c_{\{111\}} \\ c_{\{003\}} \\ c_{\{013\}} \\
      c_{\{113\}} \end{array}\right) =
\left(\begin{array}{c}
1 \\ 0 \\ 0 \\ 0 \\ 0 \\ 0 \\ 0 \end{array}\right),
\end{equation}
where
$$
A \equiv \left(\begin{array}{ccccccc}
         1 & 6 & 12 & 8 & 6 & 24 & 24\\
         1 & 2 & -4 & -8 & 2 & -8 & -24\\
         1 & -6 & 12 & -8 & -6 & 24 & -24\\
         0 & 2 & 8 & 8 & 18 & 80 & 88\\
         0 & -2 & -8 & -8 & -18 & -80 & -88\\
         0 & 2 & 0 & -8 & 18 & 0 & -88\\
         0 & -2 & 8 & -8 & -18 & 80 & -88
         \end{array}\right).
$$

\begin{figure}
\begin{center}
\scalebox{\figscale}{\scalebox{0.55}{\includegraphics{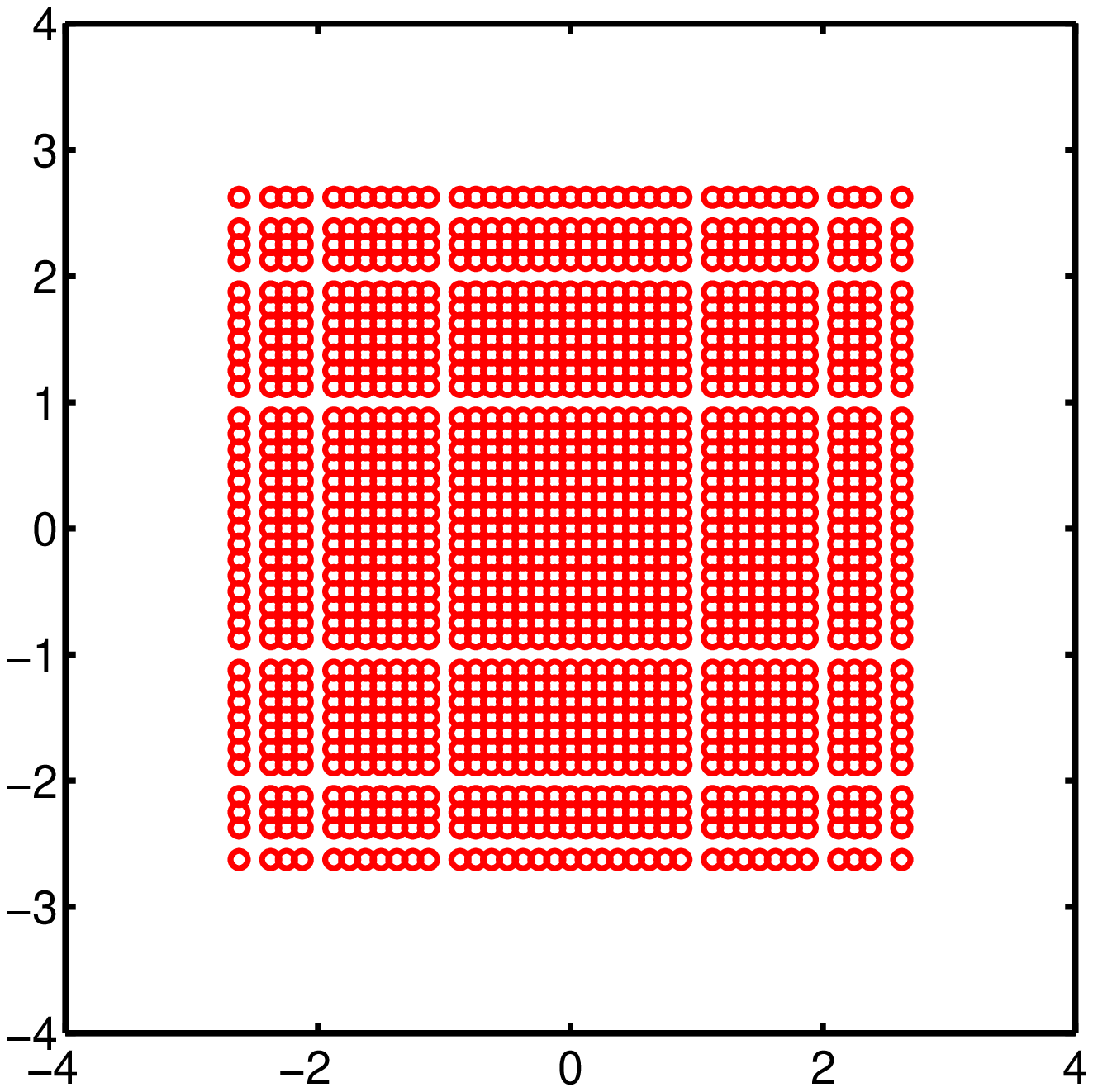}}} 
\scalebox{\figscale}{\scalebox{0.55}{\includegraphics{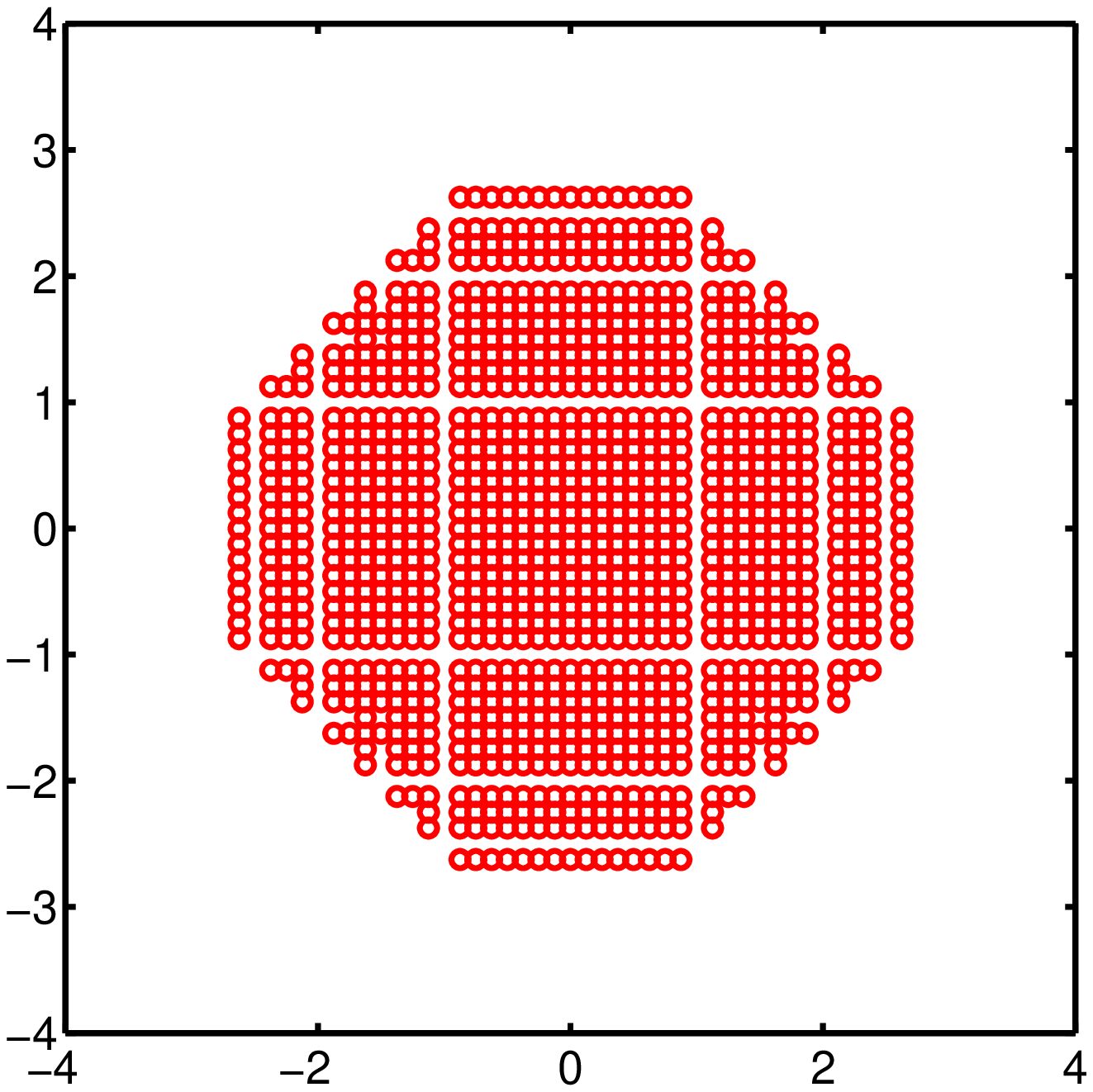}}} 
\end{center}
\caption{Support in $x_3=0$ plane of $d=3$-dimensional third-order
interpolets: product form (left), three-dimensional construction (right).}
\label{fig:spd3}
\end{figure}

\begin{figure}
\begin{center}
\scalebox{\figscale}{\scalebox{0.6}{\includegraphics{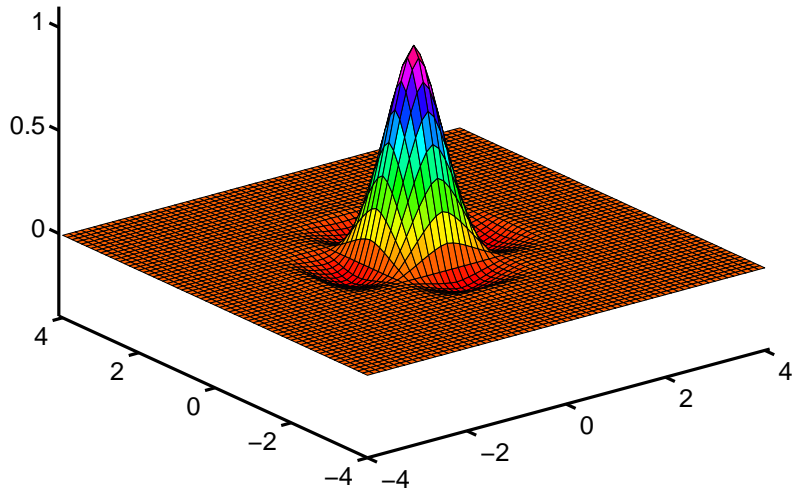}}} 
\end{center}
\caption{Minimally supported $d=3$ third-order interpolet
in the $x_3=0$ plane.}
\label{fig:d3L3}
\end{figure}

Table~\ref{tbl:d3L3} gives the resulting coefficients along with the
coefficients for the outer product of three one-dimensional, $\ci=3$
interpolets.  Figure~\ref{fig:spd3} illustrates the more compact
nature of the result of the three-dimensional construction by
comparing the supports of the two functions in the $x_3=0$ plane.  The
three-dimensional volume of the support of the new function is smaller
than that of the product form by a factor of two.  Finally,
Figure~\ref{fig:d3L3} shows the appearance of the new function in the
$x_3=0$ coordinate plane as computed from the recursion
(\ref{eqn:selfIr}).
         
\section{Multilevel Methods} \label{sec:semicbases:fastalgs}

As Section~\ref{subsubsec:semicops} discusses, the direct approach of
applying the linear operators $\cI$, $\cJ$, $\cI^\dagger$,
$\cJ^\dagger$, $\cO$, $L$ by multiplying by the corresponding matrices
is relatively costly, requiring thousands of operations per
coefficient in the restricted multiresolution analyses typical in the
calculation of electronic structure.  For {\em unrestricted}
multiresolution analyses, more methods more efficient than direct
matrix multiplication exist which exploit the structure imposed by the
two-scale relation to expend only $O(1)$ floating point operations per
expansion coefficient.  Many of these methods, however, process
information on the unrestricted grid and, thus, in a restricted
multiresolution analysis of $n_r$ functions will expend $O(n/n_r)$
floating point operations per expansion coefficient, where $n$ is the
number of points in the unrestricted multiresolution analysis.  For
all-electron calculations, $n/n_r \ge 10^4$, and these approaches are
also extremely inefficient.

The recent resolution of this problem has been to find new methods which
require only $O(1)$ operations per coefficient but which give results
which are {\em unchanged} when ignoring the
$n-n_r$ coefficients associated with functions removed in the
restriction of the multiresolution analysis.  These methods then
attain the goal of expending only $O(n_r)$ operations in a restricted
multiresolution analysis of $n_r$ basis functions.  We refer to such
methods as {\em restrictable}.

This section reviews the approaches used in the calculation of
electronic structure, which included both
unrestrictable\cite{chou,tymczak} and restrictable\cite{BCR,JCP}
methods.
For clarity of presentation, we consider in this section
only operations in {\em unrestricted} multiresolution analyses, and
dedicate Section~\ref{sec:inhomogridops} specifically to the effects
of restriction.

  \subsection{Transforms: $\cI$, $\cI^\dagger$, $\cJ$, $\cJ^\dagger$} \label{sec:xforms}

    \subsubsection{General and orthonormal bases} \label{sec:gentrans}

\paragraph{Forward transform}

The forward transform of
Sec.~\ref{sec:physops} converts the multiscale expansion coefficients
$\vec F_{N:M}$ into the values of the function on a set of points in
real-space.  In a multiresolution analysis, the product $\cI_{N:M}
\vec F_{N:M}$ gives the coefficients $\vec F_N$ of the
single-scale expansion of the function in $V_N$.  The values of the
function on the points $p$ of the real-space grid $G$ are then
$$
(\vec f)_p = \sum_{\alpha \in C_N} \phi(2^N(p-\alpha)) (\cI_{N:M} \vec F_{N:M})_\alpha,
$$
or in matrix language,
$$
\vec f = \Phi \cI_{N:M} F_{N:M},
$$
where $\Phi_{p\alpha} \equiv \phi(2^N(p-\alpha))$.  The forward transform
operator $\cI$ of (\ref{def:cI}) is thus
\begin{equation} \label{eqn:fullI}
\cI=\Phi \cI_{N:M}.
\end{equation}

In general, the invariance of the single-scale basis for $V_N$ under the
translations of the lattice $C_N$ means that multiplication by the
matrix $\Phi$ takes the form of a convolution.  Multiplication by
$\Phi$ may thus be carried out using fast Fourier transform techniques
with approximately $10 \, n \log_2 n$ floating-point
operations\cite{dau,chou}.  Alternately, when the $\phi(x)$ are
products of one-dimensional functions, $\Phi$ may be factored into
one-dimensional convolutions along each of the coordinate directions,
which expend a total of $d (\Vol \supp \phi(x))^{1/d}\,n$ operations where $d$ is
the dimension of space and $\Vol \supp \phi(x)$ is the volume of the
region of space over which the function $\phi(x)$ is non-zero.
For multiresolution analyses with cardinal scaling functions (both
semicardinal bases\cite{mgras,dicle} and lifted
bases\cite{goedecker}), multiplication by $\Phi$ is far simpler:
cardinality of the scaling functions implies $\Phi=I$, where $I$ is
the identity.

\begin{figure}
\begin{center}
\scalebox{\figscale}{\scalebox{0.85}{\includegraphics{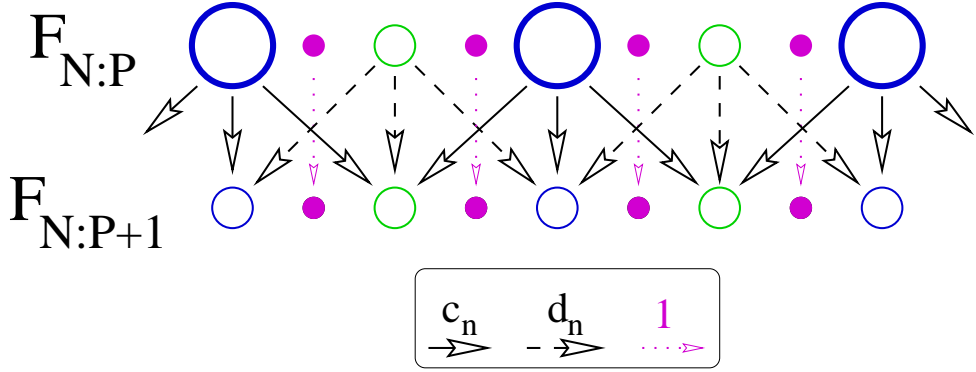}}}
\end{center}
\caption{Information flow of the operation $F_{N:P+1}=\cI_{P+1,P}
F_{N:P}$ (general multiresolution analysis).}
\label{fig:Ipp}
\end{figure}

\begin{figure}
\begin{center}
\scalebox{\figscale}{\scalebox{0.85}{\includegraphics{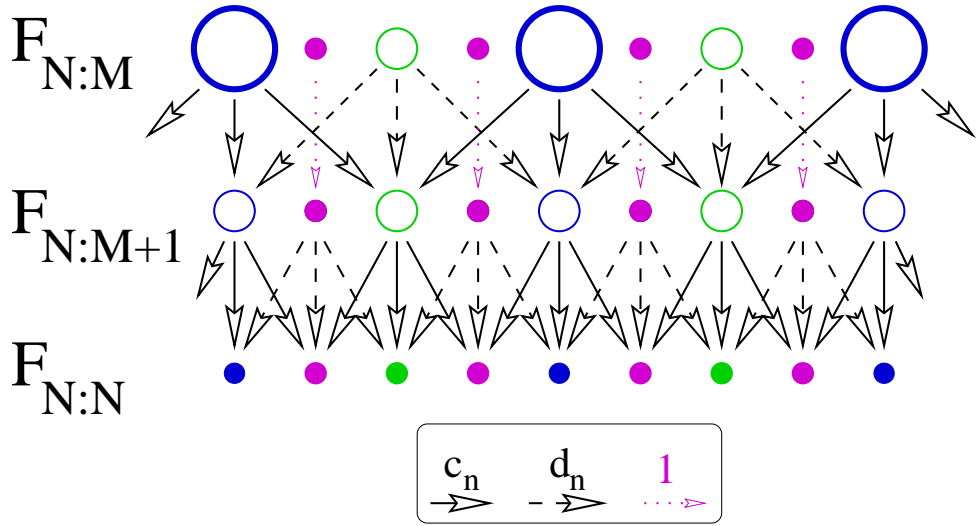}}}
\end{center}
\caption{Information flow of the multiscale operation $F_{N:N}=\cI_{N:M}
F_{N:M}$ (general multiresolution analysis).}
\label{fig:MIpp}
\end{figure}

Figure \ref{fig:Ipp} illustrates the flow of information corresponding
to each factor $\cI_{P+1,P}$ in the definition (\ref{eqn:matMRA}) of
$\cI_{N:M}$.  The circles of the upper and lower rows represent
coefficients of the multiscale expansions $\vec F_{N:P}$ and $\vec
F_{N:P+1}$, respectively.  On the upper row, the larger and smaller
open circles represent the coefficients for the scaling functions of
scale $P$ and detail functions of scale $P+1$, respectively, and the
open circles on the lower row represent coefficients for the scaling
functions of scale $P+1$.  The filled circles on both rows represent
coefficients from finer scales, which are unaffected by $\cI_{P+1,P}$.
The arrows represent the individual multiply-add operations in the
application of $\cI_{P+1,P}$ to compute $\vec F_{N:P+1} = \cI_{P+1,P}
\vec F_{N:P}$.  Each arrow multiplies the expansion coefficient at its
base by a constant factor and accumulates the result onto coefficient
at its head.  The factors which the arrows carry are the two-scale
coefficients $c_n$ and $d_n$, or unity in the case of the dotted
arrows.  Note that the coefficients $\vec F_{N:P+1}$ are taken to be
zero before the accumulation begins.  Also, for clarity, the figure
shows only the connections associated with two-scale coefficients in
the range $|n| \le 1$.  Figure~\ref{fig:MIpp} illustrates, for three
levels, the cascading data flow of (\ref{eqn:defINM}) when the
$\cI_{P+1,P}$ string together to make the final multiscale operator
$\cI_{N:M}$.

The translational symmetry of
the $c_n$ and $d_n$ coefficients, reflecting the invariance of the two-scale
basis for $V_P \oplus W_{P+1}$ under the translations of $C_P$, implies that
the data flow corresponding to $\cI_{P+1,P}$ takes the form of a set
of convolutions, one for each scaling and detail function of the
two-scale basis.  As with $\Phi$, these convolutions may be carried
out using Fourier transform techniques, or, in the case of product
basis functions, factored into sets of one dimensional convolutions
along the coordinate directions and treated all at
once\cite[Ch. 10]{dau}.

\paragraph{Conjugate forward transform}

The conjugate forward transform is
\begin{equation} \label{eqn:prodIt0}
\cI^\dagger = \cI_{N:M}^\dagger \Phi^\dagger.
\end{equation}
The implementation of this expression parallels that of
(\ref{eqn:fullI}).  In the general case, $\Phi^\dagger$ is a
convolution carried out as described above, and for the case of
cardinal scaling functions, $\Phi^\dagger=I$.  The component factors
$\cI_{P+1,P}^\dagger$ of $\cI_{N:M}^\dagger$ are again invariant under
the translations of $C_{P}$ and may be either implemented with either Fourier
techniques or, for product functions, factored into one-dimensional
convolutions.

\begin{figure}
\begin{center}
\scalebox{\figscale}{\scalebox{0.85}{\includegraphics{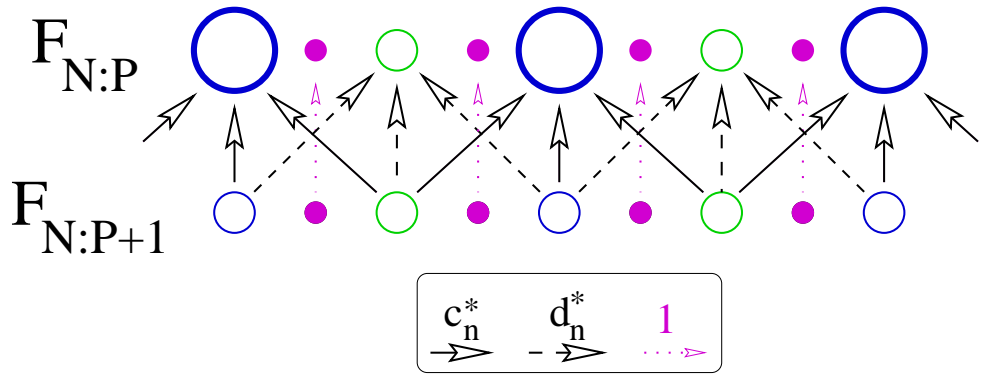}}}
\end{center}
\caption{Information flow of the operation
$F_{N:P}=\cI_{P+1,P}^\dagger F_{N:P+1}$ (general multiresolution analysis).}
\label{fig:Itpp}
\end{figure}

The information flow for multiplication by $\cI_{N:M}^\dagger$ is
again a cascade of two-scale connections $\cI_{P+1,P}^\dagger$, but
now in reverse order.  Note that, when written in terms of components,
the matrix multiplication $\vec a=M \vec b$ is $a_i = \sum_j m_{ij}
b_j$, and multiplication by the Hermitian conjugate, $\vec a=M^\dagger
\vec b$, is $a_j = \sum_i m_{ij}^* b_i$.  This means that
multiplication by the Hermitian conjugate simply reverses the
direction of and conjugates the factors associated with the arrows.
Figure \ref{fig:Itpp} illustrates this information flow for one stage
$\cI_{P+1,P}^\dagger$ of the cascade for $\cI_{N:M}^\dagger$.

\paragraph{Inverse transform}

For general unrestricted multiresolution analysis and restricted
semicardinal multiresolution analyses, the forward transform $\cI$ is
invertible, and we thus consider the case $\cJ \equiv \cI^{-1}$.
Then, the inverse transform is
\begin{equation} \label{eqn:prodJ0}
\cJ = \cI_{N:M}^{-1} \Phi^{-1}.
\end{equation}
For orthonormal wavelets, the inverse convolution $\Phi^{-1}$ must be
computed using Fourier techniques which require access to the
unrestricted grid.  The use of cardinal scaling functions removes this
complication, for then simply $\Phi^{-1}=I$.

\begin{figure}
\begin{center}
\scalebox{\figscale}{\scalebox{0.85}{\includegraphics{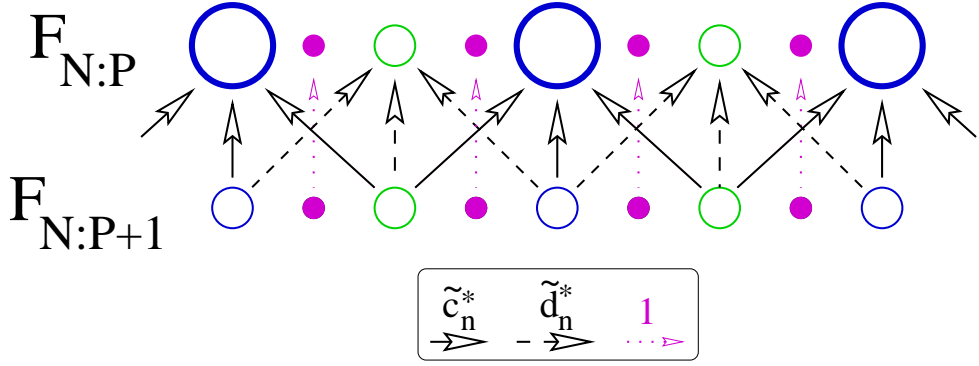}}}
\end{center}
\caption{Information flow of the operation
$F_{N:P}=\cI_{P+1,P}^{-1} F_{N:P+1}$ (general multiresolution analysis).}
\label{fig:Jpp}
\end{figure}

In multiresolution analysis, the condition (\ref{eqn:MRAdet}) ensures
that the two-scale components $\cI_{P+1,P}^{-1}$ of the cascade for
$\cI_{N:M}^{-1}$ always exist as linear operators.  The invariance of
$\cI_{P+1,P}$ under the translations of $C_P$ ensures that the
information flow for $\cI_{P+1,P}^{-1}$ always takes the form of
Figure~\ref{fig:Jpp} for some set of {\em dual coefficients}
$\tilde{c}_n$ and $\tilde{d}_n$, defined by convention with the
complex conjugation as labelled in the figure.  For orthonormal
wavelets $\cI_{P+1,P}^{-1}=\cI_{P+1,P}^\dagger$ so that, comparing
Figures~\ref{fig:Itpp} and \ref{fig:Jpp}, the two-scale coefficients
and their duals are identical.  The dual coefficients also have a
special structure in semicardinal bases, which we shall explore in
Sec.~\ref{sec:subsemic}.

\paragraph{Conjugate inverse transform}

\begin{figure}
\begin{center}
\scalebox{\figscale}{\scalebox{0.85}{\includegraphics{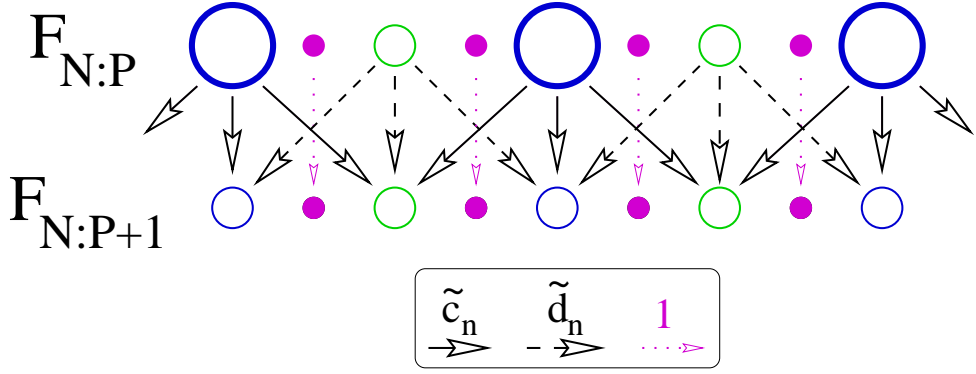}}}
\end{center}
\caption{Information flow of the operation
$F_{N:P+1}=\cI_{P+1,P}^{-\dagger} F_{N:P}$ (general multiresolution analysis).}
\label{fig:Jtpp}
\end{figure}

The remaining transform is the Hermitian conjugate of
(\ref{eqn:prodJ0}),
\begin{equation} \label{eqn:prodJt0}
\cJ^\dagger = \cI_{N:M}^{-\dagger} \Phi^{-\dagger},
\end{equation}
where for any invertible matrix, $A^{-\dagger} \equiv (A^{-1})^\dagger
= (A^\dagger)^{-1}$.  In general, the operation $\Phi^{-\dagger}$ may
be implemented either by inverting the convolution associated with
$\Phi^\dagger$ with Fourier techniques.  Again, cardinal scaling
functions are much simpler and give just $\Phi^{\dagger}=I$.  The
operations of the cascade for $\cI_{N:M}^{-\dagger}$ are obtained by
complex conjugation and reversal of the flow of Figure~\ref{fig:Jpp}.
As may be seen from Figure~\ref{fig:Jtpp}, the convention for the
definition of the dual coefficients has been established to ensure
that the conjugate inverse transform is just the forward transform
associated with the dual coefficients.  In the orthonormal case, where
the coefficients and their duals are the same, this represents the
expected result that the conjugate inverse and forward transforms are
the same.  The next section discusses the form for the
$\cI_{P+1,P}^{-\dagger}$ in semicardinal bases.

    \subsubsection{Semicardinal bases} \label{sec:subsemic}

As we saw in the preceding section, orthonormal bases have the
advantage that the {\em dual} coefficients appearing in the inverse
transforms are identical to the two-scale coefficients of the forward
transform.  The disadvantage is that access to the unrestricted grid
of finest resolution is needed to implement the inverse convolutions
$\Phi^{-1}$ and $\Phi^{-\dagger}$.  Because these later operations are
trivial for bases with cardinal scaling functions, semicardinal bases
will have a great advantage provided the dual coefficients needed for
the inverse transforms are simple in form, as we now verify.

\begin{figure}
\begin{center}
\scalebox{\figscale}{\scalebox{0.85}{\includegraphics{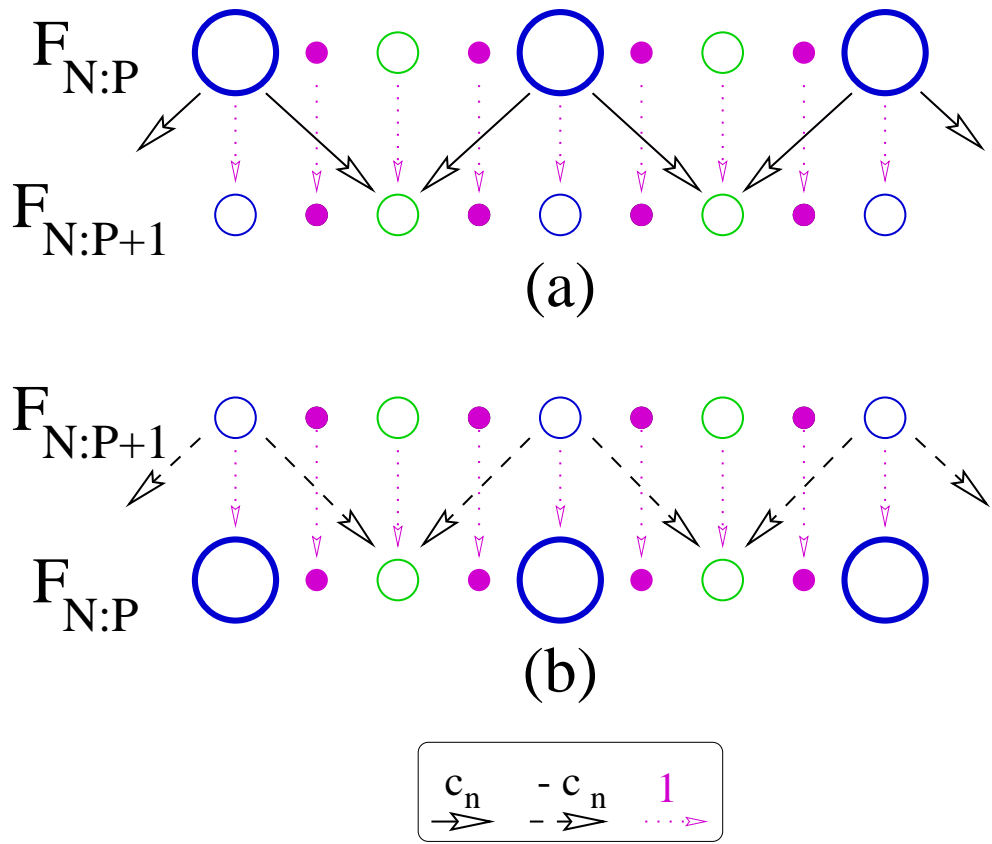}}}
\end{center}
\caption{Information flow in a semicardinal basis: (a) forward
transform, (b) inverse transform.  (For respective conjugate transforms,
conjugate and reverse the direction of the arrows.)}
\label{fig:IJsc}
\end{figure}

The particularly simple form for the dual coefficients for
semicardinal multiresolution analyses arises from the structure of the
$\cI_{P+1,P}$.  To elucidate this structure, we analyze $\cI_{P+1,P}$
into its effects on different scales, $\cI_{P+1,P} =
(\cP_{C_{P+1}}+\cP_{B_{P+1}}) \cI_{P+1,P}
(\cP_{C_{P+1}}+\cP_{B_{P+1}})$, where $B_{P+1}$ is the set of points
on scales {\em beyond} $P+1$ as defined in (\ref{def:BP}).  Because
$\cI_{P+1,P}$ does not affect functions beyond scale $P+1$, we may use
(\ref{eqn:IBQ1}) to simplify this to $\cI_{P+1,P} =
\cP_{B_{P+1}}+\cP_{C_{P+1}} \cI_{P+1,P} \cP_{C_{P+1}}$.  Next, according to
(\ref{def:semic}), the fact that the two-scale basis for $V_P+W_{P+1}$
is semicardinal on the hierarchy $C_P \subset C_{P+1}$ gives $\cP_{C_{P+1}}
\cI_{P+1,P} \cP_{C_{P}} =
\cP_{C_P}+\cP_{D_{P+1}} \cI_{P+1,P} \cP_{C_P}$ and $\cP_{C_{P}} \cI_{P+1,P}
\cP_{D_{P+1}} = \cP_{D_{P+1}}$.  Combining these
results, we have
\begin{eqnarray}
\cI_{P+1,P} 
& = & \cP_{B_{P+1}} + \left((\cP_{C_P}+\cP_{D_{P+1}} \cI_{P+1,P} \cP_{C_P})+\cP_{D_{P+1}}\right)
 \nonumber \\
& = & I + \cP_{D_{P+1}} \cI_{P+1,P} \cP_{C_P}. \label{eqn:decompIPP}
\end{eqnarray}
As Figure \ref{fig:IJsc}a illustrates, this means that, apart from the
simple copy operation $I$, the only effect of $\cI_{P+1,P}$ in the
semicardinal case is to add onto the detail points $D_{P+1}$
information coming from the coarse points $C_P$.

The full forward transform, given this structure for $\cI_{P+1,P}$ and
the fact that $\Phi=I$, therefore is
\begin{equation} \label{eqn:prodI}
\cI=\prod_{P=N-1}^M \left( I +\cP_{D_{P+1}} \cI_{P+1,P} \cP_{C_P} \right),
\end{equation}
which corresponds to the following recursive algorithm for computing
$f=\cI F$,
\begin{eqnarray}
F_{N:M} & \equiv & F; \label{alg:I} \\
F_{N:P+1} & = & F_{N:P}+\cP_{D_{P+1}} \cI_{P+1,P} \cP_{C_P} F_{N:P};
\nonumber \\
f & \equiv & F_{N:N}. \nonumber
\end{eqnarray}
As we show in Sec.~\ref{sec:inhomogridops}, this algorithm is {\em
restrictable} and gives the same results even when completely ignoring
points and coefficients eliminated in a restricted multiresolution
analysis.  Thus, with this approach, the forward transform is
extremely efficient and may be computed with only $\Vol \supp
\cI(x)$, or $d (\Vol \supp \cI(x))^{1/d}$ in the case of a product
interpolet, operations per coefficient.

As described above, the conjugate to the forward transform
(\ref{eqn:prodIt0}) must simply reverse the direction of the
information flow relative to Figure~\ref{fig:IJsc}a.  In product form,
it is
\begin{equation}  \label{eqn:prodIt}
\cI^\dagger =  \prod_{P=M}^{N-1} \left(I+  \cP_{C_P} \cI_{P+1,P}^\dagger
\cP_{D_{P+1}} \right),
\end{equation}
corresponding to the following algorithm for computing $F=\cI^\dagger f$,
\begin{eqnarray}
F_{N:N} & \equiv & f; \label{alg:It} \\
F_{N:P} & = & F_{N:P+1} + \cP_{C_P} \cI_{P+1,P}^\dagger \cP_{D_{P+1}} F_{N:P+1}; \nonumber \\
F & \equiv & F_{N:M}. \nonumber
\end{eqnarray}
Despite the fact that information appears to flow in from missing points when
reversing the arrows in Figure~\ref{fig:IJsc}a, we show in the next
section that this algorithm is also restrictable and thus
also efficient in restricted multiresolution analyses.

Determining the dual coefficients needed to invert each stage
$\cI_{P+1,P}$ amounts to finding a prescription for undoing the
effects of the forward transform.  As Figure
\ref{fig:IJsc}a shows, semicardinality ensures that the forward
transform has no effect on the data sitting on the points of $C_P$, so
that in the inverse transform too, these data may be left unmodified.  The
information flowing diagonally from the points of $C_P$ in the forward
transform, however, does corrupt the values on the points of $D_P$.
Because the information which flowed along these links sits on the
points of $C_P$ at the start of the inverse transform, to determine
the values originally sitting on $D_P$, one need only carry
information along the same diagonal links but with coefficients the
opposite sign.  Figure~\ref{fig:IJsc}b illustrates the resulting data
flow for the inverse transform.  Note that Figures~\ref{fig:IJsc}a-b
are constructed so that the entire diagram may be viewed from top to
bottom as the process which first performs the forward transform and
then inverts it.  Note that, to accomplish this, the sequence of
spaces in Figure~\ref{fig:IJsc}b is reversed relative to that of the
preceding diagrams.

Comparing Figure~\ref{fig:IJsc}b with Figure~\ref{fig:Jpp} we see that
in semicardinal multiresolution analyses, the dual coefficients are
$\tilde{c}^*_n=d_n=\delta_{n0}$ and $\tilde{d}^*_n=2 d_n-c_n$.
Expressed in our matrix language, Figure \ref{fig:IJsc}b
represents
\begin{equation} \label{eqn:compJ}
\cI^{-1}_{P+1,P} = I - \cP_{D_{P+1}} \cI_{P+1,P} \cP_{C_P},
\end{equation}
which may be verified algebraically by multiplication with
(\ref{eqn:decompIPP}) and use of the identity $\cP_{C_P}
\cP_{D_{P+1}}=0$.  The full inverse transform is thus
\begin{equation} \label{eqn:Jdef}
\cJ \equiv \cI^{-1} = \prod_{P=M}^{N-1}
\left(I - \cP_{D_{P+1}} \cI_{P+1,P} \cP_{C_P}\right), 
\end{equation}
which may be cast into a recursion to
compute $F=\cJ f$:
\begin{eqnarray}
F_{N:N} & \equiv & f; \label{alg:J} \\
F_{N:P} & = & F_{N:P+1}-\cP_{D_{P+1}} \cI_{P+1,P} \cP_{C_P} F_{N:P+1};
\nonumber \\
F & \equiv & F_{N:M}. \nonumber
\end{eqnarray}
Section~\ref{sec:inhomogridops} verifies that this inverse
transform is also restrictable.

\begin{figure}
\begin{center}
\scalebox{\figscale}{\scalebox{0.85}{\includegraphics{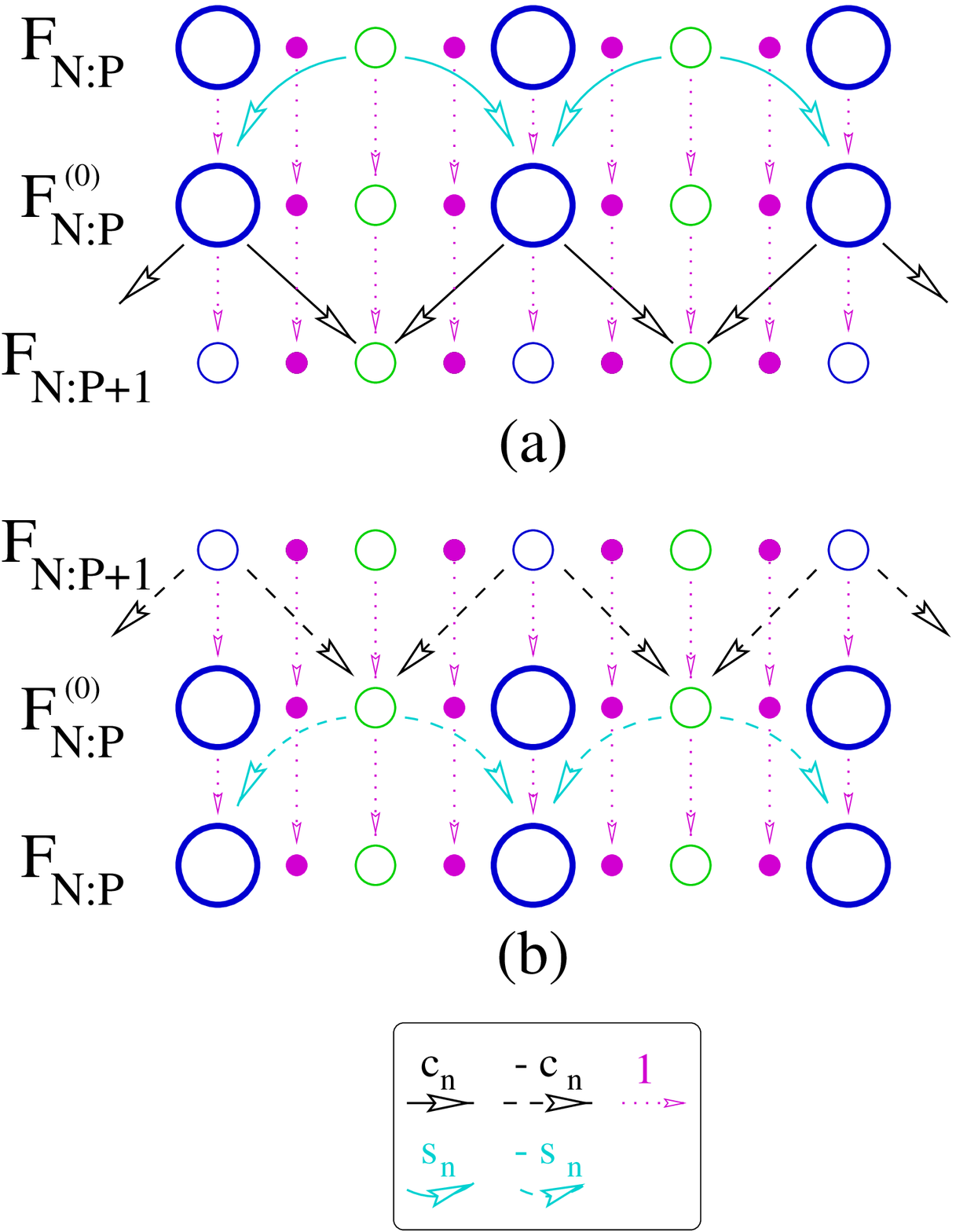}}}
\end{center}
\caption{Information flow for a lifted semicardinal basis:
(a) forward transform, (b) inverse transform.  (For respective conjugate
transforms, conjugate and reverse the direction of the arrows.)}
\label{fig:IJlsc}
\end{figure}

Finally, the conjugate inverse transform is
\begin{equation} \label{eqn:prodJt}
\cJ^\dagger = \prod_{P=N-1}^{M}
\left(I - \cP_{C_P} \cI_{P+1,P}^\dagger \cP_{D_{P+1}} \right),
\end{equation}
which may be computed with the restrictable algorithm,
\begin{eqnarray}
F_{N:M} & \equiv & F; \label{alg:Jt} \\
F_{N:P+1} & = & F_{N:P}-\cP_{C_P} \cI_{P+1,P}^\dagger \cP_{D_{P+1}} F_{N:P};
\nonumber \\
f & \equiv & F_{N:N}, \nonumber
\end{eqnarray}
which simply reverses the direction of information flow from that of
Figure~\ref{fig:IJsc}b.

    \subsubsection{Lifted bases}

We close this section with a brief consideration of bases constructed
by lifting semicardinal bases through the scheme of \cite{sweldens:96}
as in \cite{goedecker}.  In the lifting scheme, each detail
function $\psi_\alpha(x)$ consists of an original un-lifted detail
function $\psi^{(0)}_\alpha(x)$ plus an arbitrary linear combination of
scaling functions from the next coarser scale,
$$
\psi_\alpha(x-\eta_\alpha/2) \equiv \psi^{(0)}_\alpha(x-\eta_\alpha/2)+\sum_n s_{\alpha,n} \phi(x-n),
$$
where the $\eta_\alpha$ are the decoration points on which the detail
functions are centered.  (See Eq. (\ref{eqn:MRW}).)  Although a
non-zero choice for the coefficients $s_{\alpha,n}$ destroys
semicardinality, these coefficients may be chosen to optimize the
basis for other purposes.  In \cite{goedecker}, for instance, this
combination is taken to produce detail functions with zero integral
and zero multipole moments up to some order for use in solving
Poisson's equation.

With detail functions of this form, each stage of the cascade for the
full forward transform may then proceed in two phases: (1) the lifted
multilevel description $F_{N:P}$ of the function is converted to the
un-lifted representation $F^{(0)}_{N:P}$ by transferring the weights
$s_{\alpha,n}$ from the detail functions of scale $P+1$ to the scaling
functions of scale $P$; (2) the un-lifted forward transform is carried
out on $F^{(0)}_{N:P}$ to produce $F_{N:P+1}$.  Figure \ref{fig:IJlsc}a
shows this two-stage process for the component operation $\cI_{P+1,P}$
in the forward transform of a lifted a semicardinal basis.  The
conjugate forward transform just reverses this information flow and
conjugates the factors carried by the arrows.

One may apply precisely the same logic as used in generating
Figure~\ref{fig:IJsc}b to invert each of these two stages in sequence
to produce the lifted version of $\cI_{P+1,P}^{-1}$.  Figure
\ref{fig:IJsc}b shows the resulting information flow.  The conjugate
to this inverse reverses the information flow and conjugates the factors
carried by the arrows in the figure.

Finally, we note that to compute the full transforms, one cascades the
above component operations according to
(\ref{eqn:fullI}-\ref{eqn:prodJt0}) and uses the fact that, because
the scaling functions remain the same, $\Phi=I$ is still true.

  \subsection{Operators: $\cO$, $\cL$} \label{sec:homoops}

When working with {\em unrestricted} multiresolution analyses, the operation
of multiplying by matrices of the form
\begin{equation} \label{eqn:multimatel}
\cM_{\alpha\beta} \equiv \int d^dx\,b^*_\alpha(x) \hat M b_\beta(x), 
\end{equation}
where $\hat M$ is some integral-differential operator, is most easily
performed in the finest single-scale representation ($V_N$) where the
matrix elements are invariant under
translations of the lattice $C_N$ and thus represent a simple
convolution.  Substituting the construction formula
$$
b_\alpha(x) \equiv \sum_{p \in C_N} (\cI_{N:M})_{p\alpha} \phi\left(2^N(x-p)\right)
$$
into (\ref{eqn:multimatel}) gives the 
overlaps in the multiresolution representation as
\begin{equation} \label{eqn:singlescaledecomp}
\cM=\cI_{N:M}^\dagger M \cI_{N:M},
\end{equation}
where $M$ is the
single-scale overlap matrix $M$ on the finest scale defined in
(\ref{eqn:defM}) of Sec.~\ref{sec:semicbases:matels}.
This immediately gives an $O(n)$ method for applying $\cM$, where $n$
is the number of points on the finest grid $C_N$: transform to the
single-scale representation on level $N$, apply the convolution $M$,
and transform back to the multiresolution representation with
$\cI^\dagger$.  The main disadvantages of this approach, which is used
in \cite{chou,tymczak}, are that (a) it is not restrictable and (b) it
requires storage and processing of data on the unrestricted grid
$C_N$.

The non-standard multiply approach of Beylkin, Coifman and
Rokhlin\cite{BCR} provides one way of applying operators in
multiresolution bases with linear computational effort while
circumventing the need to work directly with the $V_N$ representation.
A more general approach\cite{JCP}, which includes the non-standard
multiply as one special case, is to follow the strategy which
underlies Greengard and Rokhlin's fast multipole method\cite{GR}.  The
idea is to eliminate the redundancies inherent in the matrix elements
of $\cM$ by lumping information together into coarse and fine
contributions.  Only, rather than using multipole moments to
summarize fine-scale information in an approximate way, one may
exploit the two-scale relation to gather together the fine-scale
overlaps {\em exactly}.  And, rather than using Taylor series to
summarize smooth information approximately, one may use the two-scale
relation to represent smooth-scale information {\em exactly} in terms
of coarse scaling functions.  The remaining information on proximate
scales then may be handled efficiently by direct, and thus also exact,
multiplication.

To define precisely the separation into ``fine,'' ``proximate,'' and ``smooth''
contributions, we first decompose the operator $\cM$ into contributions
affecting the results on the different scales $M \le Q \le N$ of the
multiresolution analysis,
\begin{equation}
\cM = (\cP_{C_M} + \ldots + \cP_{D_Q} + \ldots + \cP_{D_N}) \cM. \label{eqn:d1}
\end{equation}
For a given term specified by its scale $Q$, there are contributions
which originate from scales $R$ which are either ``proximate'' ($|R-Q|
< \ell$), ``finer'' ($R \ge Q+\ell$) or smoother ($R \le Q-\ell$),
where $\ell$ is a measure of scale separation.  Following this
classification, for a given $Q$, we partition the set of all points in
the multiresolution analysis as
\begin{equation} \label{eqn:part}
C_N=S_Q \cup P_Q \cup F_Q.
\end{equation}
Although the interpretation of $S_Q$, $P_Q$ and $F_Q$ is clear, the
precise nature of these sets of points involves a variety of minor
details as $Q$ approaches the coarse and fine limits of the
multiresolution analysis.  In particular, when $Q -\ell < M$, all of
the smoother grids (including $C_M$) are proximate and $S_Q$ is empty.
On the other hand, when $Q+\ell > N$, all of the finer contributions
are proximate and $F_Q$ is empty.  An additional notational
complication is the fact that the points on scale $Q=M$ are the
lattice $C_Q$ whereas the points on the finer scales $Q>M$ make up the
detail crystals $D_Q$.  To ease the burden of constantly making this
distinction, we define
$$
L_Q \equiv \left\{\begin{array}{cr}C_Q & \mbox{if $Q=M$} \\ D_Q &
\mbox{if $Q>M$}	  \end{array}\right.,
$$
as all points on the ``level'' $Q$.  With these considerations,
the precise definitions for $S_Q$, $P_Q$ and $F_Q$ are
\begin{eqnarray*}
S_Q & \equiv & \left\{
\begin{array}{lc}
\emptyset & \mbox{if $Q-\ell < M$}\\ C_{Q-\ell} & \mbox{if $Q-\ell \ge M$}
\end{array}\right.
\\
F_Q & \equiv &  \left\{
\begin{array}{lc}
\emptyset & \mbox{if $Q+\ell > N$}\\ 
B_{Q+\ell-1} & \mbox{if $Q+\ell \le N$}
\end{array}\right.
\\
P_Q & \equiv & C_N-S_Q-F_Q = \bigcup_{R=\max{M,Q-\ell}+1}^{\min{N,Q+\ell-1}} L_R,
\end{eqnarray*}
where $\emptyset$ is the empty set.
With these definitions, the full decomposition of $\cM$ is
\begin{equation} \label{eqn:spfdecomp} 
\cM  =  \sum_{Q=M}^N \cP_{L_Q} \cM (\cP_{S_Q}+\cP_{P_Q}+\cP_{F_Q}) 
\end{equation}
Our discussion for the implementation of these various terms will be
limited to the semicardinal case, which has been shown to be
restrictable\cite{JCP}, and so we return to the specialized notation
$\cI(x)$ for the scaling functions, in place of the generic $\phi(x)$.

To evaluate the proximate contributions we first define the {\em
inter-scale matrices} of overlaps connecting the scaling functions of
$V_P$ with the scaling functions of $V_R$,
\begin{equation} \label{eqn:id1}
(M_{P,R})_{pr}=\left\{\begin{array}{cl}
\int d^dx\,\,\cI^*(2^P(x-p)) \hat M \cI(2^R (x-r)) & \mbox{if $p
\in C_P$ and $r \in C_R$} \\
0 & \mbox{otherwise}
		      \end{array}\right.,
\end{equation}
which, following the same argument which led to
(\ref{eqn:singlescaledecomp}), take the form
\begin{equation} \label{def:Mms}
M_{P,R} \equiv \cP_{C_P} \cI_{N:P}^\dagger M \cI_{N:R} \cP_{C_R}.
\end{equation}
Note that the $M_{P,R}$ are invariant under the translations of the
lattice $C_{\min{P,R}}$ and thus may be implemented as convolutions.

The proximate contributions all involve terms of the form $\cP_{L_Q} \cM
\cP_{L_R} \equiv \cP_{L_Q} \cI^\dagger M \cI
\cP_{L_R}$.  To simplify this expression, we use the identity
\begin{equation} \label{eqn:detailisscale}
\cI_{N:M} \cP_{L_R} = \cI_{N:R} \cP_{L_R},
\end{equation}
which may be proven using (\ref{eqn:matMRA},\ref{eqn:decompIPP}).
Alternately, Eq.~(\ref{eqn:detailisscale}) may been seen immediately
as the algebraic statement that in a semicardinal multiresolution
analysis the basis functions associated with the points $p$ of scale
$R$, whose values are the columns picked out of $\cI_{N:M}$ by
$\cP_{L_R}$ on the left of (\ref{eqn:detailisscale}), are just the
scaling functions from the single-scale representation of scale $R$
associated with the same points $p$.  With (\ref{eqn:detailisscale}),
the proximate scale terms are just
\begin{eqnarray} 
\cP_{L_Q} \cM \cP_{L_R} & = &\cP_{L_Q} \cI_{N:M}^\dagger M \cI_{N:M}
\cP_{L_R} \label{eqn:comparable} \\
& = & \cP_{L_Q} \cP_{C_Q} \cI_{N:Q}^\dagger M \cI_{N:R} \cP_{C_R} \cP_{L_R}
\nonumber \\
& = & \cP_{L_Q} M_{Q,R} \cP_{L_R}, \nonumber
\end{eqnarray}
where on both sides of $M$ in the second line we have used
(\ref{eqn:detailisscale}) and the fact that always $L_R \subset C_R$
so that $\cP_{L_R} = \cP_{C_R}
\cP_{L_R}$.

\begin{figure}
\begin{center}
\scalebox{\figscale}{\scalebox{0.80}{\includegraphics{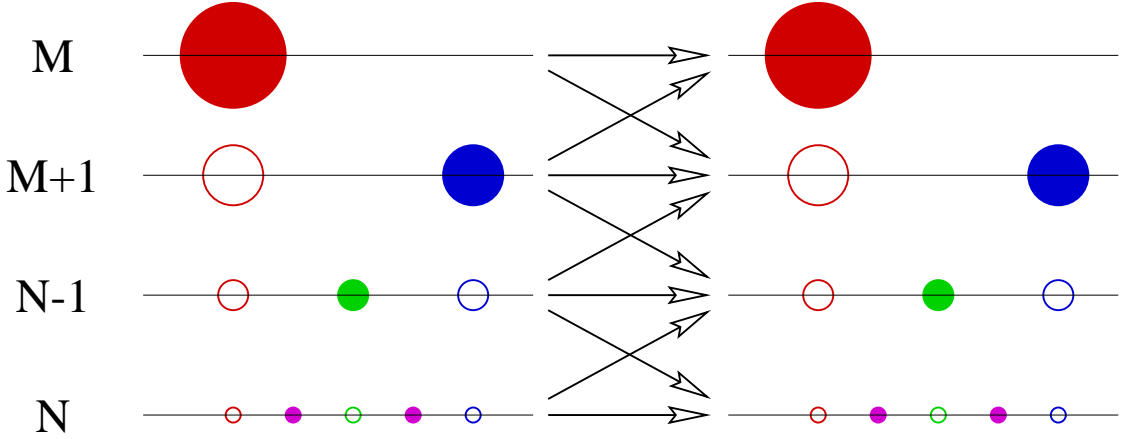}}}
\end{center}
\caption{Information flow for proximate contributions to $\cM$: inter-scale convolutions $M_{R,Q}$ (arrows).}
\label{fig:MP}
\end{figure}

Unlike the stages of the transforms, the inter-scale matrices here
used to apply operators do not convert one multiscale representation
into another but rather link together different single-scale
representations.  We thus introduce a second useful diagrammatic
representation for the coefficients of a multiresolution analysis.
The horizontal rows of Figure~\ref{fig:MP} represent the scaling
functions for each single-scale representation $V_Q$.  The basis for a
semicardinal multiresolution analysis consists of a subset of these
functions and includes for each point only the scaling function of the
coarsest scale containing that point.  These scaling functions are
indicated in the figure as filled circles.  In practical
implementations it is convenient to maintain data structures which
include coefficients for the redundant scaling functions from finer
scales, which are represented in the figure as open circles.  We refer
to this as the {\em redundant representation}.  This redundant
representation simplifies significantly the implementation of the
$M_{Q,R}$ as convolutions and represents only a small increase in
storage by a factor of $2^d/(2^d-1)$, which is just 8/7 in $d=3$
dimensions.  (See \cite{JCP}.)

The arrows in Figure~\ref{fig:MP} represent the calculation of the
proximate contributions according to
(\ref{eqn:spfdecomp},\ref{eqn:comparable}).  The calculation begins
with the coefficients of the vector to which the operator is applied
residing on the left of the diagram and with all coefficients residing
on the right being initialized to zero.  Then, each arrow pointing
from a level $R$ on the left to a level $Q$ on the right accumulates
onto level $Q$ the result of the convolution $M_{Q,R}$ applied to the
information present on level $R$.  Note that the factors $\cP_{L_R}$
in (\ref{eqn:comparable}) imply that the coefficients associated the
redundant functions on the left of the figure should be set to zero
before performing these convolutions.  After the accumulation of the
convolutions, the proximate contribution to $\cM$ appears on the
filled circles on the right side of the diagram.  As with all diagrams
in this section, the factors $\cP_{L_Q}$ in (\ref{eqn:comparable})
indicate that the final values computed on the empty circles are
extraneous and simply may be ignored, or not computed at all.

Next, we turn to the smooth contributions, which need only be computed
when $Q-\ell \ge M$.  The basic strategy for these terms is to expand
the smooth contributions exactly in terms of scaling functions on
scale $Q-\ell$.  To do this, note that with the expression
(\ref{eqn:singlescaledecomp}) for $\cM$ and the definition $S_Q \equiv
C_{Q-\ell}$, $\cP_{D_Q}
\cM \cP_{S_Q}$ ends in a product of the form $\cI_{N:M}
\cP_{S_Q} \equiv \cI_{N:M} \cP_{C_{Q-\ell}}$, which becomes
\begin{eqnarray} 
\cI_{N:M} \cP_{C_{Q-\ell}} & = & \cI_{N:{Q-\ell}} (\cP_{C_{Q-\ell}} \cI_{N:M}), \label{eqn:id1a} 
\end{eqnarray}
after some manipulation of (\ref{eqn:prodI}).  The result
(\ref{eqn:id1a} ) also may be seen immediately as the algebraic
statement that in a semicardinal multiresolution analysis the
multiscale basis functions from scale ${Q-\ell}$ or coarser, whose
values appear in $\cI_{N:M}
\cP_{C_{Q-\ell}}$, may be expanded exactly in the single-scale representation
on level ${Q-\ell}$ with coefficients equal to the values of these
basis functions at the corresponding points of $C_{Q-\ell}$.  Using
(\ref{eqn:detailisscale}) on the left, (\ref{eqn:id1a}) on the right,
and the result (\ref{def:Mms}), the smooth contributions become
\begin{eqnarray}
\cP_{L_Q} \cM \cP_{S_Q} & = & \cP_{L_Q} \cI_{N:Q}^\dagger M \cI_{N:Q-\ell}
\cP_{C_{Q-\ell}} \cI_{N:M} \label{eqn:coarser} \\
& = & \cP_{L_Q} M_{Q,Q-\ell} \cI_{N:M}. \nonumber
\end{eqnarray}

\begin{figure}
\begin{center}
\scalebox{\figscale}{\scalebox{0.80}{\includegraphics{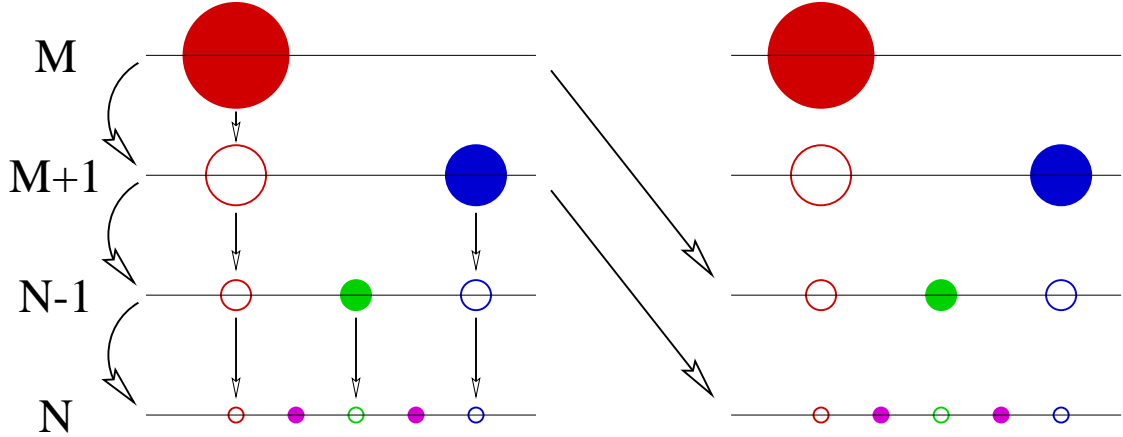}}}
\end{center}
\caption{Information flow for smooth contributions to $\cM$:
cascade algorithm for $\cI_{N:M}$ (curved arrows), access to data on
multiple scales (vertical arrows), inter-scale convolutions $M_{R,Q}$
(diagonal arrows).}
\label{fig:MS}
\end{figure}

Figure~\ref{fig:MS} illustrates the information flow for the smooth
contributions (\ref{eqn:coarser}).  First, the forward transform
$\cI_{N:M}$ is carried out using the cascade algorithm (\ref{alg:I}),
as represented by the curved arrows.  (In practice, the cascade need
not be carried out completely as the results on the finest few levels
will not be used.)  Next, the vertical arrows indicate that one must
ensure that the resulting coefficients are replicated on all finer
levels, for they will be accessed also from these levels by the
$M_{Q,Q-\ell}$.  The final step in computing the smooth contribution
is to perform the inter-scale convolutions $M_{Q,Q-\ell}$ represented
by the diagonal arrows as in Figure~\ref{fig:MP}.

The strategy for the finer contributions, present whenever $Q - \ell
\le N$, is to gather overlap information from finer scales in the form
of the matrix elements,
\begin{equation} \label{eqn:defcF}
\cF_Q \equiv \cP_{C_Q} \cM \cP_{F_Q}.
\end{equation}
Note that the finer scale contributions in (\ref{eqn:spfdecomp}) for any
scale $Q$ may be extracted from these objects as simply $\cP_{L_Q} \cF_Q$.
To gather the effects of finer scales on these matrix elements, we
use the decomposition $F_Q= D_{Q+\ell} \cup F_{Q+1}$ and make the
replacement $\cI_{N:Q} \cP_{C_Q} = \cI_{N:Q+1}
\cP_{C_{Q+1}} \cI_{Q+1,Q} \cP_{C_Q}$ to convert
(\ref{eqn:defcF}) into the recursion
\begin{eqnarray}
\cF_Q  & = & \cP_{C_Q} \cM \cP_{D_{Q+\ell}} + \cP_{C_Q} \cM \cP_{F_{Q+1}} \label{eqn:finerrecurse} \\
& = & M_{Q,Q+\ell} \cP_{D_{Q+\ell}} + \cP_{C_Q} \cI_{Q+1,Q}^\dagger
\cF_{Q+1}, \nonumber
\end{eqnarray}
which may be solved for $\cF_Q$ with $\cF_{N-\ell} \equiv \cP_{C_Q}
\cM \cP_{D_N}$ as the starting point of the recursion.

\begin{figure}
\begin{center}
\scalebox{\figscale}{\scalebox{0.80}{\includegraphics{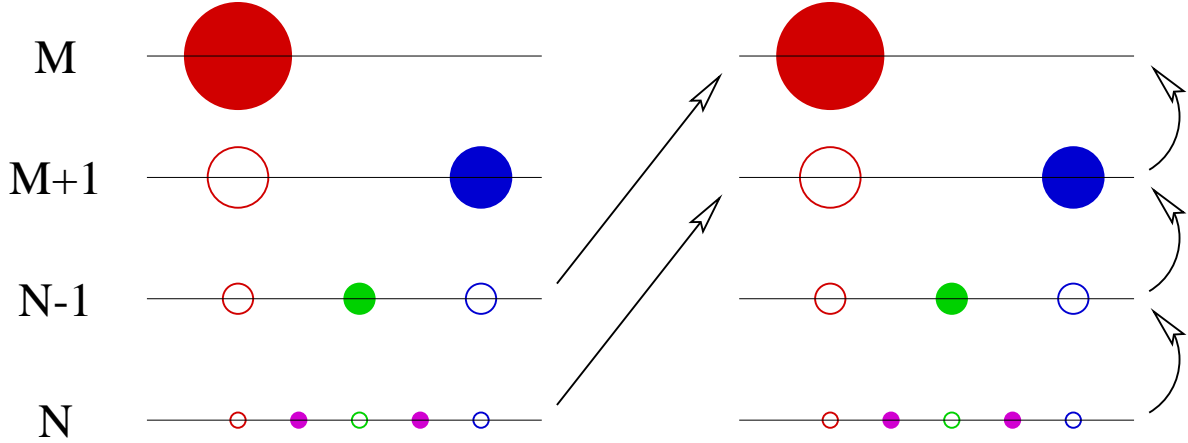}}}
\end{center}
\caption{Information flow for fine contributions to $\cM$:
inter-scale convolutions $M_{R,Q}$ (diagonal arrows),
$\cI_{P+1,P}^\dagger$ stages proceeding in sequence from finest to
coarsest (curved arrows).}
\label{fig:MF}
\end{figure}

Figure \ref{fig:MF} illustrates the information flow for this process.
The $M_{Q,Q+\ell}$ convolutions may be first computed and then
``folded-in'' as the recursion (\ref{eqn:finerrecurse}) proceeds.
These convolutions are indicated by the diagonal arrows.  As with the
proximate contributions, the projections $\cP_{D_{Q+\ell}}$ imply that
the redundant points associated with the open circles on the left
carry zero value for these convolutions.  Once the convolutions are
precomputed in this manner, the finer scale contributions are computed
from the recursive accumulation of the action of $\cI_{Q+1,Q}^\dagger$
from each scale onto the next coarser scale in sequence, from the
bottom to the top of the figure, as represented by the curved arrows.
In practice, the operations $\cI_{Q+1,Q}^\dagger$ on the finest few
scales work with data with zero value and so need not be computed.

The sum of the contributions illustrated in Figures
\ref{fig:MP}-\ref{fig:MF} gives the final result of the application
of the operator $\cM$.  The complete procedure for computing
$H = \cM G$ for any vector $G$ for the non-trivial cases $N-M
\ge \ell$ which involve more than just proximate
contributions is thus
\begin{equation} \label{alg:M}
\cF_{N-\ell} = M_{N-\ell,N} \cP_{D_N} G, \ g \equiv \cI G; 
\end{equation}
$$
\cF_Q= M_{Q,Q+\ell} \cP_{D_{Q+\ell}} G + \cP_{C_Q} \cI_{Q+1,Q}^\dagger
\cF_{Q+1} \mbox{\ \ \ (for $Q < N-\ell$)};
$$
\begin{eqnarray*}
H & = &
\sum_{Q=M+\ell}^N \cP_{L_Q} M_{Q,Q-\ell} \cdot g \\
& &+  \sum_{Q=M}^N \left( 
\sum_{R=\max{M,Q-\ell+1}}^{\min{N,Q+\ell-1}} \cP_{L_Q} M_{Q,R}
\cP_{L_R} \right) \\
& &+ \sum_{Q=M}^{N-\ell} \cP_{L_Q} \cF_Q.
\end{eqnarray*}
The non-standard multiply approach of Beylkin, Coifman and
Rokhlin\cite{BCR} corresponds to the special case $\ell=1$ case of
this approach.  For the restrictions usually encountered in atomic
calculations, the case $\ell=2$ has been shown to improve the speed of
calculations by at least a factor of four\cite{JCP,rossphd}.

\section{Impact of Restriction} \label{sec:inhomogridops}

We now consider how the prescriptions
(\ref{alg:I},\ref{alg:It},\ref{alg:J},\ref{alg:Jt},\ref{alg:M}) for
operations in semicardinal multiresolution analyses are affected when
the analysis is restricted.  We show that under relatively mild
conditions on how quickly the restriction changes resolutions, it is
possible to simply ignore the missing coefficients and yet obtain
exactly the results of an unrestricted multiresolution analysis
and thus achieve the ultimate goal of expending $O(1)$ operations per
expansion coefficient while, remarkably, introducing no
additional approximations beyond the choice of basis set.  These
issues were first explored in \cite{JCP}, which prompted the study and
development of the concept of ``S-trees'' to appear in a coming
work\cite{CohenDanchin:98}.  We now present the discussion of
\cite{JCP} using the explicit matrix language introduced in
Sec.~\ref{sec:matrep}.

We begin by introducing the notation $[*]$ as the restriction of the
scope of the object ``$*$'' to the functions and grid points {\em
surviving} the restriction of a multiresolution analysis.  The
restriction $[S]$ of a set of points $S$, for example, is defined as
the set of those points of $S$ associated with functions which survive
the restriction.  The set of all surviving points is the restriction
of the finest grid in the multiresolution analysis, $[C_N]$.  Thus, in
general, for any set of points $S$,
$$
[S] \equiv [C_N] \cap S,
$$
or in terms of projections,
\begin{equation} \label{eqn:algResProj}
\cP_{[S]} = \cP_{[C_N]} \cP_{S} = \cP_{S} \cP_{[C_N]}.
\end{equation}

The restriction of any linear operator $O$ is defined so that its
matrix elements ignore coefficients that have been restricted from the
multiresolution analysis.  Algebraically, this is achieved by
projection,
\begin{equation} \label{def:restrO}
[O] \equiv \cP_{[C_N]} O \cP_{[C_N]}.
\end{equation}
Consistent with this definition, the matrices $\cO$ and $L$ used in
Sec.~\ref{sec:physops} involve only overlaps between functions present
in the basis.

This simple prescription for the restriction of linear operators,
however, is not appropriate in general for the transforms, which
relate expansion coefficients on one side with function values at
points in real space on the other.  Even a smooth function will have
significant values in real space on the grid points associated with
extremely fine functions.  Thus, great care is needed to ensure that
the lack of access to this information does not adversely affect the
results of expressions involving the transforms.  For semicardinal
bases, as we shall see, very mild conditions on the restriction of the
multiresolution analysis can ensure that the unknown expansion
coefficients will have absolutely no impact on the function values
computed by the forward transform on the surviving points, so that
\begin{equation} \label{eqn:resI}
[\cI] \equiv \cP_{[C_N]} \cI \cP_{[C_N]} = \cP_{[C_N]} \cI,
\end{equation}
and, most critically, that the missing real-space information has
absolutely no impact on the expansion coefficients computed by the
inverse transform for the surviving functions,
\begin{equation} \label{eqn:resJ}
[\cJ] \equiv \cP_{[C_N]} \cJ \cP_{[C_N]} = \cP_{[C_N]} \cJ.
\end{equation}
Taken together, conditions (\ref{eqn:resI},\ref{eqn:resJ}) have the
remarkable consequence, to which we first alluded in
Sec.~\ref{sec:sysbasisappr}, that, even in a restricted basis, we may
compute the expansion coefficients for any non-linear combination of
fields, for example the exchange-correlation energy density
$\epsilon_{xc}(r)$, and obtain results identical to what would be
obtained by working with an {\em unrestricted} multiresolution
analysis of {\em arbitrarily} high resolution.

  \subsection{Transforms}

For the transforms, we begin by considering the implications of
condition (\ref{eqn:resJ}) for the allowable restrictions of
semicardinal multiresolution analyses.  The inverse transform in
semicardinal bases (\ref{eqn:Jdef}) has a particularly simple form
which is useful for understanding the impact of restriction.  Because
$\cP_{C_P} \cP_{D_{Q+1}} = 0$ for all $P<Q$, the fact that the product
(\ref{eqn:Jdef}) orders terms from coarsest to finest from left to
right, implies that all but the zero and first order terms in
$\cP_{D_{P+1}} \cI_{P+1,P} \cP_{C_P}$ vanish from the product, leaving
only the {\em sum}
$$ 
\cJ = I - \sum_{P=M}^{N-1} \cP_{D_{P+1}} \cI_{P+1,P} \cP_{C_P}.
$$
Inserting this into (\ref{eqn:resJ}) and using (\ref{eqn:algResProj})
gives
\begin{eqnarray*}
\cP_{[C_N]}-\sum_{P=M}^{N-1} \cP_{[D_{P+1}]} \cI_{P+1,P} \cP_{[C_P]} & = & \cP_{[C_N]} -\sum_{P=M}^{N-1}
\cP_{[D_{P+1}]} \cI_{P+1,P} \cP_{C_P}.
\end{eqnarray*}
Because the $[D_{P+1}]$ are disjoint, the sums must be equal term by
term in $\cP_{[D_{P+1}]}$, leading to the condition
\begin{equation} \label{cond:goodgrid}
\cP_{[D_{P+1}]} \cI_{P+1,P} (\cP_{C_P} - \cP_{[C_P]}) =
\cP_{[D_{P+1}]} \cI_{P+1,P} \cP_{C_P-[C_P]}=0. 
\end{equation}
Given that all the points in $[C_P]$ must appear in $[C_{P+1}]$,
the simplest geometric interpretation of (\ref{cond:goodgrid}) is
that no single-scale basis function associated with a point
dropped from scale $P$ may have a non-zero two-scale expansion coefficient
for any scaling function maintained in the basis on level $P+1$.  This
is referred to as the {\em good-grid condition}.

\begin{figure}
\begin{center}
\scalebox{\figscale}{\scalebox{1.0}{\includegraphics{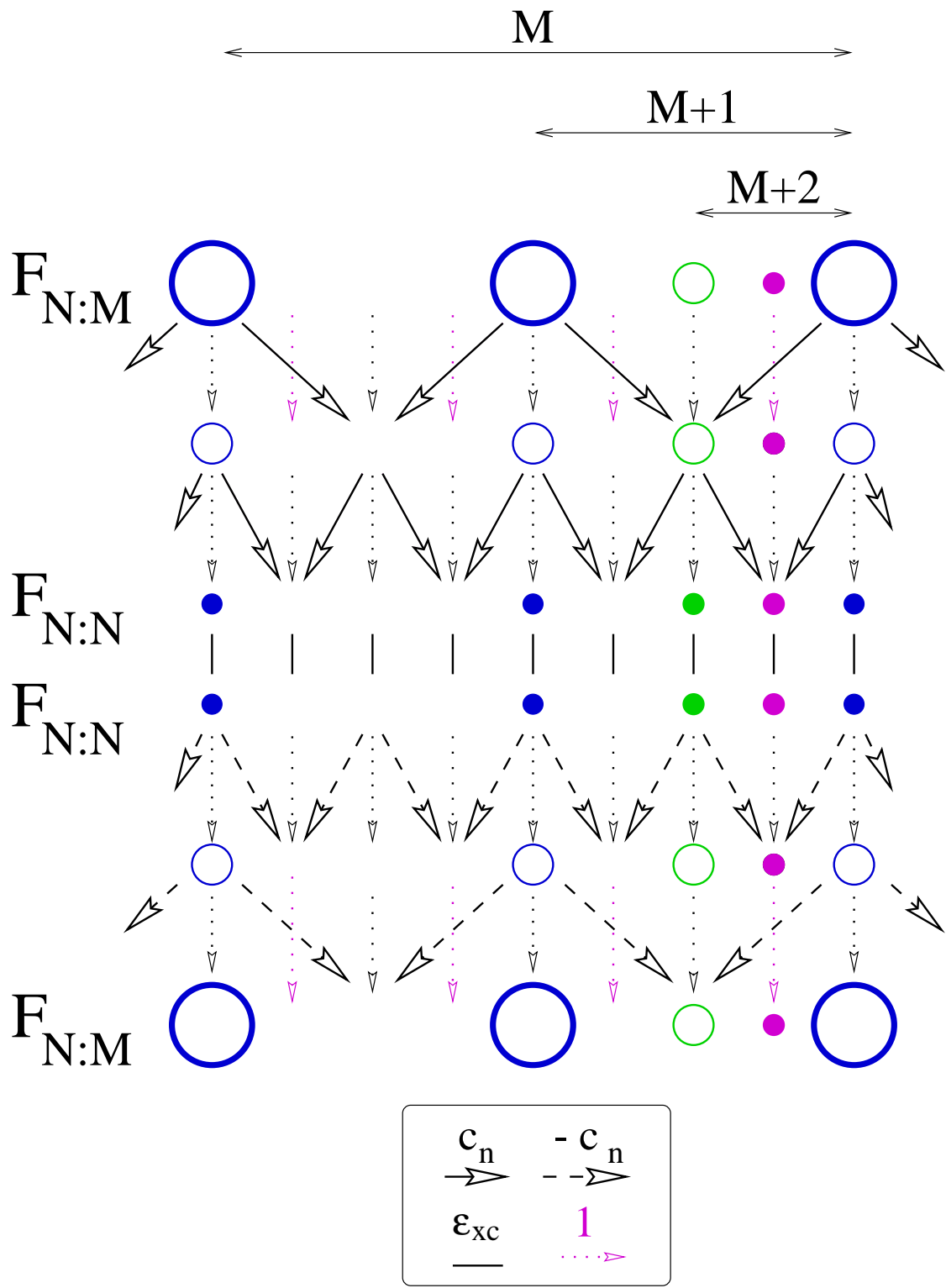}}} 
\end{center}
\caption{Information flow in the
calculation of expansion coefficients for $\epsilon_{xc}$ in a
restricted semicardinal multiresolution analysis on a good grid:
forward transform (upper half), non-linear local interaction
(vertical connections in center of figure), inverse transform (lower
half).}
\label{fig:rtrans}
\end{figure}

Figure~\ref{fig:rtrans} illustrates the effects of the good-grid
condition on the progress of a calculation.  The figure follows the
same conventions as the figures of Sec.~\ref{sec:xforms}, but now the
multiresolution analysis is restricted to provide the resolution of
$V_M$ in the left half of the figure, to make a brief transition
through $V_{M+1}$, and finally to attain the resolution of
$V_{M+2}=V_N$ on the right.  Although the figure shows the information
flow which would occur in the full multiresolution analysis, it shows
only those coefficients which survive the restriction.  Geometrically,
the condition (\ref{cond:goodgrid}) states that no arrow in the
forward transform (upper half of the figure) which originates from a
missing point may terminate on a surviving point.  This ensures that
the surviving coefficients computed at each stage of the restricted
transform are unaffected by the missing coefficients.  For
semicardinal multiresolution analyses, the pattern of the information
flow for the inverse transform (lower half of the figure) is the same
as for the forward transform.  Thus, the inverse too is unaffected by
ignoring coefficients on the restricted points.  Finally, the figure
illustrates the observation made above that so long as the good-grid
condition is maintained, the process of computing expansion
coefficients for the results of a local non-linear interaction
(represented by the vertical solid connections in the center of the
figure) such as $\epsilon_{xc}(\sum_i f |\psi_i(x)|^2)$ remains
unaffected by coefficients
restricted from the multiresolution analysis.

\begin{figure}
\begin{center}
\scalebox{\figscale}{\scalebox{.85}{\includegraphics{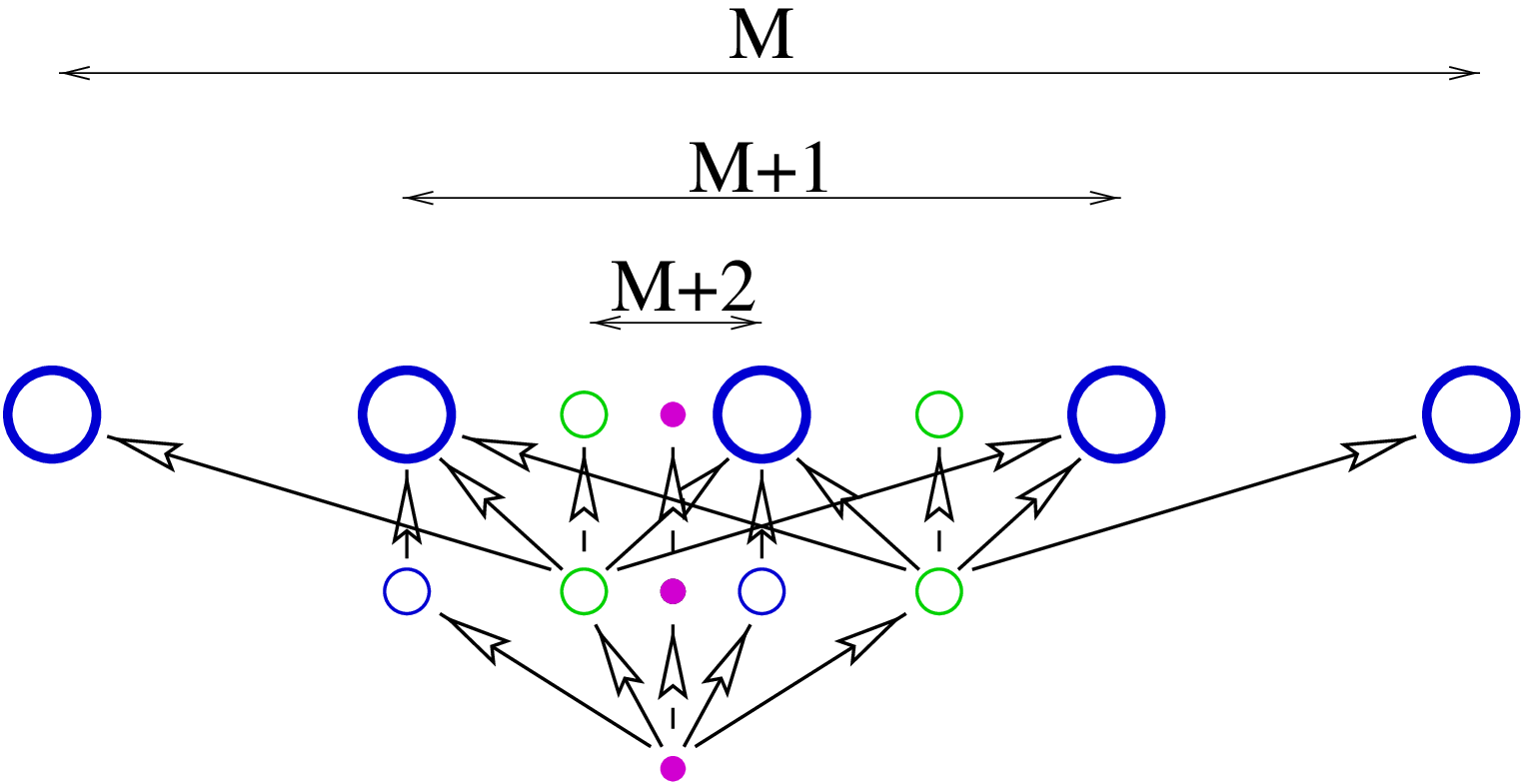}}} 
\end{center}
\caption{Implications of good grid condition on scales $(M:M+2)$
stemming from a single point (solid circle) on scale $M+2$ for
functions of support $\pm 2$: requirements of $\cP_{[D_{P+1}]}
\cI_{P+1,P} \cP_{C_P-[C_P]}=0$ (solid arrows), points already required
from lower levels (dashed arrows).}
\label{fig:goodgrid}
\end{figure}

Figure~\ref{fig:goodgrid} illustrates the constraints of the good grid
condition on allowable restrictions for functions of support $\pm 2$.
The presence of the one point on scale $M+2$ in the figure requires
the survival, in the restriction of the multiresolution analysis $V_M
\oplus W_{M+1} \oplus W_{M+2}$, of all of the other points in the
figure.  As the figure illustrates, the physical requirement of the
good-grid condition is just that the restriction pass through finite
regions representing each level of resolution in sequence and that it
not not skip discontinuously from level to level.  Because the
expansion coefficients of physical functions exhibit this behavior
naturally (Figure~\ref{fig:wfsI}), the good-grid condition generally
represents little or no additional burden in the restriction of the
basis.

To establish algebraically that on good grids the procedures
(\ref{alg:I},\ref{alg:J},\ref{alg:It},\ref{alg:Jt}) give correct
results when working only with data on the surviving points, we first
note that on a good grid the separate stages of the transforms
individually obey the restriction condition,
\begin{eqnarray} 
\left[ \cI_{P+1,P} \right] & = & \cP_{[C_N]} \cI_{P+1,P} \label{eqn:goodgridII} \\
\left[ \cI_{P+1,P}^{-1} \right] & = & \cP_{[C_N]} \cI_{P+1,P}^{-1},
\end{eqnarray}
as follows directly from the defining good-grid condition
(\ref{cond:goodgrid}) and the decompositions
(\ref{eqn:decompIPP},\ref{eqn:compJ}).  Next we note that working with
only data on the surviving points amounts to ignoring at each step of
a procedure the input from and output onto the restricted points.
Algebraically, this corresponds to replacing the factors in the
products making up the transforms
(\ref{eqn:prodI},\ref{eqn:Jdef},\ref{eqn:prodIt},\ref{eqn:prodJt})
with their restricted counterparts, $[\cI_{P+1,P}]$,
$[\cI_{P+1,P}^{-1}]$, $[\cI_{P+1,P}^\dagger]$,
$[\cI_{P+1,P}^{-\dagger}]$, respectively.

From these considerations, we see that the forward transform
(\ref{alg:I}) executed with only surviving coefficients gives
\begin{eqnarray}
\prod_{P=N-1}^M [\cI_{P+1,P}] 
& = & \cP_{[C_N]} \cI_{N,N-1} \cP_{[C_N]} \ldots \cP_{[C_N]} \cI_{P+2,P+1} \cP_{[C_N]} \cI_{P+1,P}\cP_{[C_N]} \nonumber\\
& = & \cP_{[C_N]} \cI_{N,N-1} \cP_{[C_N]} \ldots \cP_{[C_N]} \cI_{P+2,P+1} (\cP_{[C_N]} \cI_{P+1,P}\cP_{[C_N]}) \nonumber\\
& = & \cP_{[C_N]} \cI_{N,N-1} \cP_{[C_N]} \ldots (\cP_{[C_N]} \cI_{P+2,P+1} \cP_{[C_N]})\cI_{P+1,P} \nonumber\\
& = & \cP_{[C_N]} \cI, \label{eqn:rI}
\end{eqnarray}
where we have collapsed the product telescopically using
(\ref{eqn:goodgridII}).  Directly analogous considerations for the
inverse transform lead to
\begin{eqnarray}
\prod_{P=M}^{N-1} [\cI^{-1}_{P+1,P}] & = & \cP_{[C_N]} \cJ. \label{eqn:rJ}
\end{eqnarray}
Thus, we see that applying the algorithms (\ref{alg:I},\ref{alg:J}) to
only the surviving coefficients leads to results on the surviving
points which are {\em identical} to what would be obtained with the
unrestricted transforms $\cI$ and $\cJ$.  Note that because the
left-hand sides of (\ref{eqn:rI},\ref{eqn:rJ}) are unchanged by
post-multiplication by $\cP_{[C_N]}$, these results also confirm
directly that (\ref{eqn:resI},\ref{eqn:resJ}) obtain on good grids.

Finally, we note that the conjugate transforms defined in
Sec.~\ref{sec:KohnSham} for any basis are the Hermitian conjugates of
the associated forward and inverse transforms.  From
(\ref{eqn:resI},\ref{eqn:resJ}) and (\ref{eqn:rI},\ref{eqn:rJ}), these
are thus simply
$$
[\cI]^\dagger = \prod_{P=M}^{N-1}{[\cI_{P+1,P}^\dagger ]}
$$
$$
[\cJ]^\dagger = \prod_{P=N-1}^{M} [\cI^{-\dagger}_{P+1,P}].
$$
Composed entirely of restricted component operations, these expressions are
precisely what the algorithms (\ref{alg:It},\ref{alg:Jt}) compute when
working only with data on surviving points.

  \subsection{Operators}

The procedure (\ref{alg:M}) for applying physical operators divides
the contributions to $\cM$ into a sum of three classes of
contribution: smooth, proximate, and fine.  The proximate
contributions call upon $M_{Q,P}$ to collect data from functions
present in the basis onto coefficients present in the basis.  Thus,
regardless of the form of the restriction, the restriction of these
parts of $\cM$ amounts to ignoring coefficients restricted from the
basis, which is precisely what is accomplished when carrying out
(\ref{alg:M}) on data associated with only the surviving points.
Algebraically, this corresponds to the fact that restricting
(\ref{eqn:comparable}) leads to corresponding expressions with all
component operations replaced by their restricted counterparts.  The
proximate contributions in (\ref{alg:M}) are therefore always
restrictable.

\begin{figure}
\begin{center}
\scalebox{\figscale}{\scalebox{1.0}{\includegraphics{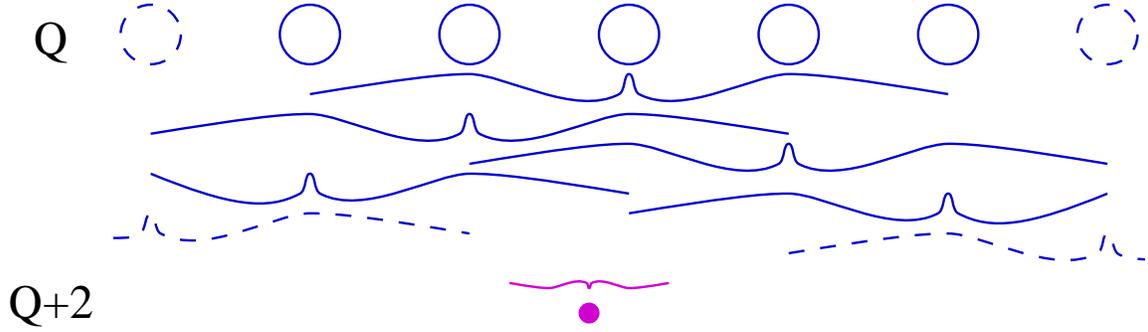}}} 
\end{center}
\caption{Requirements of the
grid-touching condition (\ref{eqn:1}) on scale $Q$ which stem from the
presence of a detail function (filled circle) on scale $Q+2$: required
points (solid circles), optional points (dashed circles), support of
functions (braces).  (Illustrated for the $\ell=2$-level algorithm.)}
\label{fig:blanket}
\end{figure}

The smooth contributions, on the other hand, do impose conditions on
the restriction.  The contributions (\ref{eqn:coarser}) connect 
the single-scale basis associated with $C_{Q-\ell}$
through the operator $\cM$ onto the multiscale basis functions
associated with the points of $L_Q$.  To ensure
that no information is lost when computing these contributions, the
restriction must be such that all scaling functions associated with
the lattice of the scale
$\ell$ levels coarser which overlap with, or {\em touch}, a function
in the basis must also appear in the basis,
\begin{equation}\label{eqn:touchgridI}
\cP_{[L_Q]} M_{Q,Q-\ell} \cP_{C_{Q-\ell}-[C_{Q-\ell}]}=0.
\end{equation}
Algebraically, this condition allows the transition
from the first to the second line in the derivation
\begin{eqnarray}
[\cP_{L_Q} M_{Q,{Q-\ell}} \cI_{N:M}] 
& = & \cP_{[L_Q]} M_{Q,{Q-\ell}} \cP_{C_{{Q-\ell}}} \cI_{N:M}
\cP_{[C_N]} \label{eqn:coarseworks}\\
& = & \cP_{[L_Q]} M_{Q,{Q-\ell}} \cP_{[C_{{Q-\ell}}]} \cI_{N:M}
\cP_{[C_N]} \nonumber \nonumber\\
& = & [\cP_{L_Q}] [M_{Q,{Q-\ell}}] [\cI_{N:M}], \nonumber
\end{eqnarray}
which establishes that the coarse contributions to $[\cM]$ may be
computed correctly while processing data on the surviving points
only.  Figure~\ref{fig:blanket} illustrates the requirements of the
this condition for functions of support $\pm 2$ when using
the $\ell=2$ version of the algorithm.

Finally, we turn to the finer contributions.  The first term in the
recursion (\ref{eqn:finerrecurse}) for these contributions is of the
same generally restrictable form as the proximate contributions and so
impose no condition on the restriction.  However, proper computation
of the second term of (\ref{eqn:finerrecurse}) requires a condition on
the grid which is slightly stronger than (\ref{eqn:touchgridI}).  Now
we must require that all scaling functions from the scales {\em at
least} $\ell$ levels coarser which touch a function in the basis must
also survive the restriction,
\begin{equation} \label{eqn:1}
\cP_{C_{Q}-[C_{Q}]} \cM \cP_{[F_{Q}]} = 0,
\end{equation}
which, from the definition (\ref{eqn:defcF}) of $\cF$, is equivalent
to $\cF_Q \cP_{[C_N]} = [\cF_{Q}]$.  Combining the $Q+1$ case of this
condition with the previously established good-grid condition
(\ref{cond:goodgrid}), we see that when computing on only surviving
coefficients, the second term in (\ref{eqn:finerrecurse}) gives
\begin{eqnarray}
[\cP_{C_Q}] [\cI_{Q+1,Q}^\dagger] [\cF_{Q+1}]
& = & \cP_{[C_Q]} \cdot \cI_{Q+1,Q}^\dagger \cP_{[C_{N}]} \cdot [\cF_{Q+1}]\label{eqn:fineworks}\\
& = & \cP_{[C_Q]} \cI_{Q+1,Q}^\dagger  [\cF_{Q+1}]\nonumber\\
& = & \cP_{[C_Q]} \cI_{Q+1,Q}^\dagger \cF_{Q+1} \cP_{[C_N]}\nonumber\\
& = & [\cP_{C_Q} \cI_{Q+1,Q}^\dagger  \cF_{Q+1}],\nonumber
\end{eqnarray}
precisely the correct restricted result.

We refer to condition (\ref{eqn:1}), which contains
(\ref{eqn:touchgridI}) as a special case, as the {$\ell$-level
grid-touching condition}.  Combined with the good-grid condition
(\ref{cond:goodgrid}), this condition ensures that the procedure given
in Sec.~\ref{sec:homoops} gives correct results for
integral-differential operators even in restricted multiresolution
analyses.  Note that the $\ell$-level grid-touching condition requires
that the the situation illustrated in Figure~\ref{fig:blanket} holds
not only for the scaling functions on scale $Q$ but also for all
coarser scales $R<Q$ as well.

As the scale separation $\ell$ increases, the detail function on the
lower scales appear relatively smaller, and the
grid-touching condition becomes weaker.  In the limiting case of very
large $\ell$, the support of the detail functions appears like a
single point and the touching condition approaches the good-grid
condition.  For low $\ell$, the differences in the conditions on the
restriction can have important consequences.  For example, going from
$\ell=1$, which corresponds to the non-standard multiply of
\cite{BCR}, to $\ell=2$ allows for quicker changes in resolution that
have been shown to accelerate calculations by over a factor of four
for the restrictions typically encountered in atomic
calculations\cite{JCP}.

As a closing note, the alternative restriction conditions
\begin{eqnarray*}
\cP_{C_{{Q-\ell}}-[C_{{Q-\ell}}]} \cI_{N:M} \cP_{[C_N]} & = & 0\\
\cP_{[C_Q]} \cI_{Q+1,Q}^\dagger \cP_{C_{Q+1}-[C_{Q+1}]} & = & 0
\end{eqnarray*}
would also allow for proper computation of the smooth and fine
contributions in (\ref{eqn:coarseworks},\ref{eqn:fineworks}),
respectively.  These conditions, however, are ``finer-looking'' in the
sense that each function in the basis requires the presence of
functions on even finer scales.  This leads to an infinite chain of
conditions that ultimately requires the multiresolution analysis to be
of infinite resolution.  As a result, the $\ell$-level touching
condition is the appropriate condition to impose in real computations.
\section{Concluding Remarks} \label{sec:concs}

Mallat and Meyer's development of multiresolution analysis provides a
tool not only of great use in functional analysis and signal
processing but also of great potential for the study of physical
systems exhibiting behavior on multiple length scales.  In particular,
multiresolution analysis provides the first practical possibility for
a unified, systematic treatment of core and valence behavior in the
electronic structure of molecular and condensed-matter systems.  The
first all-electron density-functional calculations of atoms and
molecules carried out with this approach (\cite{mgras,aps},
Sec.~\ref{sec:intro}.) have demonstrated that multiresolution analysis
provides an extremely efficient means of representing the core and
valence electrons simulataneously.  Latter work underscores the
promising possibilities of combining multiresolution analysis with
pseudopotential theory\cite{chou}, developing dynamic restriction
schemes\cite{tymczak}, and using lifted bases for some phases of the
calculations\cite{goedecker}.  Finally, the recent development of the
theory of fast restrictable algorithms for semicardinal bases
(\cite{JCP}, Sec.~\ref{sec:inhomogridops}) now paves the way for the
first application of multiresolution analysis to the calculation of
the electronic structure of large, complex systems.

In closing, the author would like to leave the reader with the
understanding that many opportunities remain for making significant
contributions to the field of the multiresolution analysis of electronic
structure.  The basic groundwork is now in place, but many interesting
and important open questions remain.  For example, the differing
stiffnesses of the core and valence degrees of freedom clearly call
for some new form of preconditioning.  The significant expense of solving
Poisson's equation at each electronic iteration indicates that 
methods which search directly for the saddle point of the LDA
Lagrangian will be more efficient than present approaches.  Also,
procedures are needed to anticipate accurately the changes in the
expansion coefficients of the wave functions as the nuclei move.  Finally,
there is the intriguing possibility that the weakness of the coupling
between the core and valence degrees of freedom will allow the atomic
cores to be treated independently and in parallel.

\def\thesection{}
\section{Acknowledgments}

This work was supported in part by the MRSEC Program of the National
Science Foundation (DMR 94-00334) and the Alfred~P.~Sloan Foundation
(BR-3456).  The author is greatly indebted to M.P.~Teter for his
insights into the utility of cardinal functions, which eventually led
to the concept of semicardinality, to S.~Ismail-Beigi for his
recognition of the restrictability of semicardinal transforms and his
most careful reading of the final manuscript, to Dicle Ye\c{s}illeten
for our first implementation and verification of the restrictable
overlap operators, to T.D.~Engeness for his reading of the earlier
drafts of this work, and to all the others on whose constant
intercession the author has relied for help during the
preparation of this review.

\appendix
\section{Cofactors of the semicardinal two-scale decomposition matrix}

In this appendix, we determine the cofactors of the matrix
$$
P_{ij}\equiv (-1)^{\eta_i \cdot \eta_j},
$$
where, as defined in the text, the $\eta_i$ range over the set
$\{0,1\}^d$.  

The central result we use to determine these cofactors is that
$P^2=2^d I$, which we show by explicit computation,
\begin{eqnarray*}
(P^2)_{ij} & = & \sum_k P_{ik} P_{kj} \\
& = & \sum_{\{\eta_k\}}  \prod_{e=1}^d (-1)^{ (\eta_i)_e (\eta_k)_e +
(\eta_k)_e (\eta_j)_e } \\
& = & \prod_{e=1}^d\  \sum_{\{{(\eta_k)}_e=0,1\}} (-1)^{ {(\eta_k)}_e (
{(\eta_i)}_e + {(\eta_j)}_e ) } \\
& = & \prod_{e=1}^d \left( 1+(-1)^{ {(\eta_i)}_e + {(\eta_j)}_e  }
\right) \\
& = & \prod_{e=1}^d \left( 2 \delta_{{(\eta_i)}_e,{(\eta_j)}_e  } \right)
\\
& = & 2^d \delta_{ij}.
\end{eqnarray*}

With $P^2=2^d I$ established, we have $P^{-1} =  P/2^d =
(\cof P)^T/{\det P}$, where $\cof P$ is the matrix of the cofactors of
the matrix $P$.  To determine $\det P$, we note  that
$\det P^2 = {(2^d)}^{2^d}$, so $|\det P|={(2^d)}^{2^{d-1}}$.  Thus, 
$$
\cof P=(\det P) (P^{-T}) = \left( \pm (2^d)^{2^{d-1}} \right) \left(
\frac{1}{2^d} P^T\right) = \pm (2^d)^{2^{d-1}-1} P^T.
$$
Because all entries in the top row of $P$ are unity,
the elements of the first column of the cofactor matrix, needed at the
end of Sec.~\ref{subsubsec:semicmra}, all have a single constant value, $\pm
(2^d)^{2^{d-1}-1}$.
\bibliography{biblio}\bibliographystyle{alpha}\end{document}